\documentclass{aa}  
\usepackage{graphicx}
\usepackage{siunitx}
\usepackage{txfonts}
\usepackage{subfig}

\usepackage{hyperref}

\hypersetup{breaklinks=true, colorlinks=true, urlcolor=blue, linkcolor=blue, citecolor=blue}

\newcommand {\asp}{\mbox{$.\!\!^{\prime\prime}$}}

\usepackage{color}

\begin{document} 

   \title{ISPY - NACO Imaging Survey for Planets around Young stars\thanks{ESO program IDs \mbox{096.C-0679}, \mbox{097.C-0206}, \mbox{198.C-612}, and \mbox{199.C-0065}.}}

   \subtitle{Survey description and results from the first 2.5 years of observations}

   \author{R. Launhardt \inst{1}
       \and Th. Henning \inst{1}
       \and A. Quirrenbach \inst{2}
       \and D. S\'egransan \inst{3}
       \and H. Avenhaus \inst{4,1}
       \and R. van Boekel \inst{1}
       \and S.\,S. Brems \inst{2}
       \and A.\,C. Cheetham \inst{1,3}
       \and G. Cugno \inst{4}
       \and J. Girard \inst{5}
       \and N. Godoy \inst{8,11}
       \and G.\,M. Kennedy \inst{6}
       \and A.-L. Maire \inst{1,10}
       \and S. Metchev \inst{7}
       \and A. M\"uller \inst{1}
       \and A. Musso Barcucci \inst{1}
       \and J. Olofsson \inst{8,11}
       \and F. Pepe \inst{3}
       \and S.\,P. Quanz \inst{4}
       \and D. Queloz \inst{9}
       \and S. Reffert \inst{2}
       \and E.\,L. Rickman \inst{3}
       \and H.\,L. Ruh \inst{2}
       \and M. Samland \inst{1,12}}

   \institute{Max-Planck-Institut f\"ur Astronomie, K\"onigstuhl 17, 69117 Heidelberg, Germany
     \and
    Landessternwarte, Zentrum für Astronomie der Universit\"at Heidelberg, K\"onigstuhl 12, 69117 Heidelberg, Germany
     \and
    Observatoire Astronomique de l'Universit\'e de Gen\`eve, 51 Ch. des Maillettes, 1290 Versoix, Switzerland
     \and
    ETH Z\"urich, Institute for Particle Physics and Astrophysics, Wolfgang-Pauli-Str. 27, 8093 Z\"urich, Switzerland
     \and
     Space Telescope Science Institute, Baltimore 21218, MD, USA 
        \and
    Department of Physics \& Centre for Exoplanets and Habitability, University of Warwick, Coventry, UK
     \and
    The University of Western Ontario, Dept. of Physics and Astronomy, 1151 Richmond Avenue, London, ON N6A 3K7, Canada
     \and
    Instituto de Física y Astronomía, Facultad de Ciencias, Universidad de Valparaíso, Av. Gran Bretaña 1111, Playa Ancha, Valparaíso, Chile
     \and
    Cavendish Laboratory, J J Thomson Avenue, Cambridge, CB3 0HE, UK
     \and
    STAR Institute, University of Li\`ege, All\'ee du Six Août 19c, 4000, Li\`ege, Belgium
     \and
    N\'ucleo Milenio Formaci\'on Planetaria - NPF, Universidad de Valpara\'iso, Av. Gran Breta\~na 1111, Valpara\'iso, Chile
     \and
    Department of Astronomy, Stockholm University, Stockholm, Sweden
             }

   \date{Received 28 October 2019 / 2 February 2020}

 
  \abstract
   {The occurrence rate of long-period ($a\gtrsim50$\,au) giant planets around young stars is highly uncertain since it is not only governed by the protoplanetary disc structure and planet formation process, but also reflects both dynamical re-structuring processes after planet formation as well as possible capture of planets not formed in-situ. Direct imaging is currently the only feasible method to detect such wide-orbit planets and constrain their occurrence rate.}
   {We aim to detect and characterise wide-orbit giant planets during and shortly after their formation phase within protoplanetary and debris discs around nearby young stars.} 
   {We carry out a large $L^{\prime}$-band high-contrast direct imaging survey for giant planets around 200 young stars with protoplanetary or debris discs using the NACO instrument at the ESO Very Large Telescope on Cerro Paranal in Chile. We use very deep angular differential imaging observations with typically $>$\,60\degr\ field rotation, and employ a vector vortex coronagraph where feasible to achieve the best possible point source sensitivity down to an inner working angle of about 100\,mas. This paper introduces the NACO Imaging Survey for Planets around Young stars ("NACO-{\it ISPY}"), its goals and strategy, the target list, and data reduction scheme, and presents preliminary results from the first 2.5 survey years.}
   {We achieve a mean $5\,\sigma$\ contrast of \mbox{$\Delta L^{\prime} = 6.4\pm0.1$\,mag} at 150\,mas and a background limit of \mbox{$L^{\prime}_{\rm bg} = 16.5\pm0.2$\,mag} at >1\farcs5. Our detection probability is >50\% for companions with $\gtrsim$\,8\,M$_{\rm Jup}$\ at semi-major axes of  \mbox{80\,--\,200\,au} and \mbox{>13\,M$_{\rm Jup}$} at \mbox{30\,--\,250\,au}. It thus compares well to the detection space of other state-of-the-art high-contrast imaging surveys. We have already contributed to the characterisation of two new planets originally discovered by VLT/SPHERE, but we have not yet independently discovered new planets around any of our target stars. We have discovered two new close-in low-mass stellar companions around R\,CrA and HD\,193571 and report in this paper the discovery of close co-moving low-mass stellar companions around HD\,72660 and HD\,92536. Furthermore, we report $L^{\prime}$-band scattered light images of the discs around eleven stars, six of which have never been imaged at $L^{\prime}$-band  before.}
   {The first 2.5\,years of the NACO-{\it ISPY} survey have already demonstrated that VLT/NACO combined with our survey strategy can achieve the anticipated sensitivity to detect giant planets and reveal new close stellar companions around our target stars.} 


   \maketitle


\section{Introduction} \label{sec:intro}

We are currently experiencing a golden era of exoplanet research that has led us from the first discovery of a planet orbiting another Sun-like star \mbox{\citep{MQ1995}}\footnote{The first confirmed discovery of an extrasolar planet was actually published by \mbox{\citet{wolszczan1992}}, but this planet orbits a neutron star.} to the realisation that planetary systems are a natural by-product of star formation. Indeed, our immediate Galactic environment is richly populated with stars that harbour planetary systems. We are already aware of many stars with known exoplanets within only a few parsecs around the Sun, such as for example, Proxima Centauri at 1.3\,pc \mbox{\citep{anglada2016}}, Barnard's Star at 1.8\,pc \mbox{\citep{ribas2018}}, Teegarden's Star at 3.8\,pc \mbox{\citep{zechmeister2019}}, Fomalhaut at 7.7\,pc \mbox{\citep{kalas2008}}\footnote{The nature of the companion candidate to Formalhaut is still debated \mbox{\citep{janson2012}}.}, Trappist-1 at 12\,pc \mbox{\citep{gillon2017}}, or $\beta$\,Pictoris at 19\,pc \mbox{\citep{lagrange2010,lagrange2019}}. Most stars in Sun-like environments\footnote{We still know very little about planets around stars in very different environments, such as for example close to the Galactic centre or in very low-metallicity environments.} seem to host planets. We have also learnt that the variety of planets and the architecture of planetary systems can be very different from what we know from our own Solar System. However, significant gaps remain in our knowledge of the occurrence rate and architecture of planetary systems in general and of the origin and evolution of our own Solar System in particular.

\begin{figure}[htb]
 \centering
 \includegraphics[width=9cm]{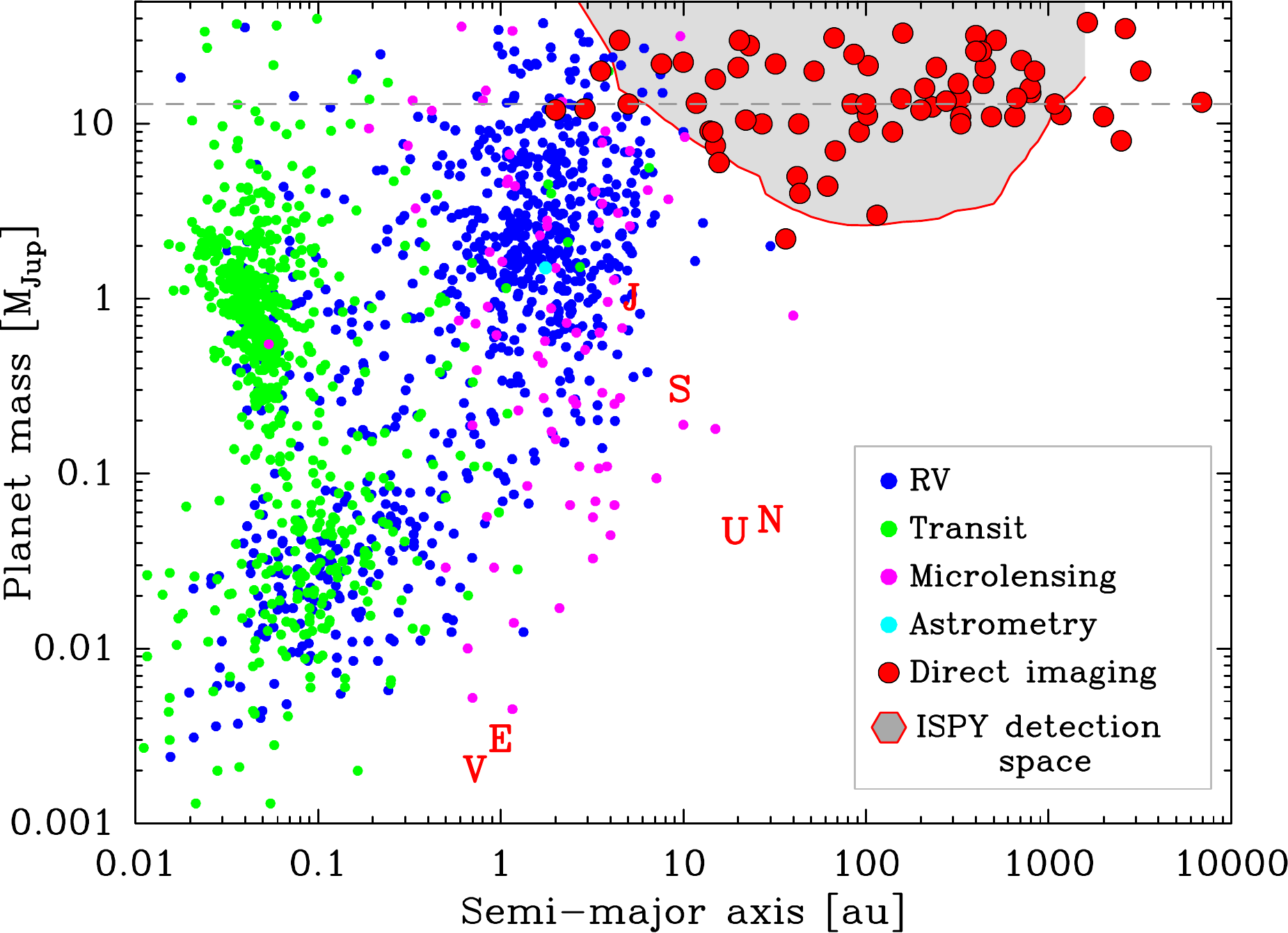}
 \caption{\label{fig:fig1}
  Distribution of planet mass vs. orbital separation of known exoplanets and candidates as listed on exoplanet.eu \mbox{\citep[][]{schneider2011}}. 
  The main detection methods are marked by different colours.
  Solar system planets are marked by red letters.
  The dashed horizontal line marks the approximate deuterium burning mass limit. 
  The grey-shaded area marks the parameter space probed by our NACO-{\it ISPY} survey (10\% detection probability, cf. Fig.\,\ref{fig-detspace}).}
\end{figure}

The most successful exoplanet discovery techniques thus far are the radial velocity (RV) and transit methods (see also Fig.\,\ref{fig:fig1}). Although very successful in terms of discoveries and providing statistics, these methods nevertheless have non-negligible detection biases and limitations meaning that important questions remain unanswered. Both methods are intrinsically biased toward short-period planets and both usually avoid young stars because their intrinsic photospheric and chromospheric activity can mask the subtle signals induced by the presence of a planet (\mbox{\citealt{saar1997}}, \mbox{\citealt{schrijver2008}}, \mbox{\citealt{barnes2017}}). Astrometry on the other hand, and in particular with the anticipated final release of the {\it Gaia} data (individual measurements), will provide us with new discoveries that are expected to reveal the still incompletely known giant planet (GP) population in the 3\,--\,5\,au separation range \mbox{\citep[][]{casertano2008,perryman2014}} in the near future.

However, we still know very little about the frequency of rocky planets in the habitable zone or the occurrence rate of GPs at orbital separations beyond 5\,au. Both are important pieces of the puzzle needed for constraining the uncertain ends in planet formation and evolution models (e.g. \mbox{\citealt{baraffe2003}}, \mbox{\citealt{fortney2008}}, \mbox{\citealt{sb2012}}, \mbox{\citealt{mordasini2012}}, \mbox{\citealt{allard2013}}).
The occurrence rate of long-period GPs is only poorly constrained since it is not only governed by the structure of the protoplanetary discs in which they form and by the physics of the planet formation process, that is, classical core accretion \mbox{\citep{pollack1996,idalin2004}}, gravitational instability \mbox{\citep{boss1997}}, or pebble accretion (\mbox{\citealt{johansen2010}}, \mbox{\citealt{ormel2010}}, \mbox{\citealt{bitsch2015}}), but also reflects dynamical re-structuring processes taking place well after planet formation. Both migration processes during the gas-rich protoplanetary disc phase and dynamical interactions between planets well after the clearing of the disc can change the architecture of planetary systems dramatically between birth and maturity \mbox{\citep[e.g.][]{davies2014,morbidelli2018}}. Furthermore, an unknown fraction of the giant planet population  on wide orbits may actually not originate from the system they are in now, but could have been captured free-floating in space, for example in dense star clusters \mbox{\citep{perets2012}}.

Therefore, if we want to understand how mature planetary systems like our own are formed, there is an explicit scientific demand to find and characterise GPs
in wide orbits around young stars, which cannot be satisfied by the successful indirect detection techniques mentioned above. Because of the long orbital timescales involved, the only currently feasible way to explore the GP population in wide orbits is by direct imaging (DI).

Direct imaging with the currently most advanced instruments allows us to probe exoplanets at separations from the host star down to a few astronomical units (1\,--\,10\,au, depending on distance and other parameters) and planet masses down to about 1\,M$_{\rm Jup}$\ (see Fig\,\ref{fig:fig1}). In terms of separations, DI is thus truly complementary to the RV and transit methods. However, DI is currently only sensitive to gaseous, self-luminous (i.e. young) GPs, while smaller (and rocky) planets are too faint to be detectable with current instrumentation in the glare of their host stars. 

Nevertheless, DI with high-contrast techniques has already provided us with some spectacular discoveries (e.g. \mbox{\citealt{marois2008,marois2010}}; \mbox{\citealt{lagrange2010}}; \mbox{\citealt{rameau2013c}}; \mbox{\citealt{kuzuhara2013}}; \mbox{\citealt{macintosh2015}}; \mbox{\citealt{keppler2018}}).
Moreover, the recent advancement of very large high-contrast DI planet surveys with several hundred target stars (see Table\,\ref{tab:lssurveylist}) 
has provided us with first constraints on the occurrence rate and distribution of massive (1-13\,M$_\mathrm{Jup}$) GPs in wide orbits (20-300\,au) around young stars. The derived values for the substellar companion frequency still show a large scatter as they range from $<1$\% up to 10\%, depending on sample parameters and models used 
(e.g. \mbox{\citealt{lafreniere2007}}; \mbox{\citealt{biller2013}}; \mbox{\citealt{brandt2014b}}; \mbox{\citealt{chauvin2015}}; \mbox{\citealt{nielsen2016,nielsen2019}}; \mbox{\citealt{bowler2016}}; \mbox{\citealt{galicher2016}}; \mbox{\citealt{vigan2017}}; \mbox{\citealt{meshkat2017}}; \mbox{\citealt{stone2018}}). Overall, these results show that such wide-orbit GPs are rare. 

\begin{figure}[htb]
 \centering
 \includegraphics[width=8.5cm]{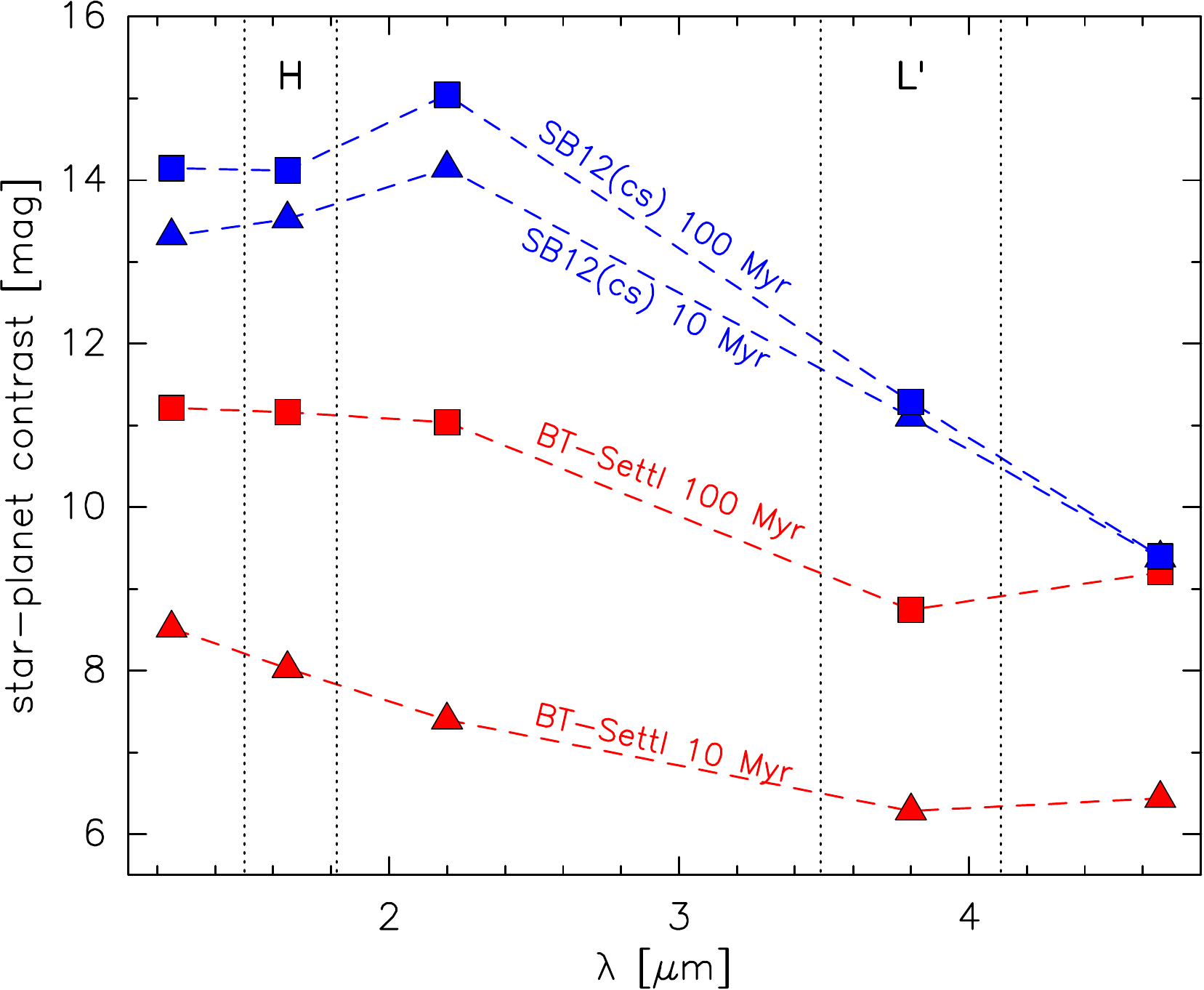}
 \caption{\label{fig:fig2}
  Star-planet contrast for a 1\,M$_{\odot}$\ star and a 10\,M$_\mathrm{Jup}$\ GP as a function of wavelength for two different ages (10 and 100\,Myr) and two evolutionary models (blue: \mbox{\citet{sb2012}}, cold start; red: \mbox{\citet{allard2014}}, hot start), showing the 2\,--\,3\,mag contrast advantage of the $L^{\prime}$-band vs. $H$-band which is used for most other large DI surveys.}
\end{figure}

However, most of these surveys operate at wavelengths $\lambda\le$\,\SI[]{2.2}{\micro\meter} and are thus optimised for hot (young and massive) planets. 
Very deep observations at longer wavelengths are needed to also reveal both very young and still embedded (i.e. reddened) planets and protoplanets as well as more evolved (i.e. cooler) planets around somewhat older (few hundred\,Myr) stars (Fig.\,\ref{fig:fig2}). Furthermore, it is also important to probe physical separations down to 5\,--\,10\,au to bridge the gap between the RV-constrained GP population around mature stars and the very wide-orbit GP population around very young stars revealed by these DI surveys.

To address these needs, we initiated a large ($\approx$\,200 stars) $L^{\prime}$-band Imaging Survey for Planets around Young stars with NACO (see Sect.\,\ref{sec:obs}) at the ESO Very Large Telescope (VLT) on Cerro Paranal in Chile ("NACO-{\it ISPY}"). The survey was launched in December 2015, carried out as guaranteed time observations (GTO) with a total budget of 120 observing nights, and was completed after 4\,years in late 2019. The NACO-{\it ISPY} survey is complementary to the largest and most recent other $L^{\prime}$-band imaging survey, the Large Binocular Telescope (LBT) Interferometer Exozodi Exoplanet Common Hunt ({\it LEECH}; \mbox{\citealt{skemer2014}}; \mbox{\citealt{stone2018}}), because it covers the southern sky as opposed to the northern sky accessible from the LBT located in Arizona. 
It is also complementary to most of the other large imaging surveys listed in Table\,\ref{tab:lssurveylist} because it is carried out in the $L^{\prime}$-band (\SI[]{3.8}{\micro\meter}) as opposed to the most widely used $H$- and $K$-bands. Last but not least, our survey is also different from most of the other surveys in that we focus exclusively on nearby stars that are surrounded by either protoplanetary transition discs (PPDs) or debris discs (DEBs).

In this paper, we give a general overview of the NACO-{\it ISPY} survey, its goals, strategy and targets, and present preliminary results from the first 2.5 survey years.
In Section\,\ref{sec:goals} we describe the scientific goals and strategy of the survey.
Section\,\ref{sec:targets} gives an overview of the target selection, the master target sample, and the list and properties of targets observed during ESO periods 96 through 100. Observations and data reduction are described in Sections\,\ref{sec:obs} and \ref{sec:red}.
The first survey results are presented in Section\,\ref{sec:res} and discussed in Section\,\ref{sec:dis}.
Section\,\ref{sec:con} summarises and concludes the paper.


\section{Survey goals and strategy} \label{sec:goals}

Our NACO-{\it ISPY} survey is tailored to characterise the wide-separation GP population around nearby young stars during and shortly after the planet formation phase. 
We exploit the $L^{\prime}$-band capability of NACO (see Sect.\,\ref{sec:obs}) and focus on young stars, thus making NACO-{\it ISPY} complementary to most other current large DI surveys which mostly employ the $H$-band.
In particular, we address the following main scientific questions:
\begin{enumerate}

\item Can we detect and characterise GPs during their formation phase within PPDs, where indirect methods for shorter-period planets fail?
\item What are the properties (luminosity and temperature), location (separation), and occurrence rate of such planets?
\item What is the occurrence rate and luminosity\,and/or\,separation distribution of long-period GPs in DEBs?
\item What is the relation between DEB properties and the presence of wide-separation GPs? Can we constrain stirring models and establish disc properties as planet tracers (see below)?
\item How frequent are GPs at separations $\ge5$\,au around young ($\lesssim100$\,Myr) solar-type stars, i.e. closer to what has already been partially constrained by previous imaging surveys?
\item Does the distribution change with age and at what age do the most significant changes occur?
\item Are the occurrence and properties of wide-orbit GPs correlated with shorter-period GPs such that in-situ formation and outward scattering can be disentangled?
\end{enumerate}

Consequently, our survey is focused on only two types of nearby young stars with circumstellar discs:  {\it (a)} young ($<$10\,Myr) stars with gas-rich PPDs with direct or indirect evidence for gaps that could be carved by (proto)-planets (transition discs), and {\it (b)} somewhat older (up to a few\,hundred\,Myr) stars with well-characterised DEBs, where the primordial dust has been processed completely, and the current dust population results from the collisional grinding of larger planetesimals \mbox{\citep{wyatt2008}}. 

We plan to observe approximately 200 stars in total over the planned survey duration of 4\,years and within the time budget of 120 nights (which includes weather losses and second-epoch observations).
This total number of survey targets is a compromise between the goal to go deeper than previous surveys, the need for a sufficiently large target list to ensure a robust statistical interpretation of the expected low-number discovery outcome, and the GTO time budget.
The entire survey is carried out with the NACO instrument \mbox{\citep[][see Sect.\,\ref{sec:obs}]{rousset2003,lenzen2003}} and the $L^{\prime}$\ filter at the ESO VLT on Cerro Paranal in Chile.
Every target star is observed once with very deep angular differential imaging \mbox{\citep[ADI;][]{marois2006}} observations (2\,--\,4\,hr with typically $>$\,60\degr\ field rotation) and a coronagraph employed (where feasible) to achieve the best possible point-source sensitivity down to the smallest possible separations (see Sect.\,\ref{sec:obs}). Second epoch observations are only scheduled where interesting companion candidates discovered in the first epoch image require confirmation.

For the young PPD stars, we exploit the advantage of the $L^{\prime}$-band for embedded planets with possible large circumplanetary extinction over the shorter-wavelength near-infrared (NIR) bands. Furthermore, the suspected circumplanetary discs were shown to be very bright at $\lambda>$\,\SI[]{3}{\micro\meter} (e.g. \mbox{\citealt{zhu2015}}; \mbox{\citealt{eisner2015}}; \mbox{\citealt{szulagyi2019}}), further helping $L^{\prime}$-band observations even where the embedded planets may not be directly detectable. Indeed, $L^{\prime}$-band observations were key for detecting the few forming planet candidates that we know of today, namely 
LkCa\,15\,b \mbox{\citep{kraus2012}}, 
HD\,100546\,b \mbox{\citep{quanz2015b}}, 
HD\,169142\,b \mbox{\citep{biller2014,reggiani2014}}, 
MWC\,758\,b \mbox{\citep{reggiani2018}}, and 
PDS\,70\,b  \mbox{\citep{keppler2018},} 
even though all but the last one are still debated (\mbox{\citealt{currie2019}}; \mbox{\citealt{rameau2017}}; \mbox{\citealt{ligi2018}}; \mbox{\citealt{wagner2019}}). Further advantages of the $L^{\prime}$-band are, for example, that scattering from circumstellar disc material, which increases the noise and the probability of false positives, and contamination by background objects are less severe than at shorter wavelengths.

For the older DEB stars, we also exploit the $L^{\prime}$-band advantage, but here for cooler, that is, lower-mass and/or older planets. More importantly, our NACO-{\it ISPY} survey is clearly distinguished from the SPHERE-SHINE survey \mbox{\citep{chauvin2017a} for example,} in that we explicitly target a large sample of well-characterised DEB stars and employ a target selection and survey strategy that will allow us to constrain the relation between DEB properties (e.g. signatures of dynamical excitation, see below) and the occurrence of GPs.


\section{Targets} \label{sec:targets}

\subsection{Target selection} \label{ssec:targetsel}

The targets for the {\it ISPY} survey were selected from various databases and the literature. In particular, the PPD targets were compiled from catalogues and studies of Herbig Ae/Be stars \mbox{\citep[e.g.][]{the1994,menu2015}} and then complemented with additional lower-mass objects for which high-spatial resolution observations showed substructures in their circumstellar discs possibly indicative of ongoing (gas giant) planet formation \mbox{\citep[e.g.][]{andrews2011,andrews2018}}. Here, we could only add objects that were not blocked from other ongoing observing programs. 
The primary focus on Herbig Ae/Be stars was motivated by the idea that in these systems (more massive) gas giant planets might be forming at larger radial separations \mbox{\citep[cf.][]{quanz2015a}}. Indeed, thus far most candidates for young, embedded planets have been detected around such objects
(e.g. \mbox{\citealt{quanz2013,quanz2015b}}; \mbox{\citealt{brittain2014}}; \mbox{\citealt{biller2014}}; \mbox{\citealt{currie2017}}; \mbox{\citealt{reggiani2014,reggiani2018}}), 
although not all of these candidates were confirmed in follow-up studies and the origin of some of the detected signals has recently been questioned (e.g. \mbox{\citealt{rameau2017}}; \mbox{\citealt{cugno2019b}}; \mbox{\citealt{ligi2018}}; \mbox{\citealt{huelamo2018}}).

The DEB targets were mainly compiled from the {\it Spitzer} IRS catalogue of \mbox{\citet{chen2014}}, the first comprehensive DEB catalogue based on spectroscopic data. To ensure that we select only stars with significant DEB signals, we re-compiled and re-evaluated the complete spectral energy distributions (SEDs) of all target candidates (see Sect.\,\ref{ssec:targetsobs}), and applied primary cutoffs for the fractional disc luminosity at \mbox{$L_\mathrm{IR}/L_\mathrm{bol}>3\times10^{-6}$} 
and for the blackbody temperature of the disc excess emission at \mbox{$T_\mathrm{d1}<220$\,K}. The threshold values are chosen to avoid outliers with non-significant excess or suspiciously high fit temperatures. We also inspected all SED fits individually to identify suspicious cases that could be affected by confusion with background or other objects. The new SED fits were also used to re-derive other disc and stellar parameters (see Sect.\,\ref{ssec:targetsobs} and Fig.\,\ref{fig-hd203sed}).
Additional DEB targets were selected from the results of the {\it Herschel} DEB surveys {\it DUNES} \mbox{\citep{eiroa2013,maldonado2015}} and {\it DEBRIS} \mbox{\citep{matthews2010,booth2013}}. In particular, we selected those targets which show indications not only of cold outer debris belts, but also of an additional hot exozodiacal dust belt closer to the host star. The motivation for selecting the latter derives from the speculation that hot inner debris belts are related to the formation of rocky planets and that in such systems more massive gas giant planets might also be forming at larger radial separations. Most of the approximately $30$\ DEB targets older than 1\,Gyr (see Fig.\,\ref{fig-histo1}) originate from the {\it Herschel} detections. 

In addition to the specific selection criteria mentioned above, we use the following general selection criteria for compiling our master list of target candidates:
\begin{enumerate}
\item $-70\deg \le {\rm DEC}\le +15\deg$, to ensure sufficient observability from the location of Paranal observatory ($>$\,4\,hr at air mass\,$<$\,1.5; with a few exceptions);
\item distance\,$\le 160$\,pc (DEB) and $\le 1000$\,pc (PPD) to achieve reasonable spatial resolution;
\item avoid extreme (later than M4, earlier than B8) and uncertain spectral types;
\item $K <10$\,mag to ensure good adaptive optics (AO) correction for the NACO NIR wavefront sensor;
\item no known close ($<$1\arcsec) binaries and multiples around bright stars (data archives, catalogues, and literature checked) that could hamper centring of the coronagraph (see Sect.\,\ref{sec:obs});
\item no existing and sufficiently deep ADI $L^{\prime}$-band observations available (various archives searched).
\end{enumerate}

With these selection criteria, we compiled a master list consisting of 90 PPD and 300 DEB stars. Distances were inferred from {\it Gaia}\,DR2 parallaxes \mbox{\citep{gaia_mission,gaia_dr2}} with the method described by \mbox{\citet{bailer2018}} and retrieved through VizieR \mbox{\citep{vizier2000}}, unless stated otherwise in Tables\,\ref{tab:slist1} and \ref{tab:slist2}.
Age estimates were compiled from the literature. Many stars have multiple partially contradicting published age estimates derived with different methods. In Tables\,\ref{tab:slist1} and \ref{tab:slist2}, we list only the age value that we consider to be the most reliable one. The listed age reference may not necessarily point to the original age estimate, but can also refer to a paper that compiles and discusses various age estimates. $L^{\prime}$\,magnitudes and their uncertainties are derived from the WISE photometry \mbox{\citep[][]{cutri2013}} and black body interpolation between WISE filters W1 (\SI[]{3.35}{\micro\meter}) and W2 (\SI[]{4.6}{\micro\meter}) to $L^{\prime}$\ (\SI[]{3.8}{\micro\meter}).

Since the total time budget for our survey of 120 observing nights permits deep observations of only $\approx$\,200 stars (see Sect.\,\ref{sec:obs}), the target candidates in this master list were separated into three priority categories: priority\,1 targets that will be observed, priority\,2 targets that will be observed if time permits, and priority\,3 targets that will not be scheduled for observations, unless needed as fillers.

Among the 90 PPD target candidates, we gave the highest priority to nearby ($<200$\,pc) Herbig Ae/Be stars and to transition discs with known gaps and cavities with separations large enough so that $L^{\prime}$-band observations with VLT/NACO can probe for embedded sources. 
Intermediate priority was given to Herbig Ae/Be stars at distances $>200$\,pc and to lower-mass (T\,Tau) objects with less pronounced or more poorly characterised transition disc signatures.
The lowest priority was given to objects where high-contrast $L^{\prime}$\ observations were already available or published or where the discovery space was limited for other reasons. This resulted in a list of 43 priority\,1 targets, 33 priority\,2 targets, and 14 targets that were discarded from the observing list and not even used as fillers.

To prioritise the DEB target candidates in such a way that we maximise the detection probability (success-oriented) while at the same time preserving an (relatively) unbiased sample, we evaluated what the achievable planet mass detection limit at a projected physical separation of 20\,au would be. For this purpose, we used mean $5\,\sigma$\ contrast curves achieved during the first few observing runs (Tab.\,\ref{tab-obsruns}) under good weather conditions with our standard observing procedures with and without the coronagraph. These contrast curves were then scaled to the $L^{\prime}$\ magnitudes of each target candidate in such a way that at small angular separations, the contrast is preserved independently of the stellar brightness \mbox{($\Delta L^{\prime}\approx 6.4$\,mag} at $\Delta r=150$\,mas), and at large separations ($>3\arcsec$), the mean background limit of \mbox{$\Delta L^{\prime}\approx 16.7$\,mag} is preserved (see Sects.\,\ref{sec:obs} and \ref{ssec:res:overview} and Fig.\,\ref{fig-contr}). The resulting contrast curves are then converted to planet-mass detection limits using the $L^{\prime}$\ magnitudes, ages, and distances of the stars (tabs.\,\ref{tab:slist1}, \ref{tab:slist2}, \ref{tab:obsslist1}, and \ref{tab:obsslist2}) together with BT-Settl evolutionary models \mbox{\citep{allard2014,baraffe2015}}.

The highest priority was given to DEB stars around which the predicted planet-mass detection limit at 20\,au projected physical separation was $\leq10$\,M$_{\rm Jup}$.
We also gave high priority to DEBs with detectable gas (CO) content
(e.g. \mbox{\citealt{zuckerman1995}}; \mbox{\citealt{moor2011,moor2015b}}; \mbox{\citealt{marino2016,marino2017}}; \mbox{\citealt{greaves2016}}; \mbox{\citealt{lieman2016}}), although it is not a priori clear if the CO gas is of primary or secondary origin, in particular in the discs around A\,stars \mbox{\citep{moor2017}}. However, in some cases and in particular in the few later-type CO-rich discs, the gas is likely to originate from colliding or evaporating icy planetesimals \mbox{\citep{kral2016,matra2019}}.
Intermediate priority was given to DEB stars with predicted 20\,au detection limits between 10 and 20\,M$_{\rm Jup}$.
Lowest priority was given to pre-selected target candidates with predicted mass detection limits $>20$\,M$_{\rm Jup}$, which were only observed if needed as fillers.
This selection process resulted in a final list of 177 DEB targets, of which 87 are priority\,1, the same number are priority 2, plus three stars for which we re-classified the IR excess as not significant only after they were observed early in the survey.

\subsection{Target list} \label{ssec:targetsample}

Our final survey target list thus consists of 253 stars, of which 76 stars are surrounded by young PPDs, 177 stars are surrounded by DEBs, and 
(including the three targets mentioned above for which the IR excess was later classified as not significant). Of these 253 targets, we assigned 130 stars as priority\,1 (43 PPDs and 87 DEBs). 

\begin{figure}[htb]
 \centering
 \includegraphics[width=9cm]{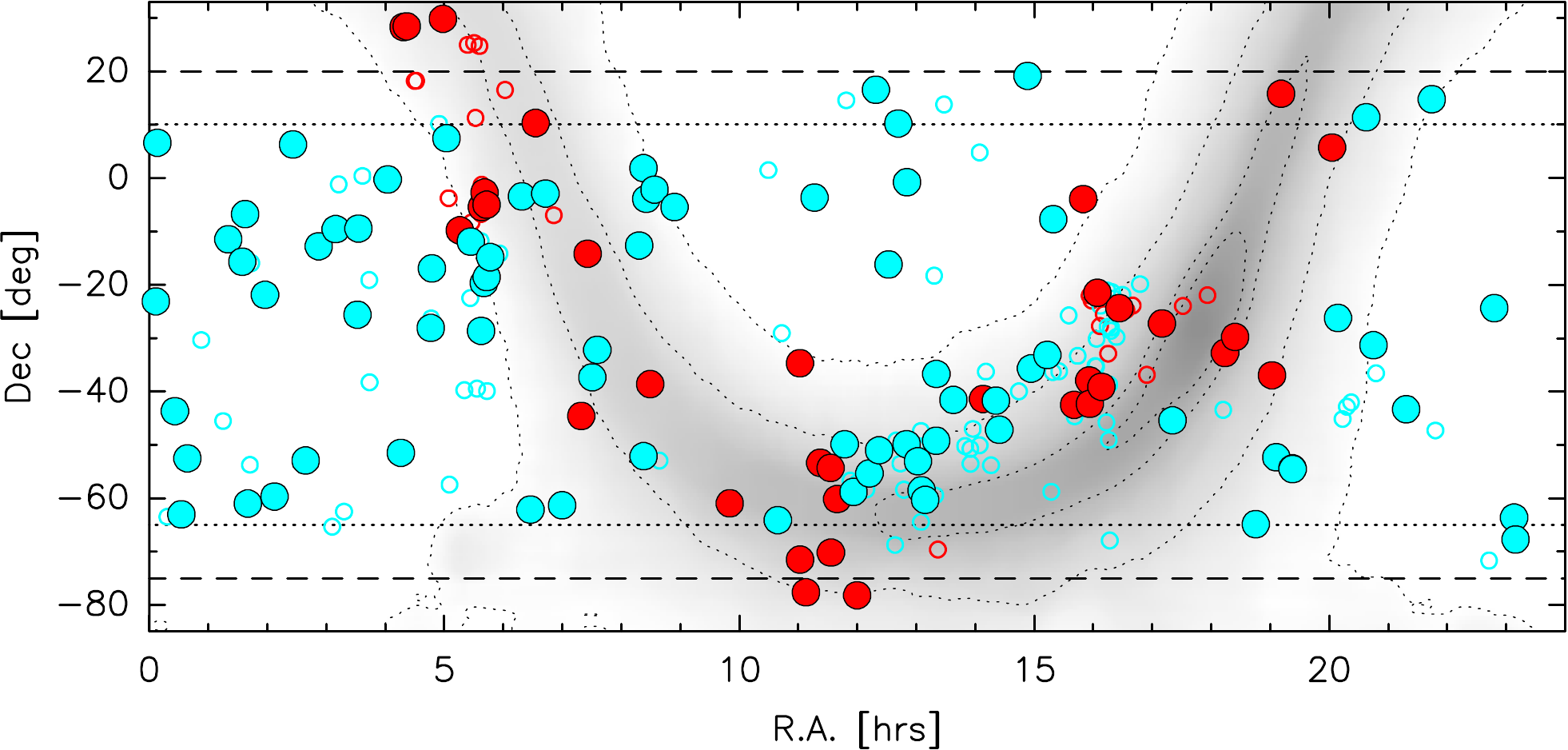}
 \caption{\label{fig-skydistr}
  Sky distribution of all 253 {\it ISPY} targets. PPD targets are shown in red, DEB targets in blue. 
  Targets observed in P\,96 through P\,100 are marked as large filled circles. Targets observed later and with the analysis not yet included in this paper are marked as small empty circles. The grey-shaded area and dotted contours outline the Milky Way disc and bulge as traced by the COBE-DIRBE band\,2 ($K$) zodi-subtracted all-sky map \mbox{\citep{hauser1989}}.
   Dotted and dashed horizontal lines mark the approximate declinations for field rotations of 60\degr\ and 50\degr, respectively, achievable with a total observing time of 3\,hr scheduled around meridian passage of the star.}
\end{figure}

\begin{figure}[htb]
 \centering
 \includegraphics[width=9cm]{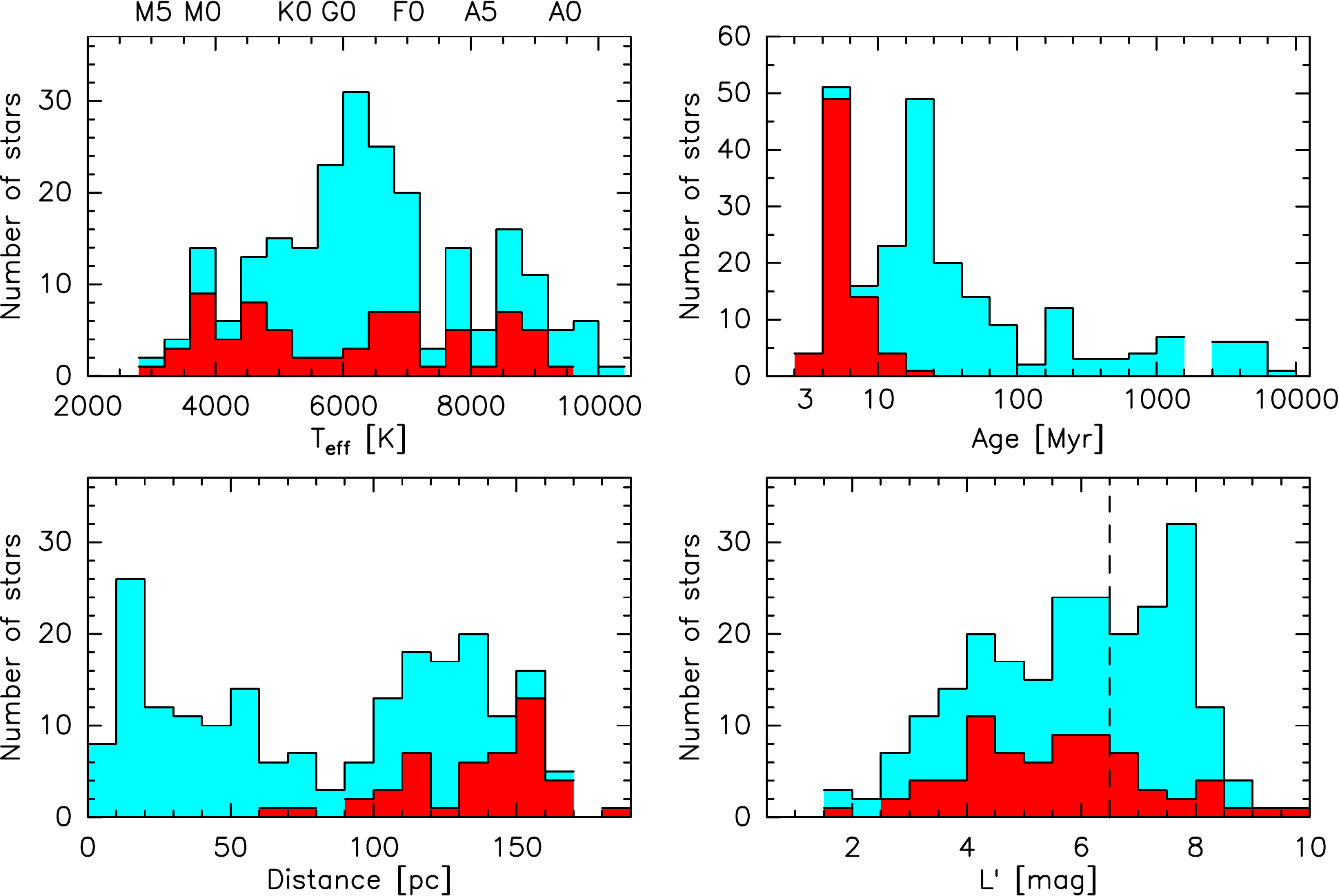}
 \caption{\label{fig-histo1}
  Histograms showing the distribution of stellar effective temperatures, ages, distances, and $L^{\prime}$\ magnitudes
  of all 253 {\it ISPY} target candidates. PPD targets are shown in red, DEB targets in blue on top of PPD histograms.
  Spectral types above the first panel correspond to main-sequence stars \mbox{\citep{cox2000}}.
  We note that 25 of the PPD targets have distances $>200$\,pc and are thus not contained in the distance distribution histogram.
  The vertical dashed line in the lower right panel marks the approximate brightness limit up to which the AGPM can be used.}
\end{figure}

\begin{figure}[htb]
 \centering
 \includegraphics[width=8cm]{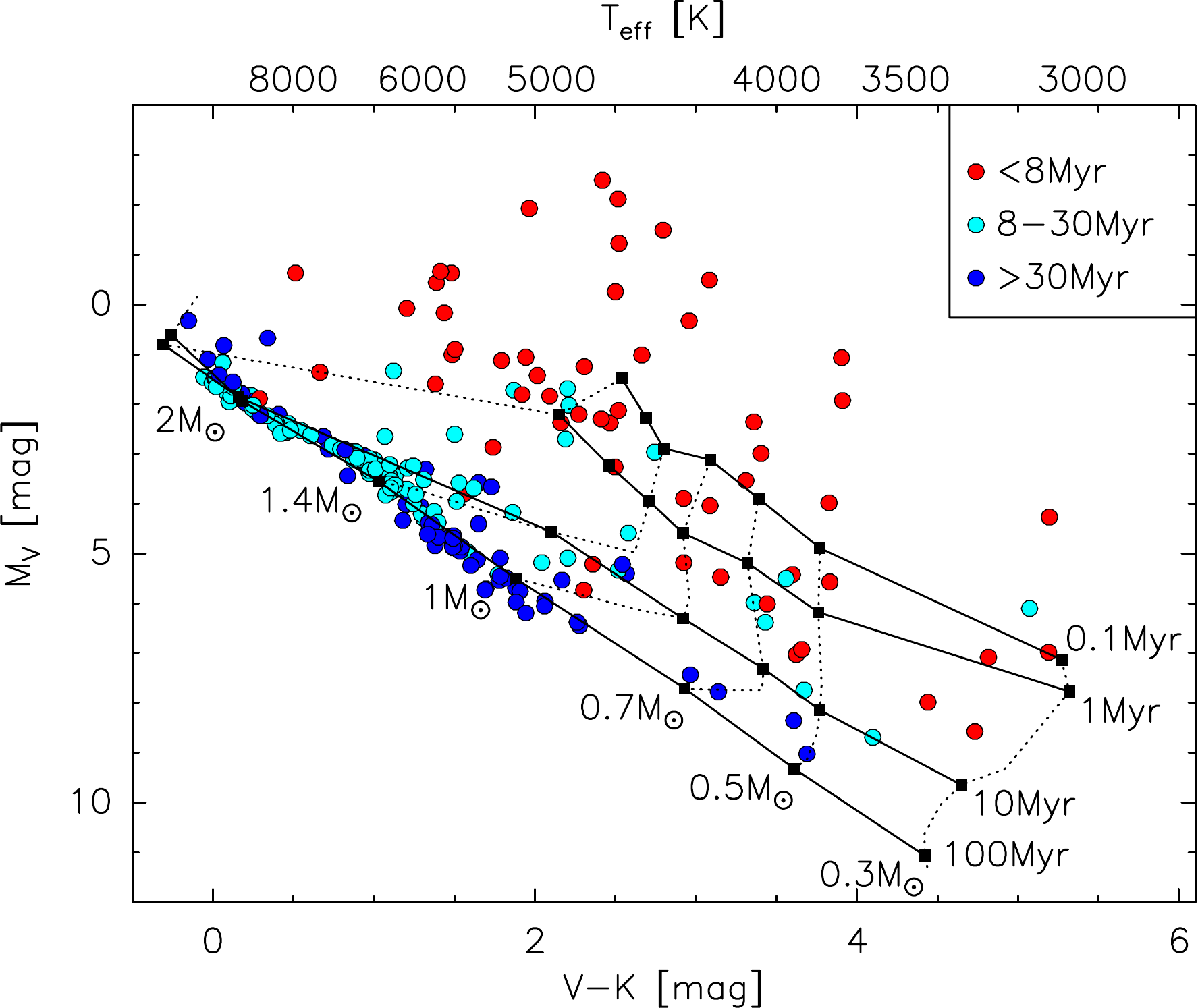}
 \caption{\label{fig-cmd}
  Colour-magnitude diagram of all NACO-{\it ISPY} target candidates. Marked on top are the effective temperatures which a main-sequence star with the corresponding \mbox{$V-K$} colour would have \mbox{\citep{cox2000}}. Stellar ages are colour-coded: red: <8\,Myr, light blue: 8\,--30\,Myr, dark blue: >30\,Myr. Isochrones \mbox{\citep{siess2000}} for four stellar ages (counting from the birthline) and for metallicity $Z=0.02$\ are overplotted.}
\end{figure}

Figure\,\ref{fig-skydistr} shows the sky distribution of all 253 {\it ISPY} GTO target candidates.
Figure\,\ref{fig-histo1} shows the distribution of spectral types, ages, distances, and $L^{\prime}$\ magnitudes.
Spectral types range from early M to late B, corresponding to $T_\mathrm{eff} \approx 3000 - 10000$\,K.
Ages of the PPD stars range from about 2 to 10\,Myr.
The majority of the DEB targets have ages of between 10 and 100\,Myr with a distribution tail out to a few gigayears (the older ones being mostly {\it Herschel} detections). Distances to the PPD stars are mostly between 100 and 170\,pc, while the DEB targets are located between 3 and 150\,pc, with two distribution peaks around 10\,--\,50\,pc and 100\,--\,140\,pc. $L^{\prime}$\,magnitudes range from 2.5 to 8.5\,mag with very few outliers towards the brighter and fainter ends.

Figure\,\ref{fig-cmd} shows the colour-magnitude diagram (CMD) of our target stars with the stellar ages colour-coded and \mbox{\citet{siess2000}} isochrones overplotted. The zero-age main sequence (ZAMS) is clearly visible as well as the many redder (cooler) and over-luminous (because of their larger radii) pre-main sequence (PMS) stars.

\begin{figure}[htb]
 \centering
 \includegraphics[width=9cm]{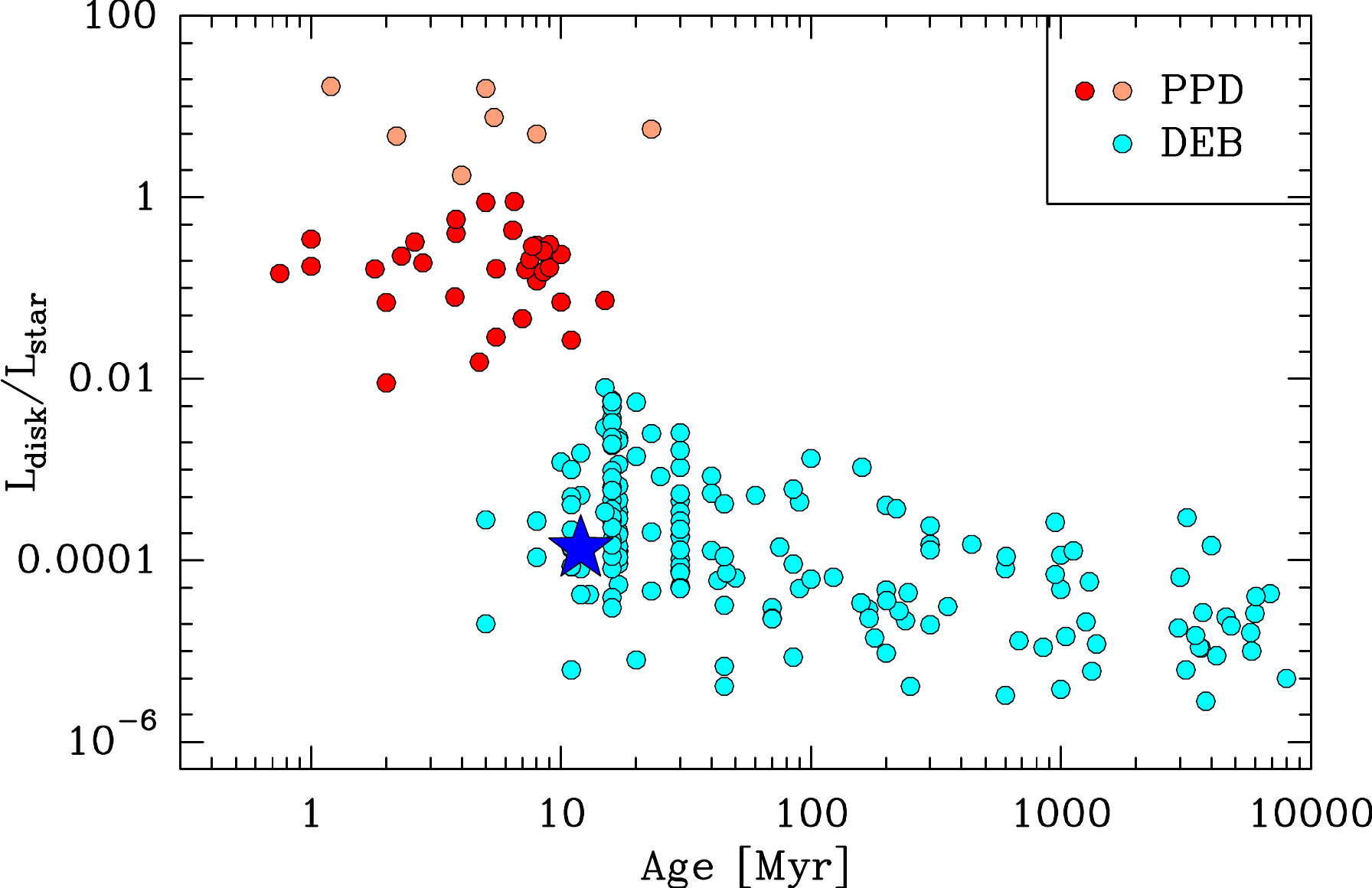}
 \caption{\label{fig-ldisc-age}
 Fractional disc-related excess luminosity vs. stellar age for all {\it ISPY} targets. PPD targets are shown in red, DEB targets in light blue. The large dark blue asterisk marks HD\,203 for which we show the SED in Fig.\,\ref{fig-hd203sed}. 
 We note that stellar luminosities were not corrected for extinction such that fractional disc luminosities are overestimated and appear unphysically large for some of the very young and embedded PPD stars ($>1$, light red).}
\end{figure}

\begin{figure}[htb]
 \centering
 \includegraphics[width=9cm]{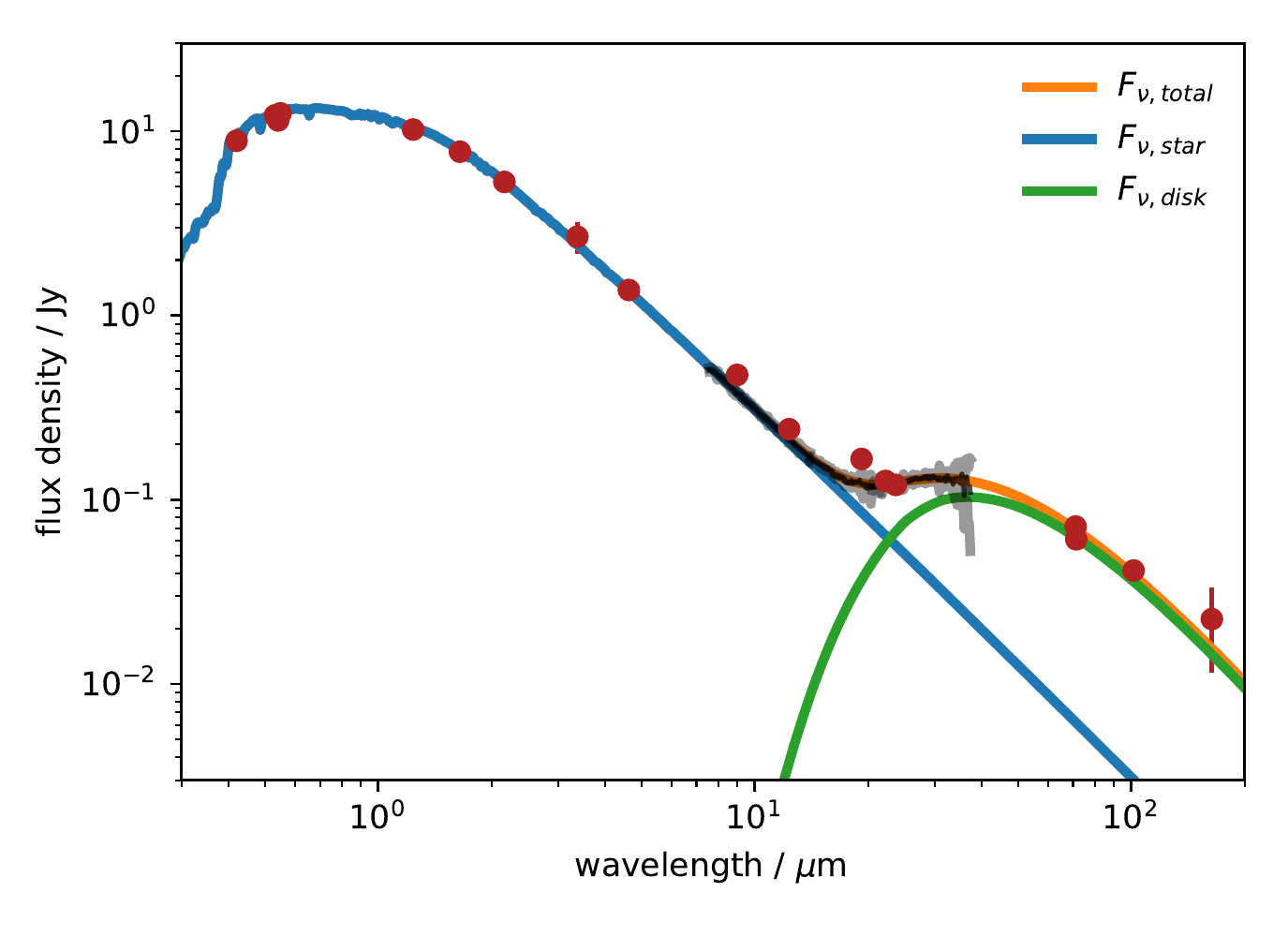}
 \caption{\label{fig-hd203sed}
 Flux density distribution for a typical DEB target (HD\,203, cf. Fig.\,\ref{fig-ldisc-age}). Photometric data are shown as red circles, and the Spitzer IRS spectrum is shown as a black solid line with uncertainties in grey. The best-fit model (orange line) has stellar (blue) and blackbody (green, for the DEBs) components (see Sect.\,\ref{ssec:targetsobs} for details).}
\end{figure}

Figure\,\ref{fig-ldisc-age} shows the fractional disc-related excess luminosity 
\mbox{$L_{\rm disc}/L_{\ast}=f_{\rm d}$} 
(see Sect.\ref{ssec:targetsobs}) versus stellar age for all {\it ISPY} targets, illustrating the dominance of the disc emission over the stellar contribution for the young PPDs and the relatively faint disc contribution from the more evolved DEBs. Error bars in both dimensions can be substantial, in particular for the ages and for the fractional disc luminosities of the young PPD targets (see below). As ages are compiled from the literature and derived with different methods and assumptions, we have no means of deriving consistent uncertainties for all sources and therefore we refrain from showing error bars here.
The PPD and DEB targets appear as two clearly separated groups in this diagram with 
\mbox{$f_{\rm d}^{\rm PPD}>0.01>f_{\rm d}^{\rm DEB}$}, albeit with some age overlap in the range 5\,--\,20\,Myr. 
Because stellar luminosities derived from the SED fitting (see Sect.\,\ref{ssec:targetsobs}) were not corrected for extinction, fractional disc luminosities for some of the very young and embedded PPD stars are significantly overestimated and appear unphysically large (>1). For the older DEB stars with typically optically thin discs, where extinction should not play a significant role, the diagram suggests a trend of fractional disc luminosity declining with age as 
\mbox{$\log(f_{\rm d})=-0.5\times \log({\rm age})-3$}, 
albeit with the note of caution that we introduced a selection bias by applying a lower limit to 
\mbox{$f_{\rm d}>3\times10^{-6}$}. 
Nevertheless, this trend also holds for the upper envelope of the distribution, which is not affected by the selection bias.
Figure\,\ref{fig-hd203sed} shows the flux density distribution of a typical DEB survey target (here HD\,203). 

\subsection{Targets observed during ESO periods 96 through 100} \label{ssec:targetsobs}

In this section, we describe the 112 target stars (34 PPDs and 78 DEBs) observed during ESO periods 96 throughout 100, which comprise approximately the first half of the survey. The analysis of the complete survey will be presented in a forthcoming paper. Tables\,\ref{tab:slist1} and \ref{tab:slist2} list the names, coordinates, and other parameters for these stars. 
The observations are described in more detail in Sect.\,\ref{sec:obs}.

Distances and mean uncertainties for all but four targets (marked in Table\,\ref{tab:slist1}) are derived by inverting {\it Gaia} parallaxes\footnote{A more appropriate way to derive distances from parallaxes for stars more distant and with larger relative parallax uncertainties than our targets is outlined by \mbox{\citet{bailer2015}}.} \mbox{\citep{gaia_dr2}}. Stellar luminosities and $T_{\rm eff}$\ of PPD targets are also taken from {\it Gaia}\,DR2 \mbox{\citep{gaia_dr2}}, although their values have rather large uncertainties owing to circumstellar extinction, which was not taken into account. Ages and outer disc radii of PPD targets are compiled from the literature with references are given in Table\,\ref{tab:slist1}.

Stellar luminosities, effective temperatures, and disc radii of DEB targets are derived by fitting stellar \mbox{\citep[PHOENIX;][]{husser2013}} and blackbody models to observed photometry and spectra. The photometry is 
obtained from multiple catalogues and publications, including 2MASS, APASS, Hipparcos/Tycho-2, Gaia, AKARI, WISE, IRAS, Spitzer, Herschel, JCMT, and ALMA. In some cases, photometry has been excluded, for example due to saturation or confusion with background or other objects. The fitting method uses synthetic photometry of grids of models to find the best-fitting models with the MultiNest code \mbox{\citep{feroz2009}}, and an example is shown in Fig.\,\ref{fig-hd203sed}. We first fit star\,+\,single disc component models, but in a few cases the fits were significantly improved by adding a second blackbody component, which we interpret as an indication that the star harbours debris belts at multiple radii \mbox{\citep{kennedy2014}}. The blackbody radius of the debris belts, $R_{\rm BB}$, is then given by \mbox{\citep{pawellek2015}}:
\begin{equation} \label{eq:rbb}
R_{\rm BB} = \Bigg( \frac{278\,K}{T_{\rm dust}}\Bigg)^2\,\Bigg(\frac{L}{L_{\odot}}\Bigg)^{1/2}
.\end{equation}
An estimate of the `true' disc radius, $R_{\rm disc}$, is then obtained by applying a stellar luminosity-dependent correction factor, $\Gamma$, which accounts for the radiation pressure blowout grain size,
\begin{equation} \label{eq:gamma}
\Gamma = a\,( L_{\ast}/{\rm L_{\odot}})^b 
,\end{equation}
\mbox{\citep{pawellek2015}}, but using the new coefficients given in \mbox{\citet{pawellek2016}}, namely
$a=7.0$\ and $b=-0.39,$\ and limiting $\Gamma$\ to $\Gamma_{\rm max}=4.0$\ for stars with $L_{\ast}\lesssim 4\,{\rm L}_{\odot}$.

\begin{figure}[htb]
 \centering
 \includegraphics[width=8cm]{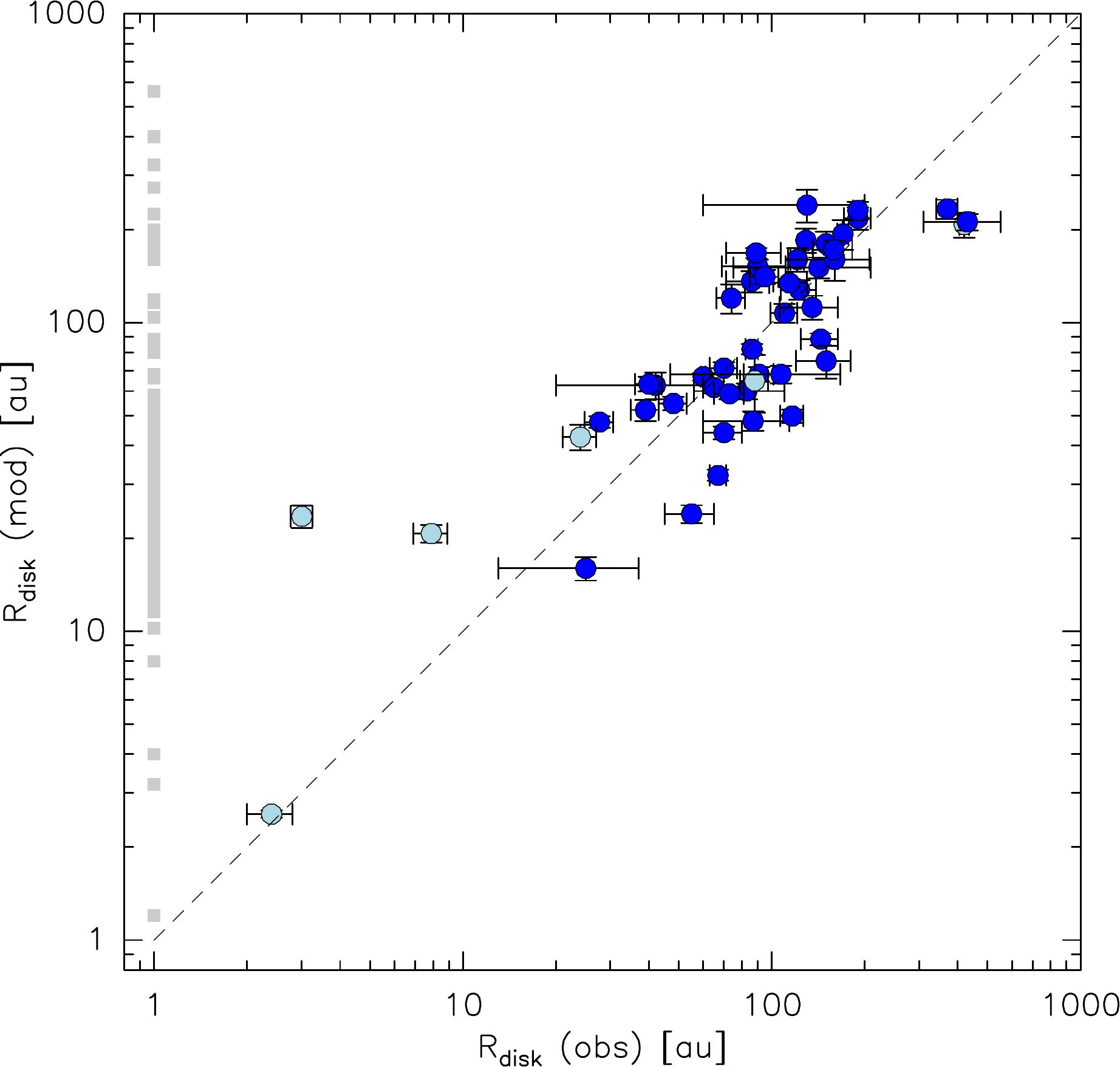}
 \caption{\label{fig-rbb-robs}
  Debris disc sizes (outer belt radii) obtained via SED fitting and corrected for the luminosity-dependent blowout grain size \mbox{\citep{pawellek2015}} vs. directly measured disc radii derived from spatially resolved observations (values and references for the stars observed in ESO periods P\,96 through 100 are in Table\,\ref{tab:slist2}). Disc sizes derived from VIS/NIR or FIR/submm observations are marked in dark blue. Disc sizes derived from MIR observations are marked in light blue. The dashed line marks equality. Marked in light grey at $R_{\rm disc}$(obs)\,=\,1\,au are those targets that do not have spatially resolved observations.}
\end{figure}

In addition, we also compile and list in Table\,\ref{tab:slist2} disc radii directly obtained from spatially resolved observations where available. These directly measured disc sizes originate from various different tracers and methods for which  properly derived uncertainties are not always available and may therefore not always reflect the `true' outer disc radius. While VIS and NIR observations tracing scattered light and FIR and submm images tracing thermal emission from cold dust should be reliable tracers of the total DEB extent, mid-IR (MIR) observations may trace only hot dust and miss the cold outer dust belts. Nevertheless, these observations provide the best measure we can currently obtain to evaluate the robustness and reliability of the model disc sizes obtained via SED fitting and available for all targets. Figure\,\ref{fig-rbb-robs} compares the SED fit-based model radii to the directly observed disc sizes for those 48 targets for which there are known and published disc size measurements available. As can be seen, there is general good agreement between SED-derived and observed disc sizes over a size range of more than two magnitudes. The mean relative discrepancy is $\approx50$\% with the most significant outlier showing a discrepancy of a factor of seven. However, this outlier (HD\,38678, see Table\,\ref{tab:slist2}) is one of the five stars for which the disc radius was derived from marginally resolved MIR observations which do not trace the cold outer dust belt and thus underestimate the disc size.
We therefore conclude that the SED-derived disc radii are, in general, a good proxy for the actual sizes of the cold outer debris belts.


\begin{table*}
\caption{\label{tab:slist1} PPD targets observed between December 2015 and March 2018 (P\,96 through P\,100)}
\begin{tabular}{llllccllll}
\hline\hline
Name\tablefootmark{a}             & 
RA(J2000)\tablefootmark{b}        & 
DEC(J2000)                        & 
dist.\tablefootmark{c}            & 
$V$\tablefootmark{d}              & 
log($L_{\ast}$)\tablefootmark{e}  &  
$T_{\rm eff}$\tablefootmark{e}    &
age\tablefootmark{f}              & 
$R_{\rm disc}$\tablefootmark{g}   & 
Refs.\tablefootmark{h}            \\
               &
[hh:mm:ss]     & 
[dd:mm:ss]     & 
[pc]           & 
[mag]          & 
[L$_{\odot}$]  & 
[K]            &                 
[Myr]          &  
[au]           &
               \\
\hline
V$^{\ast}$\,V892\,Tau & 04:18:40.61 & ~28:19:15.6 & 117$\pm$2   & 14.7 & -1.82 & 4250 & 5         & 90          & 11, 11  \\
HD\,283571    & 04:21:57.41 & ~28:26:35.5 & 444$\pm$50  & 9.3  & 1.26  & 4420 & 1.8       & 243$\pm$40  & 12, 41 \\
HD\,31648     & 04:58:46.27 & ~29:50:37.0 & 161$\pm$2   & 7.7  & 1.29  & 7860 & 8.5$\pm$2 & 634$\pm$13  & 13, 42 \\
HD\,34282     & 05:16:00.48 & -09:48:35.4 & 359$\pm$5   & 9.8  & 0.95  & 8250 & 6.4$\pm$2 & 293$\pm$15  & 14, 43 \\
HD\,37411     & 05:38:14.51 & -05:25:13.3 & 359$\pm$4  & 9.8  & 1.28  & 6810 & 9$\pm$4.5 & 123         & 3, 1, 44 \\
HD\,37806     & 05:41:02.29 & -02:43:00.7 & 423$\pm$11  & 7.9  & 2.04  & 8590 & ...       & 141         & 1, 45 \\
HD\,38120     & 05:43:11.89 & -04:59:49.9 & 402$\pm$13  & 9.1  & 1.60  & 9120 & ...       & 290$\pm$48  & 43 \\
HD\,259431    & 06:33:05.19 & ~10:19:20.0 & 711$\pm$24  & 8.8  & 2.88  & 6500 & 0.06$\pm$0.03 & 8.1     &  15, 46 \\
NX\,Pup       & 07:19:28.29 & -44:35:11.2 & 410$\pm$82 & 10.0 & 1.66 & 4760 & 0.05  & ...         & 1, 16 \\
HD\,58647     & 07:25:56.10 & -14:10:43.5 & 316$\pm$4   & 6.9  & 2.23  & 7228 & ...       & 137         & 45 \\
HD\,72106     & 08:29:34.90 & -38:36:21.1 & 279$\pm$88  & 8.6  & 1.48  & 7570 & ...       & 40          & 2, 47 \\
HD\,85567     & 09:50:28.54 & -60:58:03.0 & 1002$\pm$29 & 8.6  & 2.54  & 7880 & 0.01      & 0.5$\pm$0.2 & 16, 48 \\
TW\,Hya       & 11:01:51.91 & -34:42:17.0 & 60$\pm$0.2  & 10.9 & -0.5 & 4240 & 7.5$\pm$0.7 & 172$\pm$11 & 17, 43 \\
HD\,95881     & 11:01:57.62 & -71:30:48.4 & 1148$\pm$44 & 8.3  & 2.8  & 6740 & 8         & 200         & 16, 49 \\
HD\,97048     & 11:08:03.31 & -77:39:17.5 & 184$\pm$1   & 8.5  & 1.14  & 6750 & 6.5$\pm$1 & 866         & 13, 50 \\
HD\,98922     & 11:22:31.67 & -53:22:11.5 & 662$\pm$16  & 6.7  & 2.92  & 8800 & 0.01      & 110         & 16, 51 \\
HD\,100453   & 11:33:05.58 & -54:19:28.5 & 104$\pm$0.4  & 7.8  & 0.81  & 7270 & 10$\pm$2  & 41          & 18, 52 \\
HD\,100546   & 11:33:25.44 & -70:11:41.2 & 110$\pm$0.6  & 6.8  & 1.37  & 9180 & 3.8$\pm$0.5 & 500       & 13, 53 \\
HD\,101412   & 11:39:44.46 & -60:10:27.7 & 407$\pm$5    & 9.0  & 1.43  & 7840 & 2$\pm$1   & 92.5        & 19, 54 \\
HD\,104237   & 12:00:05.09 & -78:11:34.6 & 108$\pm$0.7  & 6.6  & 1.32  & 7250 & 5.5$\pm$0.5 & 66$\pm$14 & 13, 55 \\
PDS\,70      & 14:08:10.16 & -41:23:52.6 & 113$\pm$0.5 & 12.2 & -0.46 & 3970 & 5.4$\pm$1  & 113       & 20, 31, 56 \\
HD\,139614   & 15:40:46.38 & -42:29:53.5 & 134$\pm$1    & 8.3  & 0.90  & 8020 & 10$\pm$3   & 94$\pm$3   & 13, 43 \\
HD\,141569   & 15:49:57.75 & -03:55:16.3 & 110$\pm$0.6  & 7.1  & 1.19  & 8440 & 4.7$\pm$0.3 & 279$\pm$5 & 14, 43 \\
IM\,Lup      & 15:56:09.18 & -37:56:06.1 & 158$\pm$1.4  & 11.6 & 0.02  & 4060 & 0.75$\pm$0.25 & ...     & 21  \\
HD\,142527   & 15:56:41.89 & -42:19:23.2 & 157$\pm$1    & 8.3  & 1.04  & 5320 & 2$\pm$0.5   & 257$\pm$15 & 13, 57 \\
HD\,144668   & 16:08:34.29 & -39:06:18.3 & 160$\pm$2    & 7.1 & 1.14 & 5860 & 2.8$\pm$1    & 38$\pm$5   & 13, 45 \\
Elias2-27    & 16:26:45.03 & -24:23:07.8 & 118$\pm$15   & 20.4\tablefootmark{i} & ... & .... & 0.8  & 525  & 23, 58 \\
KK\,Oph      & 17:10:08.12 & -27:15:18.8 & 220$\pm$7    & 11.0 & 1.14  & 5140 & 8$\pm$2     & 276          & 24, 59 \\
HD\,319139   & 18:14:10.48 & -32:47:34.5 & 72.3$\pm$0.3 & 10.7 & -0.21 & 4250 & 21          & 362          & 25, 60 \\
HD\,169142   & 18:24:29.78 & -29:46:49.3 & 114$\pm$0.8  & 8.2 &  0.76  & 7320 & 7.7$\pm$2   & 194          & 13, 61 \\
R\,CrA       & 19:01:53.86 & -36:57:08.1 & 95$\pm$7     & 11.9 & 1.85  & 3960 & 1           & ...          & 26, 32 \\
T\,CrA       & 19:01:58.79 & -36:57:50.3 & 130          & 11.7 &  ...  & 5000 & 23          & ...          &  3, 16 \\
HD\,179218   & 19:11:11.25 & ~15:47:15.6 & 264$\pm$3    & 7.2 & 1.83   & 8490 & 2.3$\pm$0.3 & 131$\pm$22   & 24, 43 \\
HD\,190073   & 20:03:02.51 & ~05:44:16.7 & 872$\pm$52   & 7.8 & 2.73   & 8380 & 1           & 108$\pm$4    & 16, 44 \\
\hline
\end{tabular}
\tablefoottext{a}{Where available, we use the HD number as main source ID.}
\tablefoottext{b}{ICRS, from {\it Gaia} \mbox{\citep{gaia_dr2}} where available.}
\tablefoottext{c}{Distances and their uncertainties are inferred from {\it Gaia}\,DR2 parallaxes \mbox{\citep{gaia_mission,gaia_dr2}} with the method described by \mbox{\citet{bailer2018}} and retrieved through VizieR \mbox{\citep{vizier2000}}, unless indicated otherwise (references 1 through 3).}
\tablefoottext{d}{$L^{\prime}$\, magnitudes are listed in Table\,\ref{tab:obsslist1}.}
\tablefoottext{e}{log($L_{\ast}$) and $T_{\rm eff}$\ (rounded to full ten) adopted from {\it Gaia}-DR2 \mbox{\citep{gaia_dr2}}, unless indicated otherwise (references 31 through 32).}
\tablefoottext{f}{Ages compiled here are taken from the literature and derived with different methods and different treatment of uncertainties. Some ages refer to references which only summarise different other age estimation attempts.}
\tablefoottext{g}{Outer disc radii are compiled from the literature and originate from different methods. They are corrected for new {\it Gaia}-DR2 distances where necessary.} 
\tablefoottext{h}{References 1 through 3 refer to distance, 11 through 26 refer to age, 31 through 32 refer to $L_{\ast}$\ and $T_{\rm eff}$, and 41 through 61 refer to observed disc radius.}
\tablefoottext{i}{{The \it Gaia G} magnitude is listed for Elias\,2-27 because no {\it V} mag is available.}
\tablebib{
(1)~\mbox{\citet{fairlamb2015}};
(2)~\mbox{\citet{leeuwen2007}};
(3)~\mbox{\citet{avda2004}};
(11)~\mbox{\citet{hami2010}};
(12)~\mbox{\citet{palla2002}};
(13)~\mbox{\citet{meeus2012}};
(14)~\mbox{\citet{merin2004}};
(15)~\mbox{\citet{alecian2013}};
(16)~\mbox{\citet{manoj2006}};
(17)~\mbox{\citet{ducourant2014}};
(18)~\mbox{\citet{collins2009}};
(19)~\mbox{\citet{wade2005}};
(20)~\mbox{\citet{keppler2018}};
(21)~\mbox{\citet{mawet2012}};
(22)~\mbox{\citet{canovas2017}};
(23)~\mbox{\citet{andrews2018}};
(24)~\mbox{\citet{pascual2016}};
(25)~\mbox{\citet{kastner2014}};
(26)~\mbox{\citet{sicilia2011}};
(31)~\mbox{\citet{keppler2018}};
(32)~\mbox{\citet{cugno2019a}};
(41)~\mbox{\citet{isella2010}};
(42)~\mbox{\citet{pietu2007}};
(43)~\mbox{\citet{dent2005}};
(44)~\mbox{\citet{liu2011}};
(45)~\mbox{\citet{marinas2011}};
(46)~\mbox{\citet{kraus2008}};
(47)~\mbox{\citet{schegerer2009}};
(48)~\mbox{\citet{wheelwright2013}};
(49)~\mbox{\citet{verhoeff2010}};
(50)~\mbox{\citet{walsh2016}};
(51)~\mbox{\citet{caratti2015}};
(52)~\mbox{\citet{benisty2017}};
(53)~\mbox{\citet{tatulli2011}};
(54)~\mbox{\citet{geers2007}};
(55)~\mbox{\citet{grady2004}};
(56)~\mbox{\citet{hashimoto2012}};
(57)~\mbox{\citet{fukagawa2006}};
(58)~\mbox{\citet{perez2016}};
(59)~\mbox{\citet{kreplin2013}};
(60)~\mbox{\citet{rosenfeld2013}};
(61)~\mbox{\citet{fedele2017}}
}
\end{table*}



\section{Observations} \label{sec:obs}

\setcounter{table}{2}
\begin{table*}[htb!]
\caption{\label{tab-obsruns}GTO observing runs during P\,96 through P\,100.}
\begin{tabular}{llcll}
\hline\hline
Period / Run ID & Dates & No of  & Conditions (seeing) & Notes \\
          &             &  nights &  & \\
\hline
96 / 096.C-0679(A)   & 2015, Dec 15\, --\,19        & 4 & excellent (0.5\arcsec-1\arcsec)                                 & 11 stars\tablefootmark{a}, no losses \\
96 / 096.C-0679(B)   & 2016, Feb 17\,--\,20         & 3 & partly cloudy and windy (1\arcsec-2\arcsec)             & 4 stars, 28\% weather loss\tablefootmark{b} \\
97 / 097.C-0206(A1) & 2016, May 02\,--\,11         & 6 & mostly cloudy and windy (0.5\arcsec-2\arcsec)        & 7 stars, 60\% weather loss \\
97 / 097.C-0206(A2) & 2016, May 29\, --\,Jun 03 & 5 & mostly cloudy (0.5\arcsec-1.5\arcsec)                       & 5 stars, 70\% weather loss\\
97 / 097.C-0206(B)   & 2016, Jul 30\,--\,Aug 02           & 3 & partly cloudy and windy (0.4\arcsec-1.2\arcsec)       & 7 stars, 18\% weather loss \\
98 / 198.C-0612(A)   & 2016, Nov 06\,--13           & 5 & good, partly mixed (0.7\arcsec-1.6\arcsec)                & 10 stars, no loss \\
98 / 198.C-0612(C)  & 2016, Dec 08\,--\,13         & 5 & mostly cloudy and windy (0.7\arcsec-3\arcsec)         & 7 stars, 60\% weather loss\tablefootmark{b} \\
98 / 198.C-0612(B)   & 2017, Mar 14\,--\,18         & 4 & mixed, partly high humidity (0.4\arcsec-1.5\arcsec)  & 8 stars, 33\% weather loss \\
99 / 199.C-0065(A1) & 2017, May 01\,--\,04         & 3 & mixed, partly strong wind (0.4\arcsec-1.5\arcsec)     & 9 stars, no loss \\
99 / 199.C-0065(A2) & 2017, May 15\,--\,20         & 5 & mixed, partly strong wind (0.4\arcsec-1.5\arcsec)     & 11 stars, 10\% weather loss \\
99 / 199.C-0065(B)   & 2017, Jun 16\,--Sep 01\tablefootmark{c}    & 8 & mixed, partly high humidity (0.4\arcsec-1.7\arcsec)  & 15 stars, 15\% weather loss \\
100 / 199.C-0065(C)   & 2017, Oct 29\ --\,Nov 04  & 6 & good, partly windy (0.4\arcsec-1.2\arcsec)                & 14 stars, no loss \\
100 / 199.C-0065(D)   & 2018, Feb 22\,--\,Mar 03  & 7 & excellent (0.3\arcsec-1.3\arcsec)                               & 14 stars, no loss \\
\hline
\end{tabular}
\tablefoottext{a}{Number of stars for which good, i.e. useable, data were obtained (one target was observed twice).}\\
\tablefoottext{b}{Weather loss includes both closed-dome and bad-weather data with strongly variable photometry.}\\
\tablefoottext{c}{Six short, scattered subruns, of which two two-night runs were carried out in VM and four single-night runs in dVM.}
\end{table*}

During 
13 observing campaigns and 64 observing nights (all in Visitor mode) between December 2015 and March 2018 (see Table\,\ref{tab-obsruns}), we observed and obtained useable data for 112 out of the 253 NACO-{\it ISPY} target candidates. In total, we obtained 140 datasets of which 105 were obtained with a coronagraph (see below).

For all observations in this survey, we use the AO assisted imager NACO, which is mounted in the Nasmyth\,A focus at UT1 (Antu) of ESO's VLT on Cerro Paranal in Chile. NACO is equipped with the AO front end, NAOS \mbox{\citep{rousset2003}}, and the NIR imaging camera CONICA \mbox{\citep{lenzen2003}}. All stars are observed with the L27 camera of NACO, which provides a pixel scale of 27.2\,mas/pix, and with the $L^{\prime}$\ filter ($\lambda_0=$\,\SI[]{3.80}{\micro\meter}, $\Delta\,\lambda=$\,\SI[]{0.62}{\micro\meter}). 
The pixel scale corresponds to a sampling of $\sim$3.5\,pix per $\lambda/D$, the diffraction-limited full width at half maximum (FWHM).

To achieve high contrast between the star and its immediate surroundings, we carry out ADI observations in pupil tracking mode. For ADI, the total time on source time is governed by the need for a field rotation large enough to efficiently remove the speckle noise. 
Therefore, we observe each star for 2\,--\,4\,hr around its meridian passage, thus typically achieving field rotations of between 70$\degr$\, and 100$\degr$. In Figure~\ref{fig:obshist} we present histograms displaying the total time on target, field rotation, seeing, and coherence time for all data sets.

To further improve the contrast at small angular separations from bright stars, we use the annular groove phase mask (AGPM) vector vortex coronagraph \mbox{\citep{mawet2013,absil2014}} for all stars that are bright enough to centre the coronagraph \mbox{($L^{\prime}\lesssim6.5$\,mag)}. The AGPM effectively suppresses the on-axis starlight by re-directing it outside the pupil, where it is blocked by the Lyot stop. The typical detector integration time (DIT) with AGPM is 0.35\,s. To model and correct the thermal background emission, we switch to an offset sky position every eight minutes (13 exposures (NEXPO), 100 detector integration times (NDIT)). 
We use NACO's cube mode to store each individual image frame for frame selection and sky reconstruction. 
To properly correct for the off-axis AGPM transmission, we measured the AGPM throughput as described in Appendix\,\ref{app:agpmtrans} and obtained the radial throughput curve shown in Fig.\,\ref{fig:agpmtransmission}.

For targets fainter than $L^{\prime}\approx 6.5$\,mag, the AGPM cannot be used because they are too faint for precise centring behind the AGPM. However, we use the same total integration time to achieve similar sensitivity at larger separations. The typical detector integration time (DIT) without AGPM is 0.2\,s. The star is positioned at the centre of one quadrant of CONICA and after one exposure (NEXPO=1, NDIT=126) is moved to the centre of the next quadrant using fixed offsets. This allows for continuous observation of the star without applying additional telescope offsets for sky frames. 

With this approach, we reach a star-planet 5\,$\sigma$\ contrast of typically $\Delta L^{\prime} \sim$7\,mag at 0\farcs2\ and a background detection limit of $\sim$16\,mag at $>1.5$\arcsec\ (see also Sect.\,\ref{ssec:res:overview} and Fig.\,\ref{fig-contr}). 
With this strategy for very deep and high-contrast observations and a smallest possible inner working angle (IWA) of $\approx100$\,mas, we are able to observe two to three targets per night on average. 

The lower left quadrant of CONICA is strongly affected by bad columns with large constant offset and low sensitivity (see also Sect.\,\ref{sec:red:clean1}). Until June 2017 every eigth column was affected, but since then, the situation has deteriorated and 38\% of this quadrant is affected. Therefore, the lower left quadrant is no longer used in the observing cycle without the AGPM, while the less-severe bad-column problem of a second affected detector quadrant could be handled with post-processing (Sect.\,\ref{sec:red:clean1}).

The ADI observation is bracketed by unsaturated flux measurements. To avoid saturation, the DIT of CONICA is adjusted depending on the brightness of the star in $L'$-band, and the star is cycled through the three detector quadrants that are not affected by the bad columns.

An astrometric field is observed at least once per observing run with and without the AGPM to monitor and measure the pixel scale and true north orientation of the detector. The fields are 47~Tucanae and the region around the Orion Trapezium Cluster. The analysis of the astrometric fields is described in Appendix\,\ref{app:astrometry}, and the resulting mean astrometric parameters are listed in Table\,\ref{tbl:astrometryvalues}.


\section{Data reduction and analysis} \label{sec:red}

All data taken by the {\it ISPY} survey are homogeneously reduced using the GRAPHIC pipeline \mbox{\citep{2016MNRAS.455.2178H}}. While a summary of the philosophy of the pipeline is explained in \mbox{\citet{2016MNRAS.455.2178H}}, substantial changes have been made to many of the steps to optimise them for NACO-{\it ISPY} data. Since two types of data are taken during the survey (with and without the AGPM coronagraph), two reduction pathways are used. These steps converge after the data have been cleaned, and the final reduction steps are performed in the same way.


\subsection{Cleaning of coronagraphic data}  \label{sec:red:clean1}

The NACO detector has two quadrants with defective columns. For the first of these, three in every eight columns show a constant value. To avoid problems in subsequent reduction steps, we first correct for these by replacing them with the mean of the four closest non-affected pixels from the same row.
Dark frames are then made by median-combining each cube of sky frames. We use the PCA\footnote{Principal Component Analysis}-based sky-subtraction scheme outlined in \mbox{\citet{2018A&A...611A..23H}} to remove the background flux, subtracting the first five PCA modes from each image.
We then correct for the second quadrant with defective pixels. In this quadrant, one in every eight columns has a dark current offset that varies in time. We correct for this by subtracting the median of each affected column from that column. 

To calculate the position of the star behind the AGPM, we use a novel centring routine applied individually to each $\sim0.35$s frame. This procedure is summarised here, while a more extensive discussion and a description of its performance will be the subject of a subsequent publication (Godoy et al., in prep.). We perform a two-component fit to the area around the AGPM centre, consisting of a Gaussian for the star and a second profile for the subtraction by the AGPM, which we represent by a Gaussian with negative flux. To improve the stability of the fit, we fix the FWHM and position of the AGPM to the value obtained as described below.

Since the position of the AGPM coronagraph moves on the NACO detector, we use the edges of the 15" circular aperture around the AGPM to calculate its centre in each image, and use a fixed offset between the centre of this aperture and the position of the AGPM mask calculated from the sky frames. We find that the AGPM moves within a 50x10 pixel region, and its position often changes on hour timescales by several pixels, making this a crucial step in the reduction process.
Due to the movement of the AGPM and the presence of significant jitter on short timescales with NACO, we find that this centring procedure outperforms other methods used in the literature.

We remove frames from the reduction that show 5$\sigma$ outlying values in several criteria using the median absolute deviation as a robust estimator of the standard deviation. First, we use the distance between the measured star position and AGPM position, and then we use the total flux measured in an annulus around the star between 40 and 190\,mas, and the scatter of the values in the same annulus.


\subsection{Cleaning of non-coronagraphic data}  \label{sec:red:clean2}

In the same way as for the AGPM coronagraphic data, we first correct the defective columns in the first quadrant as described in the previous section.
To subtract the sky background for non-coronagraphic data, we use the three-position dithering pattern used in our observations. Often, the stellar point spread function (PSF) is not clearly visible in the raw frames due to strong dark-current patterns on the detector and so an iterative approach to finding the star position must be taken. We first calculate rough sky frames by taking the mean of the frames in each three-dither sequence. We use the same PCA-based sky subtraction routine as for coronagraphic data, removing the first five modes. We then calculate the star position by performing a Gaussian fit to the PSF, recording the model parameters for each individual exposure.

Using the calculated star positions, we build a robust set of sky frames by again taking the mean of the frames in each three-dither sequence, masking out an area of $40\times40$\ pixels around the calculated star position in each exposure. We then apply the sky subtraction routine a second time using these frames, and repeat the centring procedure to ensure a robust determination of the star position.

We apply automatic criteria to remove bad frames from our data, by removing $5\sigma$\ outliers in the flux and FWHM measured from the Gaussian fit, calculated across all frames of the entire observing sequence. We also reject $5\sigma$\ outliers in the star position calculated within each cube (i.e. each dither is treated separately).


\subsection{Angular differential imaging and point-source detection limits}  \label{sec:red:adi}

The remainder of the reduction procedure is common to both coronagraphic and non-coronagraphic observations. Data are binned by centring each individual exposure and taking the median over each original data cube (roughly 100 frames, or 35 seconds). We apply a phase ramp to the Fourier transform of each image and take the inverse transform in order to recentre the data.

We then use a PCA-based algorithm to subtract the flux from the star \mbox{\citep{soummer2012,amara2012}}. We apply PCA in annuli with a width of 2 FWHM. For each frame, we build a reference library using those frames where the change in parallactic angle is sufficiently high that a companion would have moved by more than 0.75\,FWHM. We perform a number of reductions by subtracting 10-50\% of the available PCA modes, using the 30\% reduction as our baseline for the detection limit calculation. The frames are then derotated using Fourier transforms and median-combined to produce a final image. By using Fourier transforms to shift and derotate the images, the amplitude and structure of the noise in each image is preserved as well as the amplitude and shape of any potential signals and the stellar PSF. In addition, both steps are reversible when done in this way. This stands in contrast to interpolation-based methods, which have the effect of a low-pass filter and are non-reversible. These benefits were explored by \mbox{\citet{larkin1997}} and \mbox{\citet{hagelberg2016}} among others.

Before calculating the detection limits, we first subtract large-scale structures by subtracting the median within a $20\times20$\ pixel box from each pixel. We estimate a 1D noise curve by taking the standard deviation of the pixel values in annuli of 1$\times$FWHM in width. 
This is then converted to a detection limit curve using the peak stellar flux corrected for small sample statistics and for the throughput of the PCA algorithm. For correcting for small sample statistics, we follow the approach proposed by \mbox{\cite{mawet2014}} such that a constant false positive probability of \mbox{$2.9\times10^{-7}$} (equivalent to 5\,$\sigma$\ for Gaussian noise) is maintained.
The throughput of the PCA algorithm is estimated by injecting the mean stellar PSF into the cleaned, binned images with a signal-to-noise ratio (S/N) of 7 and repeating the PCA reduction. The ratio of input and output flux is then recorded. This is repeated ten times in each annulus, with the azimuth of the injected PSF changed each time. The mean of these values is used to correct the 1D detection limit. The transmission of the AGPM coronagraph was measured on-sky (see Appendix\,\ref{app:agpmtrans}) and a corresponding correction applied where applicable.


\section{Results from the first 2.5 years of observations} \label{sec:res}

\subsection{Overview and detection thresholds} \label{ssec:res:overview}

\begin{figure}[htb]
 \centering
 \includegraphics[width=9cm]{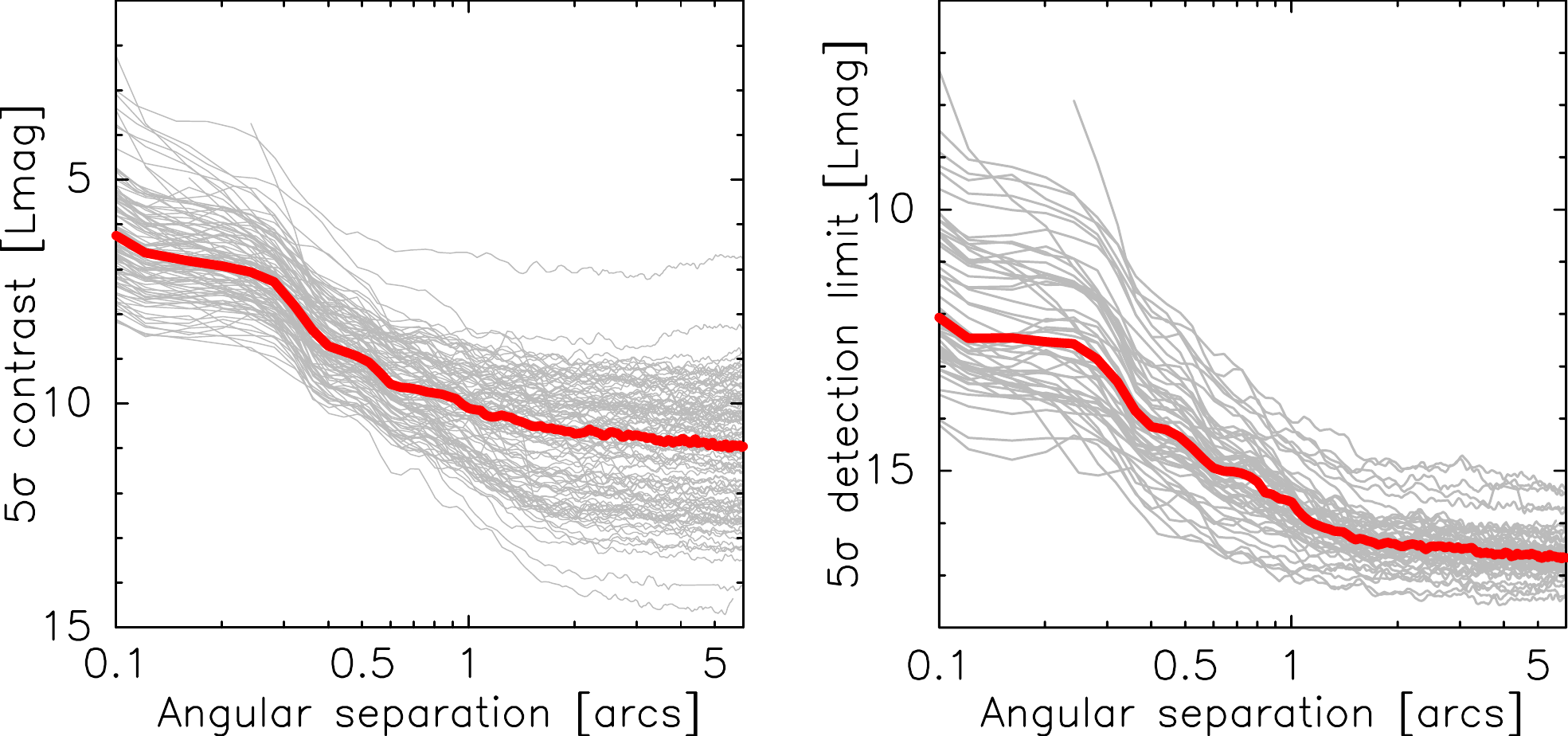}
 \caption{\label{fig-contr}
  Left: $5\,\sigma$\ $L^{\prime}$\ contrast curves for all targets observed during P96 through P100 (light grey, corrected for AGPM throughput where applicable; see Sect.\,\ref{sec:obs}) and median contrast (thick red). Right: $5\,\sigma$\ $L^{\prime}$\ detection limit curves.}
\end{figure}

Tables\,\ref{tab:obsslist1} and \ref{tab:obsslist2} list the main observing parameters, $L^{\rm \prime}$\ magnitudes, and achieved 5\,$\sigma$\ contrast values at six angular separations for all 33 PPD and 78 DEB targets observed during ESO periods 96 through 100.
We use the classical 5\,$\sigma$\ approach here, corrected for both small sample statistics \mbox{\citep{mawet2014}} and AGPM throughput, to characterise our detection thresholds and discovery space in a comparable way. At this stage of the survey, we do not employ an automated detection algorithm and therefore do not quantify our detection limits in terms of 95\% completeness like, for example, \mbox{\citet{wahhaj2013a}} or \mbox{\citet{stone2018}}. This must be kept in mind when comparing the detection spaces of different surveys.

We achieve a median $5\,\sigma$\ $L^{\prime}$\ contrast at 150\,mas of \mbox{$\Delta L^{\prime} = 6.4\pm0.1$\,mag} with best and worst values of $\sim$8\,mag\ and $\sim$4\,mag, respectively. This contrast close to the IWA is to first order independent of the brightness of the star. The scatter seen in Fig.\,\ref{fig-contr} is mainly caused by variations in the observing conditions. At separations $>3$\arcsec, we achieve a slightly stellar brightness-dependent $5\,\sigma$\ detection limit of \mbox{$L^{\prime}_{\rm bg} = (15.7\pm0.2) + L_{\ast}\times (0.16\pm0.03)$\,mag} with best and worst values of 17.5\,mag and 15\,mag, respectively (Fig.\,\ref{fig-contr}).

\begin{figure*}[htb]
 \centering
 \includegraphics[width=16cm]{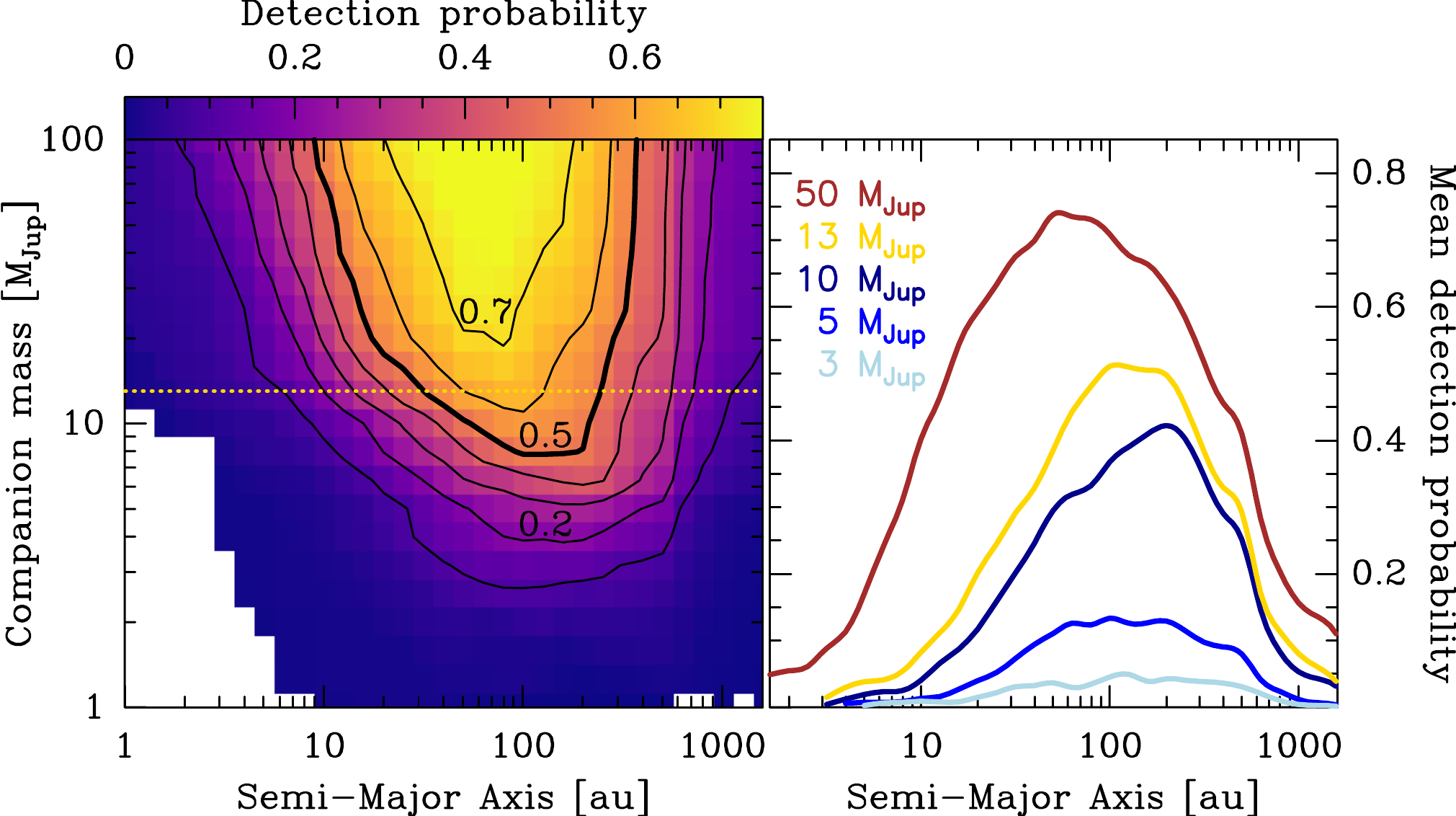}
 \caption{\label{fig-detspace}
  {\it Left:} Detection probability (colour map) as function of companion mass and semi-major axis for all 112 targets observed during P96 through P100. Contours show detection probabilities from 0.1 to 0.7 (bold: 0.5); the maximum is 0.8 (at SMA\,$\approx$\,60\,au and $M\gtrsim60$\,M$_{\rm Jup}$. The horizontal yellow dotted line marks the approximate boundary between GP and BD (13\,M$_{\rm Jup}$). {\it Right:} Probability curves for different masses as function of SMA.}
\end{figure*}

To evaluate the companion detection space of our survey, we first convert the 5\,$\sigma$\ contrast curves (Fig.\,\ref{fig-contr}) to planet mass detection limit curves using the $L^{\prime}$\ magnitudes, ages, and distances of the stars (Tables\,\ref{tab:slist1}, \ref{tab:slist2}, \ref{tab:obsslist1}, and \ref{tab:obsslist2}) together with the CIFIST2011bc BT-Settl evolutionary model isochrones for the NACO passbands \mbox{\citep{allard2014,baraffe2015}}. The BT-Settl models adopt a `hot start' and  solar abundances and employ a cloud model, which puts them in between the two models assuming atmospheric extremes with respect to dust and cloud opacity, COND \mbox{\citep[no photospheric dust opacity,][]{baraffe2003}} and DUSTY \mbox{\citep[maximal dust opacity,][]{chabrier2000}}. We then run Monte-Carlo (MC) simulations in which we assign each star one planet with random mass, semi-major axis (SMA), orbit orientation, and orbital phase, and verify if this planet is above or below the mass detection threshold (5\,$\sigma$) at the respective projected separation. To average down the MC noise, we perform $10^6$\ mock survey runs. If probability distribution functions (PDF) for planet occurrence rate, masses, and SMA are invoked, the same setup can be used to predict how many planets (and BDs) we should have detected given an assumed underlying PDF. Eccentricities are currently set to zero, but assumptions on the eccentricity distribution will be taken into account for the complete statistical analysis of the entire survey.

Figure\,\ref{fig-detspace} shows our resulting detection space in terms of achieved detection probability versus companion mass and SMA\footnote{We note that we consider the true SMA in 3D space together with an arbitrary orbit inclination and arbitrary orbital phase of the companion, such that the detection probability always remains below one.}.
Our detection probability is $\geq$50\% for companions with \mbox{$M\geq8$\,M$_{\rm Jup}$} in the SMA range \mbox{80\,--\,200\,au} and \mbox{$M>13$\,M$_{\rm Jup}$} at \mbox{30\,--\,250\,au}, but reaches at the 10\% level down to 3\,M$_{\rm Jup}$\ at \mbox{40\,--\,500\,au} and out to \mbox{5\,--\,1000\,au} for companions with \mbox{$M>13$\,M$_{\rm Jup}$}.

\begin{table*}
\caption{\label{tab:obsslist1} Observations and achieved contrast values of PPD targets.}
\begin{centering}
\begin{tabular}{lccccccccccc}
\hline\hline
Name & obs date & Seeing  & field    & AGPM & L$^{\prime}$\tablefootmark{a} & \multicolumn{6}{c}{-----------------\ 5\,$\sigma$\ contrast\tablefootmark{b} at $r$\ -----------------} \\
           &               &               & rot.         &            &            &  0\asp25 & 0\asp5 & 0\asp75 & 1\asp0 & 2\asp0 & 3\asp0  \\
           &               & [arcsec] & [deg]  &            & [mag] & [mag] & [mag] & [mag] & [mag] & [mag] & [mag] \\
\hline
V$^{\ast}$\,V892\,Tau    & 2016-12-10 & 1.3 & 50  & y & 4.75$\pm$0.19 & 5.8 & 7.8 & 9.3 & 9.8 & 10.6 & 10.7 \\
HD\,283571               & 2016-12-09 & 1.0 & 41  & y & 5.30$\pm$0.20 & 6.1 & 8.4 & 9.1 & 10.2 & 11.6 & 11.8 \\
HD\,31648                & 2017-11-01 & 1.0 & 66  & y & 4.51$\pm$0.25 & 6.9 & 9.6 & 10.5 & 11.1 & 12.0 & 12.0 \\
HD\,34282                & 2016-11-07 & 0.8 & 118 & y & 6.59$\pm$0.07 & 5.9 & 7.6 & 8.2 & 8.7 & 9.4 & 9.6 \\
HD\,37411                & 2017-11-03 & 0.8 & 70  & y & 6.26$\pm$0.09 & 6.0 & 7.3 & 8.6 & 9.7 & 10.1 & 10.1 \\
HD\,37806                & 2017-10-30 & 0.9 & 55  & y & 4.18$\pm$0.30 & 6.9 & 9.8 & 10.5 & 11.3 & 12.0 & 12.1 \\
HD\,38120                & 2017-10-29 & 0.5 & 66  & y & 6.25$\pm$0.09 & 6.3 & 8.6 & 9.5 & 9.8 & 9.9 & 10.1 \\
HD\,259431               & 2016-12-11 & 1.6 & 38  & y & 3.95$\pm$0.35 & 5.8 & 8.3 & 9.2 & 10.1 & 11.0 & 10.8 \\
NX\,Pup                  & 2018-02-22 & 0.4 & 83  & y & 4.17$\pm$0.40 & 3.7 & 8.8 & 10.6 & 11.3 & 12.4 & 12.4 \\
HD\,58647                & 2018-02-23 & 0.6 & 127 & y & 3.85$\pm$0.41 & 7.6 & 9.9 & 11.1 & 11.7 & 12.6 & 12.7 \\
HD\,72106                & 2016-12-12 & 0.9 & 130 & n & 6.76$\pm$0.04 & 6.7 & 7.4 & 7.6 & 7.8 & 10.1 & 10.2 \\
HD\,85567                & 2018-03-01 & 0.4 & 62  & y & 4.22$\pm$0.34 & 7.9 & 10.3 & 11.1 & 11.4 & 12.5 & 12.5 \\
TW\,Hya                  & 2016-05-03 & 0.9 & 134 & n & 6.96$\pm$0.06 & 5.2 & 7.9 & 8.3 & 9.0 & 9.4 & 9.5 \\
HD\,95881                & 2017-03-14 & 0.7 & 57  & y & 4.11$\pm$0.25 & 8.6 & 10.3 & 10.7 & 11.6 & 12.5 & 12.5 \\
HD\,97048                & 2016-05-02 & 1.0 & 67  & y & 4.47$\pm$0.10 & 5.8 & 8.5 & 9.4 & 10.0 & 11.0 & 11.1 \\
HD\,98922                & 2018-02-22 & 0.5 & 86  & y & 2.90$\pm$0.20 & 6.7 & 9.8 & 11.1 & 12.3 & 13.8 & 14.0 \\
HD\,100453               & 2016-05-09 & 1.2 & 85  & y & 4.31$\pm$0.35 & 6.9 & 9.4 & 10.0 & 10.0 & 11.6 & 11.7 \\
HD\,100546               & 2017-03-15 & 1.2 & 60  & y & 3.92$\pm$0.27 & 6.5 & 8.8 & 10.3 & 11.1 & 11.8 & 12.1 \\
HD\,101412               & 2017-03-17 & 0.8 & 69  & y & 5.70$\pm$0.09 & 8.1 & 9.3 & 9.8 & 10.3 & 10.7 & 10.7 \\
HD\,104237               & 2017-05-16 & 0.8 & 45  & y & 3.44$\pm$0.45 & 7.5 & 10.1 & 11.0 & 11.2 & 13.1 & 13.3 \\
PDS\,70                  & 2016-06-01 & 0.6 & 84  & n & 7.91$\pm$0.03 & 5.2 & 7.2 & 8.2 & 8.3 & 8.6 & 8.7 \\
HD\,139614               & 2017-05-01 & 1.2 & 101 & y & 5.64$\pm$0.12 & 7.7 & 8.6 & 9.6 & 10.0 & 10.4 & 10.6 \\
HD\,141569               & 2016-05-02 & 0.7 & 50  & y & 6.07$\pm$0.08 & 6.7 & 8.0 & 8.4 & 8.7 & 9.2 & 9.3 \\
IM\,Lup                  & 2017-05-15 & 1.3 & 120 & n & 6.80$\pm$0.06 & 7.1 & 9.1 & 9.9 & 10.1 & 10.3 & 10.3 \\
HD\,142527               & 2017-05-17 & 0.8 & 107 & y & 3.92$\pm$0.39 & 7.8 & 10.0 & 10.5 & 11.2 & 12.6 & 12.7 \\
HD\,144668               & 2017-06-16 & 0.7 & 61  & y & 3.18$\pm$0.16 & 5.9 & 8.1 & 8.6 & 9.3 & 10.8 & 11.2 \\
Elias\,2-27              & 2017-05-18 & 0.5 & 173 & n & 7.20$\pm$0.04 & 5.0 & 6.9 & 7.5 & 8.3 & 9.2 & 9.1 \\
KK\,Oph                  & 2016-08-01 & 0.7 & 171 & y & 4.08$\pm$0.31 & 7.1 & 9.3 & 10.2 & 10.3 & 11.2 & 12.1 \\
HD\,319139                   & 2016-05-03 & 0.7 & 161 & n & 7.12$\pm$0.05 & 5.9 & 8.1 & 8.5 & 8.8 & 9.1 & 9.2 \\
HD\,169142               & 2017-05-18 & 0.7 & 110 & y & 5.99$\pm$0.10 & 6.2 & 8.2 & 8.9 & 9.3 & 10.1 & 10.1 \\
R\,CrA                   & 2017-05-19 & 0.7 & 36  & y & 1.78$\pm$0.05 & 3.6 & 6.1 & 7.5 & 8.8 & 11.2 & 11.7 \\
T\,CrA                   & 2017-05-15 & 1.2 & 122 & y & 6.41$\pm$0.05 & 6.4 & 8.9 & 10.3 & 10.8 & 11.8 & 11.8 \\
HD\,179218               & 2016-05-02 & 0.9 & 63  & y & 4.47$\pm$0.25 & 7.5 & 9.4 & 10.0 & 10.7 & 11.5 & 11.7 \\
HD\,190073               & 2017-07-01 & 0.7 & 58  & y & 4.34$\pm$0.31 & 8.0 & 10.2 & 10.8 & 11.3 & 12.3 & 12.4 \\
\hline
\end{tabular}
\end{centering}
\tablefoottext{a}{$L^{\prime}$\ magnitudes and their uncertainties are derived by interpolation between WISE bands W1 (\SI[]{3.35}{\micro\meter}) and W2 (\SI[]{4.6}{\micro\meter}) to \SI[]{3.8}{\micro\meter} \mbox{\citep{cutri2013}}.}
\tablefoottext{b}{Corrected for AGPM throughput where applicable (see Sect.\,\ref{sec:obs}).}
\end{table*}



\subsection{Characterisation of known companions} \label{ssec:res:comp1}

Within the survey, we have observed and characterised a few stars with previously known companions, have detected and confirmed a number of previously unknown companions, have detected several discs at $L^{\prime}$-band, and have identified several companion candidates for which analysis and confirmation is still ongoing. For this reason, we also cannot yet provide a statistical analysis with constraints on the underlying population of GPs and BDs.

Early results on individual targets have been published in separate papers.
The discovery of a planetary-mass companion within the gap of the PPD around the pre-main sequence star PDS\,70 (V$^{\ast}$\,V1032\,Cen) by VLT/SPHERE observations along with $L^{\prime}$\ characterisation by VLT/NACO observations was published by \mbox{\citet{keppler2018}} and \mbox{\citet{mueller2018}}. 
A detailed characterisation of the SPHERE-discovered warm, dusty giant planet around the young A2 star HIP\,65426 via $L^{\prime}$\ NACO and other observations was published by \mbox{\citet{cheetham2019}}.


\subsection{New and confirmed companions} \label{ssec:res:comp}

\begin{figure}[htb]
 \centering
 \includegraphics[width=9cm]{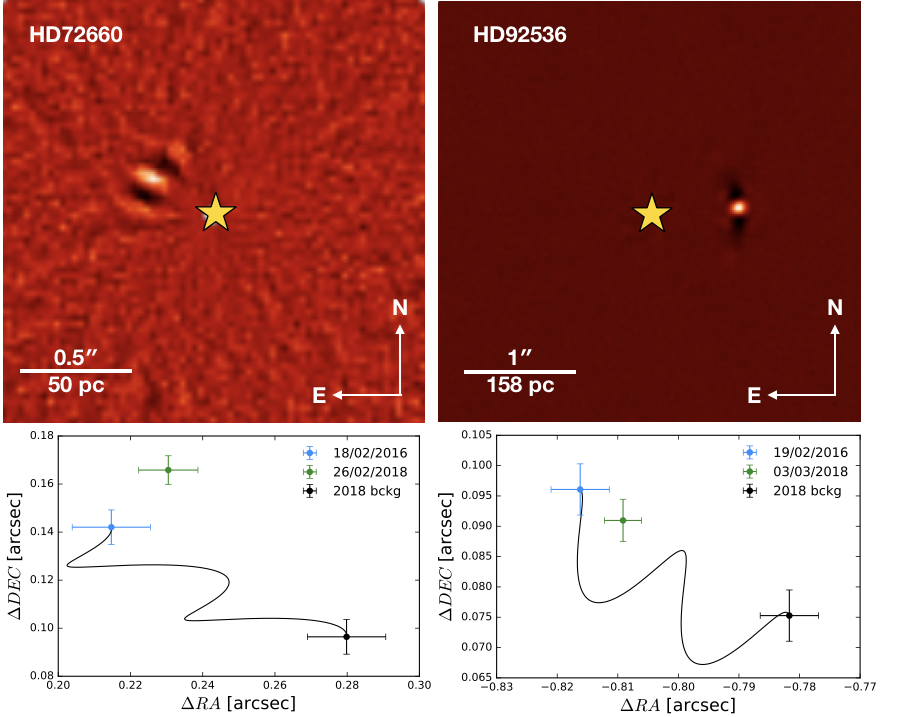}
 \caption{\label{fig-cc1}
  {\it Top:} Coronagraphic $L^{\prime}$-band images of HD\,72660 and HD\,92536, both with the newly detected low-mass stellar companions. Primary star positions are marked by asterisks.\newline
  {\it Bottom:} Proper motion analysis of the respective companion candidates. Blue and green data points with error bars show the relative positions of the companions at the two respective epochs. Black points indicate the position the companions would have had at the time of the respective second epoch if they were distant background stars with no proper motion, using the first epoch position as starting point and considering the {\it Gaia}-DR2 proper motion and parallax of the primaries, including their uncertainties (which are however too small to be noticeable).}
\end{figure}

\setcounter{table}{5}
\begin{table*}[htb!]
\caption{\label{tab-stellarcc}Astrometry and photometry of newly identified close stellar companions.}
\centering
\begin{tabular}{lccccccc}
\hline\hline
Star & Date & $\Delta$\,RA & $\Delta$\,Dec & $\rho$ & PA & $\Delta L^{\prime}$ & $m_2$ \\
 & & [mas] & [mas] & [mas] & [deg] & [mag] & [M$_{\odot}$] \\
\hline
HD\,72600\,B & 2016-02-17 & $215\pm10$ & $142\pm7$ & $257\pm10$ & $56.5\pm0.2$ & $5.36\pm0.4$ & $0.44\pm0.04$ \\
             & 2018-02-25 & $230\pm8$ & $166\pm6$ & $284\pm10$ & $54.3\pm0.2$ & $5.62\pm0.2$ & \\
\hline
HD\,92536\,B & 2016-02-18 & $-816\pm5$ & $96\pm4$ & $822\pm5$ & $276.7\pm0.3$ & $5.16\pm0.03$ & $0.33\pm0.05$ \\
             & 2018-03-02 & $-809\pm3$ & $91\pm3$ &     $814\pm3$ & $276.4\pm0.2$ &       $5.57\pm0.04$ & \\
\hline             
\end{tabular}\newline
\end{table*}

The discovery of a close (18.7\,au) low-mass stellar companion to the young (1-3\,Myr) PPD star R\,CrA was published by \mbox{\citet{cugno2019a}}. This companion was independently and simultaneously also discovered by {\it SPHERE} \mbox{\citep{mesa2019}}.
Another close (11\,au) low-mass stellar companion residing within the gap between the host star and its DEB was discovered around HD\,193571 by \mbox{\citet{musso2019}}.

Around at least two other stars (HD\,72660 and HD\,92536), we find previously unmentioned close ($<1\arcsec$) low-mass stellar companions, both shown in Fig.\,\ref{fig-cc1}. For both companions, we can already astrometrically reject the background hypothesis and confirm co-motion.
HD\,72660 is a $\sim$200\,Myr old A0 star with no significant IR excess 
at a distance of $98.8\pm0.1$\,pc (Table\,\ref{tab:slist2}). 
We find a previously unmentioned secondary source at a mean angular separation of $\rho\approx270$\,mas and P.A.\,$\approx55\degr$\ with $\Delta L^{\prime}\approx5.5$\,mag (Fig.\,\ref{fig-cc1}). The background star hypothesis could be rejected and co-motion confirmed via second-epoch observations (Table\,\ref{tab-stellarcc}). Using BT-Settl evolutionary models \mbox{\citep{allard2014,baraffe2015}} and adopting the distance of the primary, the $L^{\prime}$\ brightness and astrometry correspond to a $0.44\pm0.04\,{\rm M}_{\odot}$\ (M1) companion at a projected separation of \mbox{$R=24.4\pm0.5$\,au}. We also detect significant relative motion between primary and secondary with $\Delta\rho\approx13.5$\,mas/yr and $\Delta$\,P.A.\,$\approx1.1\degr$/yr.
Since we only have two astrometric measurements covering only a small orbital arc without any constraint on curvature, we use the prescription provided by \mbox{\citet{pearce2015}} to verify whether or not the secondary could be in a bound orbit. Adopting a primary mass of 2.4\,M$_{\odot}$\ \mbox{\citep{chen2014}}, the astrometry given in Table\,\ref{tab-stellarcc}, and the {\it Gaia} distance, we derive\footnote{http://drgmk.com/imorbel/} the dimensionless parameter 
\mbox{$B = V_{\rm sky}/V_{\rm esc}=0.14^{+0.14}_{-0.09}$} 
\mbox{\citep{pearce2015}}, indicating that the companion is very likely bound provided the SMA of its (possibly eccentric) orbit is larger than 
\mbox{$a_{\rm min} = R/2\,(1-B)^{-1}\approx 14.7$\,au.}\footnote{$B>1$\ would indicate the companion is unbound.} 
No meaningful constraints can be put on other orbital parameters such as eccentricity or inclination at this point, except that an edge-on orbit can be excluded ($i<86\degr$). A chance projection of an unbound object cannot be strictly ruled out as long as there is no constraint on the curvature of motion, for example by third epoch astrometry. However, this is extremely unlikely since no other point source is detected in this field, that is, the local stellar density must be small.

HD\,92536 is a $\sim$46\,Myr old B8 star at $157.3\pm1.1$\,pc with a debris belt at $r_{\rm DD} = 13\pm 3$\,au  (Table\,\ref{tab:slist2}). We find a previously unmentioned secondary source at $\rho\approx818$\,mas and P.A.\,$\approx276.5\degr$\ with $\Delta L^{\prime}\approx5.36$\,mag (Fig.\,\ref{fig-cc1}). The background star hypothesis could be rejected and co-motion confirmed via second-epoch observations (Table\,\ref{tab-stellarcc}). Using again the BT-Settl models, the $L^{\prime}$\ brightness and astrometry correspond to a $0.33\pm0.05\,{\rm M}_{\odot}$\ (M2-4) companion at a projected separation of $\sim129\pm1$\,au. We also detect significant relative motion between primary and secondary of $\Delta\rho\approx-3.9$\,mas/yr and $\Delta$\,P.A.\,$\approx0.15\degr$/yr.
Adopting a primary mass of 3\,M$_{\odot}$\ \mbox{\citep{chen2014}}, we derive \mbox{$B = 0.15^{+0.43}_{-0.14}$} \mbox{\citep{pearce2015}}, indicating that also this companion is very likely bound provided \mbox{$a>a_{\rm min}\approx 76$\,au.} No meaningful constraints can be put on other orbital parameters at this point, except that an edge-on orbit can be excluded ($i<85\degr$).

\subsection{Disc detections} \label{ssec:res:discs}

\begin{figure*}[htb]
 \centering
 \includegraphics[width=18.5cm]{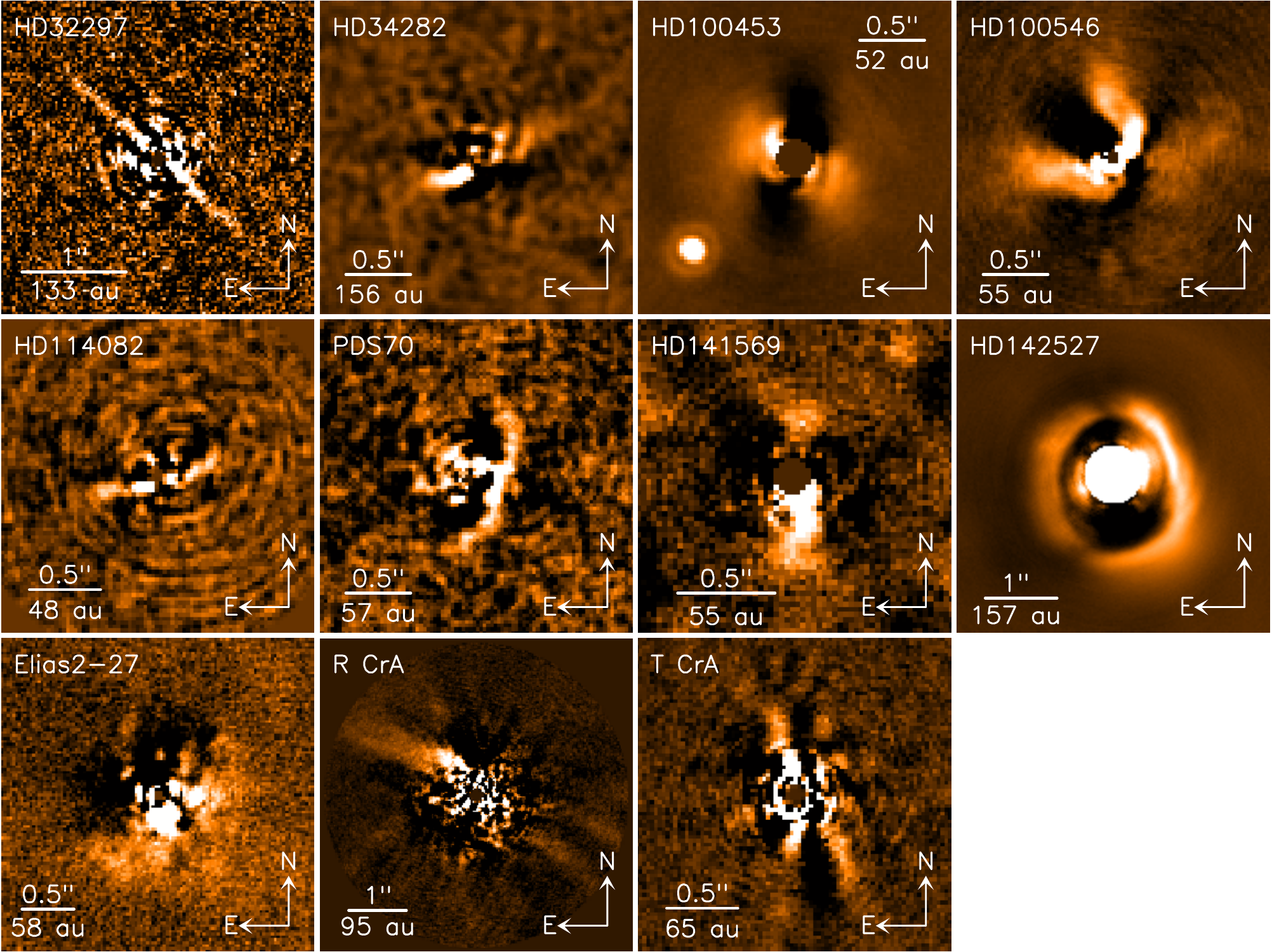}
 \caption{\label{fig-discs}
  NACO $L^{\prime}$-band images of circumstellar discs. 
  }
\end{figure*}

Although our ADI observations were not optimised to detect extended emission from circumstellar discs, we have detected scattered light at $L^{\prime}$-band from discs around 11 of the 112 target stars observed in P\,96 through 100 (Fig.\,\ref{fig-discs}). All of these discs except R\,CrA have been imaged at millimetre wavelengths with ALMA.
Five of the eleven discs have already been well-studied and also detected and imaged at $L^{\prime}$-band (and shorter NIR bands).
These are the discs around 
the 20\,Myr-old A0 DEB star HD\,32297 at 133\,pc (\mbox{\citealt{schneider2005}}; \mbox{\citealt{rodigas2014a}}; \mbox{\citealt{bhowmik2019}}),
the 3.8\,Myr-old B9 PPD star HD\,100546 at 110\,pc (\mbox{\citealt{quanz2013,quanz2015b}}; \mbox{\citealt{avenhaus2014}}; \mbox{\citealt{currie2014}}; \mbox{\citealt{sissa2018}}),
the 5.4\,Myr old K5 PPD star PDS\,70 at 113\,pc \mbox{\citep{hashimoto2012,keppler2018}}, 
the 4.7\,Myr-old B9 PPD star HD\,141569 at 111\,pc
(\mbox{\citealt{currie2016}}; \mbox{\citealt{mawet2017}}; \mbox{\citealt{perrot2016}}), and 
the 2\,Myr-old F6 PPD star HD\,142527 at 157\,pc (\mbox{\citealt{rameau2012}}; \mbox{\citealt{canovas2017}}; \mbox{\citealt{avenhaus2017}}).

In this paper, we only show the $L^{\prime}$-band images of the detected discs without attempting any kind of further analysis, which would require data at other wavelengths and radiative transfer simulations. Since the $L^{\prime}$-band scattered light does not necessarily trace the discs in their entirety and our ADI observing mode and data reduction filters out significant parts of the emission, we also do not attempt to derive disc sizes from the images.

\section{Discussion} \label{sec:dis}

\subsection{Detection limits and preliminary statistical analysis}  \label{ssec:dis:stat}

Based on the 5-$\sigma$\ contrast curves shown in Fig.\,\ref{fig-contr}, stellar properties (Tables\,\ref{tab:slist1} and \ref{tab:slist2}), and the BT-Settl evolutionary models \mbox{\citep{allard2014,baraffe2015}}, we calculate planet mass detection limit curves for all stars and derive a map of the raw detection probabilities in companion mass\,--\,SMA space (Fig.\,\ref{fig-detspace}). With the same approach, but assuming an underlying companion mass function (CMF) for GPs and BDs from \mbox{\citet{reggiani2016}} together with a normalisation based on the newest analysis of all existing DI survey results from \mbox{\citet{galicher2016}}, we evaluate how many companions we should have already detected provided the assumptions for the underlying CMF holds for our target stars and provided we would have completed all our ongoing candidate confirmation efforts. We run MC simulations of our survey in which we assign every star a companion (one or none) according to the CMF, place the companion at a random position on a randomly oriented (circular) orbit, and count the expected detections. We find that we should have detected $2.0^{+3.7}_{-1.8}$\ companions in the GP-BD mass regime ($0.5^{+1.1}_{-0.3}$\ GP and $1.5^{+1.6}_{-0.5}$\ BD). While the poisson noise accounts for an uncertainty of the predicted number of companion detections of about $\pm1.3$, the largest part of the uncertainties stated above derives from the uncertainty in our current knowledge of the underlying CMF \mbox{\citep[see][]{galicher2016}}.

Assuming both the findings reported by \mbox{\cite{meshkat2017}} as well as their assumption on the functional shape of the underlying CMF, we should have detected $8^{+3.4}_{-2.2}$\ companions in the GP-BD mass regime ($2.0^{+1.6}_{-0.9}$\ GP and $6^{+3}_{-2}$\ BD). 
Since verification efforts for several BD as well as GP candidates are still ongoing, we cannot yet draw any robust conclusions on occurrence rates or distinguish between the two different predictions. At the end of our survey (including companion verification efforts), we plan a statistical analysis that does not only include the detection thresholds and detections of our survey, but includes that of other surveys as well. 


\subsection{Comparison to other surveys}  \label{ssec:dis:comp}

\begin{table*}
\caption{\label{tab:lssurveylist} Overview of deep exoplanet DI surveys}
\begin{tabular}{lllllllccl}
\hline\hline
Survey     &
Telescope  &
Instrument &
Filt.      &
Mode\tablefootmark{a}       &
\# of      &
Spectral   &
Dist.\tablefootmark{c}  &
Age\tablefootmark{c}     &
Notes\tablefootmark{d}      \\
or Reference & & & & & Targets\tablefootmark{b} & Types & [pc] & [Myr] & \\
\hline
{\it $L^{\prime}$-band surveys:} & & & & & & & & \\
\citet{kasper2007}  & VLT  & NACO         & $L^{\prime}$ & Sat-I   & 22 (4) & FGKM   & 37  & 10-30 & -- \\
\citet{heinze2010}  & MMT  & Clio         & $L^{\prime}$ & ADI     & 54 (4) & FGKM   & 11  & 200  & -- \\
\citet{delorme2012} & VLT  & NACO         & $L^{\prime}$ & ADI     & 16 (1) &  M     & 25  & 12   & -- \\
\citet{rameau2013a} & VLT  & NACO         & $L^{\prime}$ & ADI     & 59 (30) & BAFGKM & 40  & 30   & -- \\
\citet{meshkat2015} & VLT  & NACO         & $L^{\prime}$ & ADI     & 13 (3) & AF     & 48  & 40   & -- \\
\citet{lannier2016} & VLT  & NACO         & $L^{\prime}$ & ADI     & 58 (1) & M      & 38  & 21   & 1\\
\citet{stone2018}   & LBTI & LMIRcam & $L^{\prime}$      & ADI     & 98 (8) & BAFGKM & 25  & 400  & 2\\
NACO-{\it ISPY}     & VLT  & NACO         & $L^{\prime}$ & Cor-ADI & 200    & AFGK   & 60  & 30   & -- \\
{\it Other surveys:} & & & & & & & & \\
\citet{masciadri2005} & VLT  & NACO & $H, K_{\rm S}$ & Sat-I & 30 (3) & KM & 10-60 & 10-100 & -- \\
\citet{lafreniere2007} & Gemini-N & NIRI  & $H$          & ADI     & 85 (11) & FGKM & 22   & 10-5000  & 3 \\
\citet{chauvin2010} & VLT  & NACO   & $H, K_{\rm S}$ & Cor-I   & 88 (10) & BAFGKM & 42 & $\lesssim100$ & 4 \\
\citet{wahhaj2013a} & Gemini-S & NICI  & $H$    & ADI/ASDI & 57 (31) & AFGKM   & 43  &  100   & 5, 13 \\
\citet{biller2013}  & Gemini-S & NICI  & $H$    & Cor-ASDI & 80 (17)  & BAFGKM & 40  & 10-200 & 5, 14 \\
\citet{nielsen2013}  & Gemini-S & NICI & $H$    & Cor-ASDI & 70 (23) & BA    & $\lesssim75$ &      & 5, 15 \\
\citet{janson2013a} & Subaru & HiCIAO  & $H$    & ADI    & 50 (22)  & AFGKM  & $\sim$30 & $\sim$200 & 6, 13 \\
\citet{brandt2014a}  & Subaru & HiCIAO  & $H$    & ADI    & 63 (4)  & AFGKM & $\lesssim50$  & $\lesssim500$ & 6, 14\\
\citet{chauvin2015} & VLT  & NACO      & $H$    & ADI    & 85 (4)    & FGK  & 66 & 100 & 7 \\
\citet{galicher2016}& Gemini-N      & NIRI      & $H, K$ & ADI    & 292 (33) & BAFGKM &  45  &  120  & 8\\
                    & Gemini-S      & NICI      & $H, K$ & ADI    & --      & --    & --  & -- & -- \\
                    & Keck\,II      & NIRC2     & $H, K$ & ADI    & --      & --    & --  & -- & -- \\
                    & VLT           & NACO      & $H, K$ & ADI    & --      & --    & --  & -- & 9 \\
\citet{hinkley2011} & Hale & Project\,1640 & $J, H$ & Cor-ASDI & ?  & ... & ... & ... & 10 \\
\citet{nielsen2019} & Gemini-S  & GPI  & $H$    & ASDI   & 300 {\bf (60)} & BAFGK & $\sim$30 & 10-600 & 11 \\
{\it SHINE}         & VLT  & SPHERE    & $H, K$ & ASDI   & $\sim$600  & BAFGKM & ... & ... & 12 \\
\hline
\end{tabular}
\tablefoottext{a}{Imaging modes: Sat-I: saturated imaging, ADI: angular differential imaging, Cor-ADI: coronagraphic ADI, ASDI: angular and spectral differential imaging.}
\tablefoottext{b}{In brackets the number of targets overlapping with NACO-{\it ISPY}.}
\tablefoottext{c}{A single number represents the median, otherwise the range is given.}
\tablefoottext{d}{Survey notes: (1) {\it MASSIVE},
(2) {\it LEECH}: LBTI Exozodi Exoplanet Common Hunt \mbox{\citep[see also][]{skemer2014}},
(3) {\it GDPS}: Gemini Deep Planet Survey,
(4) NACO Survey of Young Nearby Austral Stars,
(5) The Gemini/NICI planet finding campaign \mbox{\citep[see also][]{liu2010}},
(6) {\it SEEDS}: Strategic Exploration of Exoplanets with Subaru \citep[see also][]{tamura2016},
(7) {\it VLT/NACO-LP}: VLT Large Program to Probe the Occurence of Exoplanets and Brown Dwarfs at Wide Orbits \citep[see also][]{reggiani2016,vigan2017},
(8) {\it IDPS}: International Deep Planet Search \mbox{\citep{vigan2012}},
(9) {\it VLT/NACO} sub-survey: \citep{vigan2012},
(10) {\it Project\,1640}: a Coronagraphic Integral Field Spectrograph at Palomar and ongoing survey with the same name, 
(11) {\it GPIES}: Gemini Planet Imager Exoplanet Survey (600 stars planned), ongoing \mbox{\citep[see also][]{marchis2016,macintosh2018}},
(12) SpHere Infrared survEy, ongoing (\mbox{\citealt{vigan2016}}; \mbox{\citealt{chauvin2017a,chauvin2017b}}; \mbox{\citealt{keppler2018}}; \mbox{\citealt{mueller2018}}; and others).
(13) DEB stars,
(14) Young moving group stars, 
(15) Young B and A stars.}
\end{table*}

A list of deep imaging surveys of nearby (<100\,pc) stars carried out between 2003 and 2013 is conveniently provided by \mbox{\citet{chauvin2015}}. 
\mbox{\citet{stone2018}} list $L^{\prime}$-band surveys carried out between 2007 and 2017. 
\mbox{\citet{bowler2016}} provides a comprehensive overview and description of previous and ongoing large imaging surveys. 
To put our NACO-{\it ISPY} survey into context, we compile in Table\,\ref{tab:lssurveylist} a condensed overview of previous and ongoing large DI surveys dedicated to exoplanet searches, which is not complete and does not include many of the early and smaller surveys. 
In the context of this large variety of big surveys, our NACO-{\it ISPY} survey is the largest $L^{\prime}$-band survey and the largest survey (including shorter wavelengths) that exclusively targets young stars with either PPDs or DEBs.
A direct comparison between different surveys is not straightforward since there is no uniform metric, the detail of information provided by the various publications differs largely, detection limits (if provided) are derived in different ways, and a correction for small-sample statistics \mbox{\citep{mawet2014}} is not always applied. Harmonising the results regarding the two latter points would require uniform re-reduction of all data for direct comparison. 

We focus our comparison here on the following surveys:
{\it (1)} \mbox{\citet{rameau2013a}}, a deep $L^{\prime}$\ survey with NACO and with the largest target overlap,
{\it (2)} {\it LEECH}, the newest and deepest $L^{\prime}$\ survey with a small target overlap \mbox{\citep[northern sky,][]{stone2018}}, 
the two largest DEB samples from 
{\it (3)} the Gemini/{\it NICI} planet-finding campaign \mbox{\citep[$H$-band,][]{wahhaj2013a}} and
{\it (4)} {\it SEEDS} \mbox{\citep[$H$-band,][]{janson2013a}}, 
and 
{\it (5)} the first 300 stars from {\it GPIES}, which did not explicitly targets DEB stars but still has a large overlap with our DEB sample \mbox{\citep[$H$-band,][]{nielsen2019}}.
The purpose of the following mostly qualitative comparisons is to present our NACO-{\it ISPY} survey in the context of existing, comparable large DI surveys. For the reasons mentioned above and below, performance numbers can mostly not be compared directly and serve only to describe the approximate detection space of the different surveys.

Our NACO-{\it ISPY} survey shares the same instrument and filter and the largest overlap (30 targets) with the survey of young, nearby, and dusty stars conducted by \mbox{\citet{rameau2013a}}. In contrast to our observing strategy, \mbox{\citet{rameau2013a}} did not employ a coronagraph and restricted the time per target to 90\,min, irrespective of the field rotation achieved. Their median field rotation is of the order of 25\degr, while we achieve $\sim$85\degr (Fig.\,\ref{fig:histrot}). Contrast curves in \mbox{\citet{rameau2013a}} are given in the classical 5\,$\sigma$\ scheme, but are not corrected for small-sample statistics. Nevertheless, a direct comparison between the published curves reveals that we achieve on average \mbox{0.5-0.7\,mag} better contrasts at small angular separations of <300\,mas, that is, inside the wings of the PSF where the contrast is almost unaffected by the background limit (which the latter authors do not mention, but which is probably similar to what we achieve). Planet mass detection spaces as published appear similar between NACO-{\it ISPY} (Fig.\,\ref{fig-detspace}) and \mbox{\citet[][their Fig.\,6]{rameau2013a}}, but numbers cannot be compared directly because of different calibrations,  target star properties, and evolutionary models used (COND vs. BT-Settl in {\it ISPY}). For our anticipated final statistical analysis, we will  incorporate the \mbox{\citet{rameau2013a}} data after re-evaluating the significance of the DEB excess and coherent re-reduction of all data.

The most recently completed and probably deepest $L^{\prime}$-band survey, albeit with only a small overlap (eight targets), is the {\it LEECH}) survey conducted with the LBT on the northern sky \mbox{\citep{skemer2014,stone2018}}. {\it LEECH} also did not employ a coronagraph, but was able to benefit from the excellent LBTI AO system \mbox{\citep{bailey2014}}, deformable secondary mirrors, and the simultaneous but not interferometric use of two 8.4\,m mirrors. {\it LEECH} also performed ADI and achieved a mean field rotation of $\sim$65\degr. \mbox{\citet{stone2018}} show two sets of contrast curves: the classical 5\,$\sigma$\ contrasts without small sample correction and small-sample-corrected contrast curves based on 95\% completeness, which they use for the analysis. The `modern' 95\% completeness curves produced by these latter authors are on average 0.28\,mag less sensitive than their classical 5\,$\sigma$\ curves. 
A direct comparison between the {\it LEECH} 5\,$\sigma$\ curves and our small-sample-corrected median 5\,$\sigma$\ contrasts (Fig.\,\ref{fig-contr}) indicates that {\it LEECH} achieved significantly better $L^{\prime}$\ contrasts at small angular separations \mbox{($\Delta L^{\prime}\sim$\,9.0\,mag} at \mbox{$\Delta r$\,=\,300\,mas vs. $\sim$\,7.6\,mag for {\it ISPY})}, but has an IWA of $\sim$250\,mas, while the coronagraphic NACO-{\it ISPY} observations go down to 100-150\,mas.
Contrasts at larger angular separations are incomparable since they depend on the contrast between target star brightness and background limit (which we do not know for {\it LEECH}). The planet mass detection space of the entire {\it LEECH} survey as published by \mbox{\citet{stone2018}} covers a similar SMA/projected separation range to ours, but the depth cannot be directly compared without putting all data into the same model. Although {\it LEECH} did not explicitly target stars with DEBs and observed in the other hemisphere, our two surveys have eight targets in common. For our final statistical analysis, we incorporate the {\it LEECH} data not only of these eight stars, but also of other (northern) targets which prove to have a significant DEB.  

The {\it NICI} \mbox{\citep{wahhaj2013a}} and {\it SEEDS} \mbox{\citep{janson2013a}} DEB surveys targeted 57 and 50 stars, respectively, of which 31 and 22 are in common with NACO-{\it ISPY}. 
The {\it NICI} observing campaign employed a semi-transparent flat-topped Gaussian focal plane mask to reduce scattered light from the central star, which provided an effective IWA of 0\farcs32 for faint companions. \mbox{\citet{wahhaj2013a}}  achieve a mean 95\% completeness contrast (which for {\it NICI} agrees for most stars well with the traditionally used 5\,$\sigma$\ contrast; see \mbox{\citealt{wahhaj2013b}}) of \mbox{$\Delta H\approx10.5$\,mag} at 0\farcs36. In terms of planet mass detection limit \mbox{\citep[BT-Settl models,][]{allard2014,baraffe2015}}, this compares well with the median 5\,$\sigma$\ contrast of \mbox{$\Delta L^{\prime}\approx8.5$\,mag} which we achieve with NACO-{\it ISPY} at this angular separation (cf. Figs.\,\ref{fig:fig2} and \ref{fig-contr}). However, with our coronagraphic NACO-{\it ISPY} observations, we achieve a significantly smaller IWA of 100-150\,mas, which makes the two surveys truly complementary. The other two sub-surveys of the {\it NICI} campaign \mbox{\citep{biller2013,nielsen2013}} that also have certain target overlap with NACO-{\it ISPY} have similar detection limits.
The {\it SEEDS} observations were done in saturated ADI mode with the saturation extending typically out to a radius of 0\farcs3 and contrasts that typically stay below those achieved by {\it NICI}.

\mbox{\citet{nielsen2019}} recently published a statistical analysis of the first 300 stars observed by the Gemini Planet Imager Exoplanet Survey ({\it GPIES}). Although {\it GPIES} did not explicitly target DEB stars, the survey has 60 targets (mostly DEB) in common with NACO-{\it ISPY} (full sample). {\it GPIES} achieves a coronagraphic IWA at $H$-band of $0\farcs12$, which is very similar to what NACO achieves at $L^{\prime}$-band. \mbox{\citet{nielsen2019}} quantify the achieved contrasts in terms of standard deviation (8\,$\sigma$\ at $<0\farcs3$\ and 6\,$\sigma$\ further out) based on the matched-filter algorithm described by \mbox{\citet{ruffio2017}} and employ the BT-Settl models to convert these to planet detection thresholds. The detection probability \mbox{\citet{nielsen2019}} infer for {\it GPIES} with the aferomentioned thresholds and models (see their Fig.\,4) is $\geq$50\% for companions with \mbox{$\geq8$\,M$_{\rm Jup}$} in the SMA range \mbox{10\,--\,100\,au} and reaches down to 3\,M$_{\rm Jup}$\ at \mbox{3\,--\,200\,au at the 10\% level}. This compares well to our NACo-{\it ISPY} detection space albeit reaching somewhat lower SMA owing to the smaller mean distance of their targets.

Other very large ongoing exoplanet imaging surveys that have not yet published summary papers and for which we therefore cannot compare overlap and detection limits with our NACO-{\it ISPY} survey include SPHERE-{\it SHINE} \mbox{\citep{chauvin2017a}} and and {\it Project\,1640} \mbox{\citep{hinkley2011}}.
Our sensitivity map shown in Fig.\,\ref{fig-detspace} agrees very well (50\% contour) with the mean sensitivity map for FGK stars derived by \mbox{\citet{bowler2016}} from the meta-analysis of 384 stars with published high-contrast imaging observations.


\section{Summary and outlook} \label{sec:con}

We present an overview of the NACO-{\it ISPY} DI survey for planets around young stars, its scientific goals, observation strategy, targets, and data-reduction scheme. Below we summarise the performance and preliminary results from the first 2.5 years of observations.
\begin{itemize}

\item With NACO-{\it ISPY}, we target $\approx$\,200 young (median age 30\,Myr) and nearby (median distance 60\,pc) stars that are either surrounded by a gas-rich PPD with indications for inner holes or gaps ($\sim$50 stars), or by somewhat older well-characterised DEBs ($\sim$150 stars).

\item During the first 2.5 years of the survey, from December 2015 through February 2018 (ESO periods 96 through 100), we observed 112 target stars (34 PPDs and 78 DEBs).

\item All observations are carried out with the NACO L27 camera at the VLT, with $L^{\prime}$\ filter and in pupil-tracking ADI mode. For brighter stars ($L^{\prime}\lesssim6.5$\,mag), the AGPM vector vortex coronagraph is used; fainter stars are observed in saturated mode. The thermal sky background is derived by switching to an offset sky position every 8 min when the AGPM is used. Otherwise, a dither pattern is used. The ADI observation is enclosed by unsaturated flux measurements of the target star itself.

\item We typically spend 2-4\,hr on one source around meridian passage and achieve field rotations of typically $90\degr\pm20\degr$. With this approach, we reach a mean planet--star 5\,$\sigma$\ contrast of $\Delta L^{\prime}\sim7$\,mag at 0\farcs2 and a background detection limit of $\sim$16.5\,mag at >1.5\arcsec.

\item All data are homogeneously reduced using the GRAPHIC pipeline \mbox{\citep{2016MNRAS.455.2178H}}, which was optimised for NACO-{\it ISPY} data. After initial cleaning and corrections for defective detector columns and pixels, a PCA-based sky-subtraction scheme with five modes is used to remove the background from each image. For the coronagraphic data sets, we employ a novel centring procedure to determine a posteriori the (drifting) position of the AGPM on the detector and the position of the star behind the AGPM. The star position for the non-coronagraphic data is calculated by performing a Gaussian fit to the PSF in each individual exposure. Automatic criteria are used to iteratively remove bad frames and outliers. The data are then centred and median-binned ($\sim$100 frames or 35\,sec), and a PCA-based algorithm is used to subtract the star. Frames are then derotated using Fourier transforms and median combined to produce the final image. 

\item One-dimensional noise curves are estimated from the standard deviation of the pixel values in 1\,FWHM-wide annuli and converted to 5\,$\sigma$\ contrast curves using the peak stellar flux from the unsaturated flux images. The contrast curves are corrected for small sample statistics, for the transmission of the AGPM coronagraph measured on-sky, and for the throughput of the PCA algorithm.

\item We evaluate the planet mass detection space of our survey by combining the achieved 5\,$\sigma$\ $\Delta L^{\prime}$\ contrast curves with actual stellar parameters, BT-Settl evolutionary models, and MC survey simulations. The mean detection probability of our survey is >50\% for companions with \mbox{$M\gtrsim$\,8\,M$_{\rm Jup}$} in the SMA range  \mbox{80\,--\,200\,au} and \mbox{$M>13$\,M$_{\rm Jup}$} at \mbox{30\,--\,250\,au} and compares well to the detection space of other state-of-the-art high-contrast imaging surveys \mbox{\citep{bowler2016}}.

\item While we have not yet been able to independently discover and confirm a new planet in our target sample, our observations have already contributed to the characterisation of two new planets originally discovered by SPHERE: HIP\,65426\,B \mbox{\citep{chauvin2017b,cheetham2019}} and PDS\,70\,B \mbox{\citep{keppler2018,mueller2018}}. We discovered two new close-in low-mass stellar companions around the young PPD star R\,CrA \mbox{\citep{cugno2019a}} and within the DEB around HD\,193571 \mbox{\citep{musso2019}}. Around at least two other stars (HD\,72660 and HD\,92536), we find and astrometrically confirm previously unmentioned close ($<1\arcsec$) low-mass co-moving stellar companions.

\item We detected scattered light at $L^{\prime}$-band around at least nine PPD stars and two DEB stars. Six of these discs had never before been imaged at $L^{\prime}$-band.

\item As data-reduction improvements, candidate identification, and follow-up characterisation are still ongoing, further discoveries may be revealed even from the data already presented in this paper.

\item The NACO-{\it ISPY} GTO survey observations with a total budget of 120 nights will be completed in late 2019 with a total of approximately 200 stars imaged at $L^{\prime}$-band. A statistical analysis of the entire survey and the synthesis with complementary observations (see below) will be presented in an upcoming paper.

\end{itemize}

To complement our GP discovery space towards smaller separations (<1-2\,au), we are currently carrying out a complementary RV survey for planets around the DEB stars from the NACO-{\it ISPY} list ({\it RVSPY}, Zakhozhay et al., in prep.). On a timescale of 2-3\,years, the synergy of these two surveys will provide us with the most complete census of GPs around young DEB stars to date, although there will still be a mass sensitivity gap for lower-mass GPs in the DI-probed region ($\gtrsim$10\,au) and a general sensitivity gap in the 2-10\,au region. This is where astrometry will come into play. The anticipated final full release of the {\it Gaia} data (individual measurements) will provide us in the near future (no specific release time has yet been given) with the necessary data to reveal the still incompletely known GP population in the \mbox{3\,--\,5\,au} separation range \mbox{\citep[e.g.][]{casertano2008}}.


\begin{acknowledgements}
This work has made use of data from the European Space Agency (ESA) mission {\it Gaia} (\url{https://www.cosmos.esa.int/gaia}), processed by the {\it Gaia} Data Processing and Analysis Consortium (DPAC, \url{https://www.cosmos.esa.int/web/gaia/dpac/consortium}). Funding for the DPAC has been provided by national institutions, in particular the institutions participating in the {\it Gaia} Multilateral Agreement. This publication also makes use of data products from the Wide-field Infrared Survey Explorer, which is a joint project of the University of California, Los Angeles, and the Jet Propulsion Laboratory/California Institute of Technology, funded by the National Aeronautics and Space Administration. This research has made use of the SIMBAD database and the VizieR catalogue access tool, both operated at CDS, Strasbourg, France. The original description of the VizieR service was published in \mbox{\citet{vizier2000}} . This research made use of Astropy,\footnote{http://www.astropy.org} a community-developed core Python package for Astronomy \mbox{\citep{astropy2013, astropy2018}}. T.\,H. acknowledges support from the European Research Council under the Horizon 2020 Framework Program via the ERC Advanced Grant Origins 832428. A.\,M. and A.\,Q. acknowledge the support of the DFG priority program SPP 1992 ``Exploring the Diversity of Extrasolar Planets'' (MU 4172/1-1). GC and SPQ thank the Swiss National Science Foundation for financial support under grant number 200021\_169131. GMK is supported by the Royal Society as a Royal Society University Research Fellow. J.\,O. and N.\,G. acknowledge financial support from the ICM (Iniciativa Cient\'ifica Milenio) via the N\'ucleo Milenio de Formaci\'on Planetaria grant. J.\,O. acknowledges financial support from the Universidad de Valpara\'iso, and from Fondecyt (grant 1180395). N.\,G. acknowledges grant support from project CONICYT-PFCHA/Doctorado Nacional/2017 folio 21170650. Part of this work has been carried out within the framework of the National Centre of Competence in Research PlanetS supported by the Swiss National Science Foundation (SNSF). We acknowledge the financial support of the SNSF. We also acknowledge helpful discussions and direct contributions to various aspects in the survey definition phase and data handling by Ingo Stilz, Carlos Eiroa, Alexander V. Krivov, Nicole Pawellek, Attila Mo{\'o}r, and Esther Buenzli.
\end{acknowledgements}


\bibliography{rlbib}

\begin{thebibliography}{246}
\expandafter\ifx\csname natexlab\endcsname\relax\def\natexlab#1{#1}\fi

\bibitem[{{Absil} {et~al.}(2014){Absil}, {Mawet}, {Delacroix}, {Forsberg},
  {Karlsson}, {Habraken}, {Surdej}, {Absil}, {Carlomagno}, {Christiaens},
  {Defr{\`e}re}, {Gomez Gonzalez}, {Huby}, {Jolivet}, {Milli}, {Piron}, {Vargas
  Catalan}, \& {Van Droogenbroeck}}]{absil2014}
{Absil}, O., {Mawet}, D., {Delacroix}, C., {et~al.} 2014, in \procspie, Vol.
  9148, Adaptive Optics Systems IV, 91480M

\bibitem[{{Acke} \& {van den Ancker}(2004)}]{avda2004}
{Acke}, B. \& {van den Ancker}, M.~E. 2004, \aap, 426, 151

\bibitem[{{Alecian} {et~al.}(2013){Alecian}, {Wade}, {Catala}, {Grunhut},
  {Landstreet}, {Bagnulo}, {B{\"o}hm}, {Folsom}, {Marsden}, \&
  {Waite}}]{alecian2013}
{Alecian}, E., {Wade}, G.~A., {Catala}, C., {et~al.} 2013, \mnras, 429, 1001

\bibitem[{{Allard}(2014)}]{allard2014}
{Allard}, F. 2014, in IAU Symposium, Vol. 299, Exploring the Formation and
  Evolution of Planetary Systems, ed. M.~{Booth}, B.~C. {Matthews}, \& J.~R.
  {Graham}, 271--272

\bibitem[{{Allard} {et~al.}(2013){Allard}, {Homeier}, {Freytag},
  {Schaffenberger}, {}, \& {Rajpurohit}}]{allard2013}
{Allard}, F., {Homeier}, D., {Freytag}, B., {et~al.} 2013, Memorie della
  Societa Astronomica Italiana Supplementi, 24, 128

\bibitem[{{Amara} \& {Quanz}(2012)}]{amara2012}
{Amara}, A. \& {Quanz}, S.~P. 2012, \mnras, 427, 948

\bibitem[{{Andrews} {et~al.}(2018){Andrews}, {Huang}, {P{\'e}rez}, {Isella},
  {Dullemond}, {Kurtovic}, {Guzm{\'a}n}, {Carpenter}, {Wilner}, {Zhang}, {Zhu},
  {Birnstiel}, {Bai}, {Benisty}, {Hughes}, {{\"O}berg}, \&
  {Ricci}}]{andrews2018}
{Andrews}, S.~M., {Huang}, J., {P{\'e}rez}, L.~M., {et~al.} 2018, \apjl, 869,
  L41

\bibitem[{{Andrews} {et~al.}(2011){Andrews}, {Wilner}, {Espaillat}, {Hughes},
  {Dullemond}, {McClure}, {Qi}, \& {Brown}}]{andrews2011}
{Andrews}, S.~M., {Wilner}, D.~J., {Espaillat}, C., {et~al.} 2011, \apj, 732,
  42

\bibitem[{{Anglada-Escud{\'e}} {et~al.}(2016){Anglada-Escud{\'e}}, {Amado},
  {Barnes}, {Berdi{\~n}as}, {Butler}, {Coleman}, {de La Cueva}, {Dreizler},
  {Endl}, {Giesers}, {Jeffers}, {Jenkins}, {Jones}, {Kiraga}, {K{\"u}rster},
  {L{\'o}pez-Gonz{\'a}lez}, {Marvin}, {Morales}, {Morin}, {Nelson}, {Ortiz},
  {Ofir}, {Paardekooper}, {Reiners}, {Rodr{\'{\i}}guez},
  {Rodr{\'{\i}}guez-L{\'o}pez}, {Sarmiento}, {Strachan}, {Tsapras}, {Tuomi}, \&
  {Zechmeister}}]{anglada2016}
{Anglada-Escud{\'e}}, G., {Amado}, P.~J., {Barnes}, J., {et~al.} 2016, \nat,
  536, 437

\bibitem[{{Apai} {et~al.}(2008){Apai}, {Janson}, {Moro-Mart{\'{\i}}n}, {Meyer},
  {Mamajek}, {Masciadri}, {Henning}, {Pascucci}, {Kim}, {Hillenbrand},
  {Kasper}, \& {Biller}}]{apai2008}
{Apai}, D., {Janson}, M., {Moro-Mart{\'{\i}}n}, A., {et~al.} 2008, \apj, 672,
  1196

\bibitem[{{Astropy Collaboration} {et~al.}(2018){Astropy Collaboration},
  {Price-Whelan}, {Sip{\H{o}}cz}, {G{\"u}nther}, {Lim}, {Crawford}, {Conseil},
  {Shupe}, {Craig}, {Dencheva}, {Ginsburg}, {Vand erPlas}, {Bradley},
  {P{\'e}rez-Su{\'a}rez}, {de Val-Borro}, {Aldcroft}, {Cruz}, {Robitaille},
  {Tollerud}, {Ardelean}, {Babej}, {Bach}, {Bachetti}, {Bakanov}, {Bamford},
  {Barentsen}, {Barmby}, {Baumbach}, {Berry}, {Biscani}, {Boquien}, {Bostroem},
  {Bouma}, {Brammer}, {Bray}, {Breytenbach}, {Buddelmeijer}, {Burke},
  {Calderone}, {Cano Rodr{\'\i}guez}, {Cara}, {Cardoso}, {Cheedella}, {Copin},
  {Corrales}, {Crichton}, {D'Avella}, {Deil}, {Depagne}, {Dietrich}, {Donath},
  {Droettboom}, {Earl}, {Erben}, {Fabbro}, {Ferreira}, {Finethy}, {Fox},
  {Garrison}, {Gibbons}, {Goldstein}, {Gommers}, {Greco}, {Greenfield},
  {Groener}, {Grollier}, {Hagen}, {Hirst}, {Homeier}, {Horton}, {Hosseinzadeh},
  {Hu}, {Hunkeler}, {Ivezi{\'c}}, {Jain}, {Jenness}, {Kanarek}, {Kendrew},
  {Kern}, {Kerzendorf}, {Khvalko}, {King}, {Kirkby}, {Kulkarni}, {Kumar},
  {Lee}, {Lenz}, {Littlefair}, {Ma}, {Macleod}, {Mastropietro}, {McCully},
  {Montagnac}, {Morris}, {Mueller}, {Mumford}, {Muna}, {Murphy}, {Nelson},
  {Nguyen}, {Ninan}, {N{\"o}the}, {Ogaz}, {Oh}, {Parejko}, {Parley}, {Pascual},
  {Patil}, {Patil}, {Plunkett}, {Prochaska}, {Rastogi}, {Reddy Janga},
  {Sabater}, {Sakurikar}, {Seifert}, {Sherbert}, {Sherwood-Taylor}, {Shih},
  {Sick}, {Silbiger}, {Singanamalla}, {Singer}, {Sladen}, {Sooley},
  {Sornarajah}, {Streicher}, {Teuben}, {Thomas}, {Tremblay}, {Turner},
  {Terr{\'o}n}, {van Kerkwijk}, {de la Vega}, {Watkins}, {Weaver}, {Whitmore},
  {Woillez}, {Zabalza}, \& {Astropy Contributors}}]{astropy2018}
{Astropy Collaboration}, {Price-Whelan}, A.~M., {Sip{\H{o}}cz}, B.~M., {et~al.}
  2018, \aj, 156, 123

\bibitem[{{Astropy Collaboration} {et~al.}(2013){Astropy Collaboration},
  {Robitaille}, {Tollerud}, {Greenfield}, {Droettboom}, {Bray}, {Aldcroft},
  {Davis}, {Ginsburg}, {Price-Whelan}, {Kerzendorf}, {Conley}, {Crighton},
  {Barbary}, {Muna}, {Ferguson}, {Grollier}, {Parikh}, {Nair}, {Unther},
  {Deil}, {Woillez}, {Conseil}, {Kramer}, {Turner}, {Singer}, {Fox}, {Weaver},
  {Zabalza}, {Edwards}, {Azalee Bostroem}, {Burke}, {Casey}, {Crawford},
  {Dencheva}, {Ely}, {Jenness}, {Labrie}, {Lim}, {Pierfederici}, {Pontzen},
  {Ptak}, {Refsdal}, {Servillat}, \& {Streicher}}]{astropy2013}
{Astropy Collaboration}, {Robitaille}, T.~P., {Tollerud}, E.~J., {et~al.} 2013,
  \aap, 558, A33

\bibitem[{{Avenhaus} {et~al.}(2017){Avenhaus}, {Quanz}, {Schmid}, {Dominik},
  {Stolker}, {Ginski}, {de Boer}, {Szul{\'a}gyi}, {Garufi}, {Zurlo},
  {Hagelberg}, {Benisty}, {Henning}, {M{\'e}nard}, {Meyer}, {Baruffolo},
  {Bazzon}, {Beuzit}, {Costille}, {Dohlen}, {Girard}, {Gisler}, {Kasper},
  {Mouillet}, {Pragt}, {Roelfsema}, {Salasnich}, \& {Sauvage}}]{avenhaus2017}
{Avenhaus}, H., {Quanz}, S.~P., {Schmid}, H.~M., {et~al.} 2017, \aj, 154, 33

\bibitem[{{Avenhaus} {et~al.}(2014){Avenhaus}, {Quanz}, {Schmid}, {Meyer},
  {Garufi}, {Wolf}, \& {Dominik}}]{avenhaus2014}
{Avenhaus}, H., {Quanz}, S.~P., {Schmid}, H.~M., {et~al.} 2014, \apj, 781, 87

\bibitem[{{Bailer-Jones}(2015)}]{bailer2015}
{Bailer-Jones}, C. A.~L. 2015, \pasp, 127, 994

\bibitem[{{Bailer-Jones} {et~al.}(2018){Bailer-Jones}, {Rybizki}, {Fouesneau},
  {Mantelet}, \& {Andrae}}]{bailer2018}
{Bailer-Jones}, C.~A.~L., {Rybizki}, J., {Fouesneau}, M., {Mantelet}, G., \&
  {Andrae}, R. 2018, \aj, 156, 58

\bibitem[{{Bailey} {et~al.}(2014){Bailey}, {Hinz}, {Puglisi}, {Esposito},
  {Vaitheeswaran}, {Skemer}, {Defr{\`e}re}, {Vaz}, \&
  {Leisenring}}]{bailey2014}
{Bailey}, V.~P., {Hinz}, P.~M., {Puglisi}, A.~T., {et~al.} 2014, in Society of
  Photo-Optical Instrumentation Engineers (SPIE) Conference Series, Vol. 9148,
  \procspie, 914803

\bibitem[{{Baraffe} {et~al.}(2003){Baraffe}, {Chabrier}, {Barman}, {Allard}, \&
  {Hauschildt}}]{baraffe2003}
{Baraffe}, I., {Chabrier}, G., {Barman}, T.~S., {Allard}, F., \& {Hauschildt},
  P.~H. 2003, \aap, 402, 701

\bibitem[{{Baraffe} {et~al.}(2015){Baraffe}, {Homeier}, {Allard}, \&
  {Chabrier}}]{baraffe2015}
{Baraffe}, I., {Homeier}, D., {Allard}, F., \& {Chabrier}, G. 2015, \aap, 577,
  A42

\bibitem[{{Barnes} {et~al.}(2017){Barnes}, {Jeffers}, {Anglada-Escud{\'e}},
  {Haswell}, {Jones}, {Tuomi}, {Feng}, {Jenkins}, \& {Petit}}]{barnes2017}
{Barnes}, J.~R., {Jeffers}, S.~V., {Anglada-Escud{\'e}}, G., {et~al.} 2017,
  \mnras, 466, 1733

\bibitem[{{Benisty} {et~al.}(2017){Benisty}, {Stolker}, {Pohl}, {de Boer},
  {Lesur}, {Dominik}, {Dullemond}, {Langlois}, {Min}, {Wagner}, {Henning},
  {Juhasz}, {Pinilla}, {Facchini}, {Apai}, {van Boekel}, {Garufi}, {Ginski},
  {M{\'e}nard}, {Pinte}, {Quanz}, {Zurlo}, {Boccaletti}, {Bonnefoy}, {Beuzit},
  {Chauvin}, {Cudel}, {Desidera}, {Feldt}, {Fontanive}, {Gratton}, {Kasper},
  {Lagrange}, {LeCoroller}, {Mouillet}, {Mesa}, {Sissa}, {Vigan}, {Antichi},
  {Buey}, {Fusco}, {Gisler}, {Llored}, {Magnard}, {Moeller-Nilsson}, {Pragt},
  {Roelfsema}, {Sauvage}, \& {Wildi}}]{benisty2017}
{Benisty}, M., {Stolker}, T., {Pohl}, A., {et~al.} 2017, \aap, 597, A42

\bibitem[{{Bhowmik} {et~al.}(2019){Bhowmik}, {Boccaletti}, {Th{\'e}bault},
  {Kral}, {Mazoyer}, {Milli}, {Maire}, {van Holstein}, {Augereau}, {Baudoz},
  {Feldt}, {Galicher}, {Henning}, {Lagrange}, {Olofsson}, {Pantin}, \&
  {Perrot}}]{bhowmik2019}
{Bhowmik}, T., {Boccaletti}, A., {Th{\'e}bault}, P., {et~al.} 2019, \aap, 630,
  A85

\bibitem[{{Biller} {et~al.}(2013){Biller}, {Liu}, {Wahhaj}, {Nielsen},
  {Hayward}, {Males}, {Skemer}, {Close}, {Chun}, {Ftaclas}, {Clarke}, {Thatte},
  {Shkolnik}, {Reid}, {Hartung}, {Boss}, {Lin}, {Alencar}, {de Gouveia Dal
  Pino}, {Gregorio-Hetem}, \& {Toomey}}]{biller2013}
{Biller}, B.~A., {Liu}, M.~C., {Wahhaj}, Z., {et~al.} 2013, \apj, 777, 160

\bibitem[{{Biller} {et~al.}(2014){Biller}, {Males}, {Rodigas}, {Morzinski},
  {Close}, {Juh{\'a}sz}, {Follette}, {Lacour}, {Benisty}, {Sicilia-Aguilar},
  {Hinz}, {Weinberger}, {Henning}, {Pott}, {Bonnefoy}, \&
  {K{\"o}hler}}]{biller2014}
{Biller}, B.~A., {Males}, J., {Rodigas}, T., {et~al.} 2014, \apjl, 792, L22

\bibitem[{{Bitsch} {et~al.}(2015){Bitsch}, {Lambrechts}, \&
  {Johansen}}]{bitsch2015}
{Bitsch}, B., {Lambrechts}, M., \& {Johansen}, A. 2015, \aap, 582, A112

\bibitem[{{Boccaletti} {et~al.}(2012){Boccaletti}, {Augereau}, {Lagrange},
  {Milli}, {Baudoz}, {Mawet}, {Mouillet}, {Lebreton}, \&
  {Maire}}]{boccaletti2012}
{Boccaletti}, A., {Augereau}, J.-C., {Lagrange}, A.-M., {et~al.} 2012, \aap,
  544, A85

\bibitem[{{Bonfanti} {et~al.}(2015){Bonfanti}, {Ortolani}, {Piotto}, \&
  {Nascimbeni}}]{bonfanti2015}
{Bonfanti}, A., {Ortolani}, S., {Piotto}, G., \& {Nascimbeni}, V. 2015, \aap,
  575, A18

\bibitem[{{Booth} {et~al.}(2013){Booth}, {Kennedy}, {Sibthorpe}, {Matthews},
  {Wyatt}, {Duch{\^e}ne}, {Kavelaars}, {Rodriguez}, {Greaves}, {Koning},
  {Vican}, {Rieke}, {Su}, {Moro-Mart{\'{\i}}n}, \& {Kalas}}]{booth2013}
{Booth}, M., {Kennedy}, G., {Sibthorpe}, B., {et~al.} 2013, \mnras, 428, 1263

\bibitem[{{Boss}(1997)}]{boss1997}
{Boss}, A.~P. 1997, Science, 276, 1836

\bibitem[{{Bowler}(2016)}]{bowler2016}
{Bowler}, B.~P. 2016, \pasp, 128, 102001

\bibitem[{{Brandt} {et~al.}(2014{\natexlab{a}}){Brandt}, {Kuzuhara},
  {McElwain}, {Schlieder}, {Wisniewski}, {Turner}, {Carson}, {Matsuo},
  {Biller}, {Bonnefoy}, {Dressing}, {Janson}, {Knapp}, {Moro-Mart{\'\i}n},
  {Thalmann}, {Kudo}, {Kusakabe}, {Hashimoto}, {Abe}, {Brandner}, {Currie},
  {Egner}, {Feldt}, {Golota}, {Goto}, {Grady}, {Guyon}, {Hayano}, {Hayashi},
  {Hayashi}, {Henning}, {Hodapp}, {Ishii}, {Iye}, {Kand ori}, {Kwon}, {Mede},
  {Miyama}, {Morino}, {Nishimura}, {Pyo}, {Serabyn}, {Suenaga}, {Suto},
  {Suzuki}, {Takami}, {Takahashi}, {Takato}, {Terada}, {Tomono}, {Watanabe},
  {Yamada}, {Takami}, {Usuda}, \& {Tamura}}]{brandt2014a}
{Brandt}, T.~D., {Kuzuhara}, M., {McElwain}, M.~W., {et~al.}
  2014{\natexlab{a}}, \apj, 786, 1

\bibitem[{{Brandt} {et~al.}(2014{\natexlab{b}}){Brandt}, {McElwain}, {Turner},
  {Mede}, {Spiegel}, {Kuzuhara}, {Schlieder}, {Wisniewski}, {Abe}, {Biller},
  {Brandner}, {Carson}, {Currie}, {Egner}, {Feldt}, {Golota}, {Goto}, {Grady},
  {Guyon}, {Hashimoto}, {Hayano}, {Hayashi}, {Hayashi}, {Henning}, {Hodapp},
  {Inutsuka}, {Ishii}, {Iye}, {Janson}, {Kandori}, {Knapp}, {Kudo}, {Kusakabe},
  {Kwon}, {Matsuo}, {Miyama}, {Morino}, {Moro-Mart{\'{\i}}n}, {Nishimura},
  {Pyo}, {Serabyn}, {Suto}, {Suzuki}, {Takami}, {Takato}, {Terada}, {Thalmann},
  {Tomono}, {Watanabe}, {Yamada}, {Takami}, {Usuda}, \& {Tamura}}]{brandt2014b}
{Brandt}, T.~D., {McElwain}, M.~W., {Turner}, E.~L., {et~al.}
  2014{\natexlab{b}}, \apj, 794, 159

\bibitem[{{Brittain} {et~al.}(2014){Brittain}, {Carr}, {Najita}, {Quanz}, \&
  {Meyer}}]{brittain2014}
{Brittain}, S.~D., {Carr}, J.~S., {Najita}, J.~R., {Quanz}, S.~P., \& {Meyer},
  M.~R. 2014, \apj, 791, 136

\bibitem[{{Broekhoven-Fiene} {et~al.}(2013){Broekhoven-Fiene}, {Matthews},
  {Kennedy}, {Booth}, {Sibthorpe}, {Lawler}, {Kavelaars}, {Wyatt}, {Qi},
  {Koning}, {Su}, {Rieke}, {Wilner}, \& {Greaves}}]{broek2013}
{Broekhoven-Fiene}, H., {Matthews}, B.~C., {Kennedy}, G.~M., {et~al.} 2013,
  \apj, 762, 52

\bibitem[{{Bryden} {et~al.}(2009){Bryden}, {Beichman}, {Carpenter}, {Rieke},
  {Stapelfeldt}, {Werner}, {Tanner}, {Lawler}, {Wyatt}, {Trilling}, {Su},
  {Blaylock}, \& {Stansberry}}]{bryden2009}
{Bryden}, G., {Beichman}, C.~A., {Carpenter}, J.~M., {et~al.} 2009, \apj, 705,
  1226

\bibitem[{{Canovas} {et~al.}(2017){Canovas}, {Hardy}, {Zurlo}, {Wahhaj},
  {Schreiber}, {Vigan}, {Villaver}, {Olofsson}, {Meeus}, {M{\'e}nard},
  {Caceres}, {Cieza}, \& {Garufi}}]{canovas2017}
{Canovas}, H., {Hardy}, A., {Zurlo}, A., {et~al.} 2017, \aap, 598, A43

\bibitem[{{Caratti o Garatti} {et~al.}(2015){Caratti o Garatti}, {Tambovtseva},
  {Garcia Lopez}, {Kraus}, {Schertl}, {Grinin}, {Weigelt}, {Hofmann}, {Massi},
  {Lagarde}, {Vannier}, \& {Malbet}}]{caratti2015}
{Caratti o Garatti}, A., {Tambovtseva}, L.~V., {Garcia Lopez}, R., {et~al.}
  2015, \aap, 582, A44

\bibitem[{{Carpenter} {et~al.}(2009){Carpenter}, {Bouwman}, {Mamajek}, {Meyer},
  {Hillenbrand}, {Backman}, {Henning}, {Hines}, {Hollenbach}, {Kim},
  {Moro-Martin}, {Pascucci}, {Silverstone}, {Stauffer}, \&
  {Wolf}}]{carpenter2009}
{Carpenter}, J.~M., {Bouwman}, J., {Mamajek}, E.~E., {et~al.} 2009, \apjs, 181,
  197

\bibitem[{{Casagrande} {et~al.}(2011){Casagrande}, {Sch{\"o}nrich}, {Asplund},
  {Cassisi}, {Ram{\'{\i}}rez}, {Mel{\'e}ndez}, {Bensby}, \&
  {Feltzing}}]{casa2011}
{Casagrande}, L., {Sch{\"o}nrich}, R., {Asplund}, M., {et~al.} 2011, \aap, 530,
  A138

\bibitem[{{Casertano} {et~al.}(2008){Casertano}, {Lattanzi}, {Sozzetti},
  {Spagna}, {Jancart}, {Morbidelli}, {Pannunzio}, {Pourbaix}, \&
  {Queloz}}]{casertano2008}
{Casertano}, S., {Lattanzi}, M.~G., {Sozzetti}, A., {et~al.} 2008, \aap, 482,
  699

\bibitem[{{Chabrier} {et~al.}(2000){Chabrier}, {Baraffe}, {Allard}, \&
  {Hauschildt}}]{chabrier2000}
{Chabrier}, G., {Baraffe}, I., {Allard}, F., \& {Hauschildt}, P. 2000, \apj,
  542, 464

\bibitem[{{Chauvin} {et~al.}(2017{\natexlab{a}}){Chauvin}, {Desidera},
  {Lagrange}, {Vigan}, {Feldt}, {Gratton}, {Langlois}, {Cheetham}, {Bonnefoy},
  \& {Meyer}}]{chauvin2017a}
{Chauvin}, G., {Desidera}, S., {Lagrange}, A.-M., {et~al.} 2017{\natexlab{a}},
  in SF2A-2017: Proceedings of the Annual meeting of the French Society of
  Astronomy and Astrophysics, ed. C.~{Reyl{\'e}}, P.~{Di Matteo}, F.~{Herpin},
  E.~{Lagadec}, A.~{Lan{\c c}on}, Z.~{Meliani}, \& F.~{Royer}, 331--335

\bibitem[{{Chauvin} {et~al.}(2017{\natexlab{b}}){Chauvin}, {Desidera},
  {Lagrange}, {Vigan}, {Gratton}, {Langlois}, {Bonnefoy}, {Beuzit}, {Feldt},
  {Mouillet}, {Meyer}, {Cheetham}, {Biller}, {Boccaletti}, {D'Orazi},
  {Galicher}, {Hagelberg}, {Maire}, {Mesa}, {Olofsson}, {Samland}, {Schmidt},
  {Sissa}, {Bonavita}, {Charnay}, {Cudel}, {Daemgen}, {Delorme},
  {Janin-Potiron}, {Janson}, {Keppler}, {Le Coroller}, {Ligi}, {Marleau},
  {Messina}, {Molli{\`e}re}, {Mordasini}, {M{\"u}ller}, {Peretti}, {Perrot},
  {Rodet}, {Rouan}, {Zurlo}, {Dominik}, {Henning}, {Menard}, {Schmid},
  {Turatto}, {Udry}, {Vakili}, {Abe}, {Antichi}, {Baruffolo}, {Baudoz},
  {Baudrand}, {Blanchard}, {Bazzon}, {Buey}, {Carbillet}, {Carle}, {Charton},
  {Cascone}, {Claudi}, {Costille}, {Deboulbe}, {De Caprio}, {Dohlen},
  {Fantinel}, {Feautrier}, {Fusco}, {Gigan}, {Giro}, {Gisler}, {Gluck},
  {Hubin}, {Hugot}, {Jaquet}, {Kasper}, {Madec}, {Magnard}, {Martinez},
  {Maurel}, {Le Mignant}, {M{\"o}ller-Nilsson}, {Llored}, {Moulin},
  {Orign{\'e}}, {Pavlov}, {Perret}, {Petit}, {Pragt}, {Puget}, {Rabou},
  {Ramos}, {Rigal}, {Rochat}, {Roelfsema}, {Rousset}, {Roux}, {Salasnich},
  {Sauvage}, {Sevin}, {Soenke}, {Stadler}, {Suarez}, {Weber}, {Wildi},
  {Antoniucci}, {Augereau}, {Baudino}, {Brandner}, {Engler}, {Girard}, {Gry},
  {Kral}, {Kopytova}, {Lagadec}, {Milli}, {Moutou}, {Schlieder},
  {Szul{\'a}gyi}, {Thalmann}, \& {Wahhaj}}]{chauvin2017b}
{Chauvin}, G., {Desidera}, S., {Lagrange}, A.~M., {et~al.} 2017{\natexlab{b}},
  Astronomy and Astrophysics, 605, L9

\bibitem[{{Chauvin} {et~al.}(2010){Chauvin}, {Lagrange}, {Bonavita},
  {Zuckerman}, {Dumas}, {Bessell}, {Beuzit}, {Bonnefoy}, {Desidera}, {Farihi},
  {Lowrance}, {Mouillet}, \& {Song}}]{chauvin2010}
{Chauvin}, G., {Lagrange}, A.~M., {Bonavita}, M., {et~al.} 2010, \aap, 509, A52

\bibitem[{{Chauvin} {et~al.}(2015){Chauvin}, {Vigan}, {Bonnefoy}, {Desidera},
  {Bonavita}, {Mesa}, {Boccaletti}, {Buenzli}, {Carson}, {Delorme},
  {Hagelberg}, {Montagnier}, {Mordasini}, {Quanz}, {Segransan}, {Thalmann},
  {Beuzit}, {Biller}, {Covino}, {Feldt}, {Girard}, {Gratton}, {Henning},
  {Kasper}, {Lagrange}, {Messina}, {Meyer}, {Mouillet}, {Moutou}, {Reggiani},
  {Schlieder}, \& {Zurlo}}]{chauvin2015}
{Chauvin}, G., {Vigan}, A., {Bonnefoy}, M., {et~al.} 2015, \aap, 573, A127

\bibitem[{{Cheetham} {et~al.}(2019){Cheetham}, {Samland}, {Brems}, {Launhardt},
  {Chauvin}, {S{\'e}gransan}, {Henning}, {Quirrenbach}, {Avenhaus}, {Cugno},
  {Girard}, {Godoy}, {Kennedy}, {Maire}, {Metchev}, {M{\"u}ller}, {Musso
  Barcucci}, {Olofsson}, {Pepe}, {Quanz}, {Queloz}, {Reffert}, {Rickman}, {van
  Boekel}, {Boccaletti}, {Bonnefoy}, {Cantalloube}, {Charnay}, {Delorme},
  {Janson}, {Keppler}, {Lagrange}, {Langlois}, {Lazzoni}, {Menard}, {Mesa},
  {Meyer}, {Schmidt}, {Sissa}, {Udry}, \& {Zurlo}}]{cheetham2019}
{Cheetham}, A.~C., {Samland}, M., {Brems}, S.~S., {et~al.} 2019, \aap, 622, A80

\bibitem[{{Chen} {et~al.}(2014){Chen}, {Mittal}, {Kuchner}, {Forrest}, {Lisse},
  {Manoj}, {Sargent}, \& {Watson}}]{chen2014}
{Chen}, C.~H., {Mittal}, T., {Kuchner}, M., {et~al.} 2014, \apjs, 211, 25

\bibitem[{{Choquet} {et~al.}(2016){Choquet}, {Perrin}, {Chen}, {Soummer},
  {Pueyo}, {Hagan}, {Gofas-Salas}, {Rajan}, {Golimowski}, {Hines}, {Schneider},
  {Mazoyer}, {Augereau}, {Debes}, {Stark}, {Wolff}, {N'Diaye}, \&
  {Hsiao}}]{choquet2016}
{Choquet}, {\'E}., {Perrin}, M.~D., {Chen}, C.~H., {et~al.} 2016, \apjl, 817,
  L2

\bibitem[{{Churcher} {et~al.}(2011){Churcher}, {Wyatt}, \&
  {Smith}}]{churcher2011}
{Churcher}, L., {Wyatt}, M., \& {Smith}, R. 2011, \mnras, 410, 2

\bibitem[{{Close} {et~al.}(2012){Close}, {Puglisi}, {Males}, {Arcidiacono},
  {Skemer}, {Guerra}, {Busoni}, {Brusa}, {Pinna}, {Miller}, {Riccardi},
  {McCarthy}, {Xompero}, {Kulesa}, {Quiros-Pacheco}, {Argomedo}, {Brynnel},
  {Esposito}, {Mannucci}, {Boutsia}, {Fini}, {Thompson}, {Hill}, {Woodward},
  {Briguglio}, {Rodigas}, {Briguglio}, {Stefanini}, {Agapito}, {Hinz},
  {Follette}, \& {Green}}]{2012ApJ...749..180C}
{Close}, L.~M., {Puglisi}, A., {Males}, J.~R., {et~al.} 2012, \apj, 749, 180

\bibitem[{{Collins} {et~al.}(2009){Collins}, {Grady}, {Hamaguchi},
  {Wisniewski}, {Brittain}, {Sitko}, {Carpenter}, {Williams}, {Mathews},
  {Williger}, {van Boekel}, {Carmona}, {Henning}, {van den Ancker}, {Meeus},
  {Chen}, {Petre}, \& {Woodgate}}]{collins2009}
{Collins}, K.~A., {Grady}, C.~A., {Hamaguchi}, K., {et~al.} 2009, \apj, 697,
  557

\bibitem[{{Corder} {et~al.}(2009){Corder}, {Carpenter}, {Sargent}, {Zauderer},
  {Wright}, {White}, {Woody}, {Teuben}, {Scott}, {Pound}, {Plambeck}, {Lamb},
  {Koda}, {Hodges}, {Hawkins}, \& {Bock}}]{corder2009}
{Corder}, S., {Carpenter}, J.~M., {Sargent}, A.~I., {et~al.} 2009, \apjl, 690,
  L65

\bibitem[{{Cox}(2000)}]{cox2000}
{Cox}, A.~N. 2000, {Allen's astrophysical quantities}

\bibitem[{{Cugno} {et~al.}(2019{\natexlab{a}}){Cugno}, {Quanz}, {Hunziker},
  {Stolker}, {Schmid}, {Avenhaus}, {Baudoz}, {Bohn}, {Bonnefoy}, {Buenzli},
  {Chauvin}, {Cheetham}, {Desidera}, {Dominik}, {Feautrier}, {Feldt}, {Ginski},
  {Girard}, {Gratton}, {Hagelberg}, {Hugot}, {Janson}, {Lagrange}, {Langlois},
  {Magnard}, {Maire}, {Menard}, {Meyer}, {Milli}, {Mordasini}, {Pinte},
  {Pragt}, {Roelfsema}, {Rigal}, {Szul{\'a}gyi}, {van Boekel}, {van der Plas},
  {Vigan}, {Wahhaj}, \& {Zurlo}}]{cugno2019b}
{Cugno}, G., {Quanz}, S.~P., {Hunziker}, S., {et~al.} 2019{\natexlab{a}}, \aap,
  622, A156

\bibitem[{{Cugno} {et~al.}(2019{\natexlab{b}}){Cugno}, {Quanz}, {Launhardt},
  {Musso Barcucci}, {Brems}, {Cheetham}, {Godoy}, {Kennedy}, {Henning},
  {M{\"u}ller}, {Olofsson}, {Pepe}, {Quirrenbach}, {Reffert}, {Rickman}, \&
  {S{\'e}gransan}}]{cugno2019a}
{Cugno}, G., {Quanz}, S.~P., {Launhardt}, R., {et~al.} 2019{\natexlab{b}},
  \aap, 624, A29

\bibitem[{{Currie} {et~al.}(2017){Currie}, {Brittain}, {Grady}, {Kenyon}, \&
  {Muto}}]{currie2017}
{Currie}, T., {Brittain}, S., {Grady}, C.~A., {Kenyon}, S.~J., \& {Muto}, T.
  2017, Research Notes of the American Astronomical Society, 1, 40

\bibitem[{{Currie} {et~al.}(2016){Currie}, {Grady}, {Cloutier}, {Konishi},
  {Stassun}, {Debes}, {van der Marel}, {Muto}, {Jayawardhana}, \&
  {Ratzka}}]{currie2016}
{Currie}, T., {Grady}, C.~A., {Cloutier}, R., {et~al.} 2016, \apjl, 819, L26

\bibitem[{{Currie} {et~al.}(2019){Currie}, {Marois}, {Cieza}, {Mulders},
  {Lawson}, {Caceres}, {Rodriguez-Ruiz}, {Wisniewski}, {Guyon}, {Brandt},
  {Kasdin}, {Groff}, {Lozi}, {Chilcote}, {Hodapp}, {Jovanovic}, {Martinache},
  {Skaf}, {Lyra}, {Tamura}, {Asensio-Torres}, {Dong}, {Grady}, {Gerard},
  {Fukagawa}, {Hand}, {Hayashi}, {Henning}, {Kudo}, {Kuzuhara}, {Kwon},
  {McElwain}, \& {Uyama}}]{currie2019}
{Currie}, T., {Marois}, C., {Cieza}, L., {et~al.} 2019, \apjl, 877, L3

\bibitem[{{Currie} {et~al.}(2014){Currie}, {Muto}, {Kudo}, {Honda}, {Brandt},
  {Grady}, {Fukagawa}, {Burrows}, {Janson}, {Kuzuhara}, {McElwain}, {Follette},
  {Hashimoto}, {Henning}, {Kand ori}, {Kusakabe}, {Kwon}, {Mede}, {Morino},
  {Nishikawa}, {Pyo}, {Serabyn}, {Suenaga}, {Takahashi}, {Wisniewski}, \&
  {Tamura}}]{currie2014}
{Currie}, T., {Muto}, T., {Kudo}, T., {et~al.} 2014, \apjl, 796, L30

\bibitem[{{Cutri} \& {et al.}(2013)}]{cutri2013}
{Cutri}, R.~M. \& {et al.} 2013, VizieR Online Data Catalog, II/328

\bibitem[{{Davies} {et~al.}(2014){Davies}, {Adams}, {Armitage}, {Chambers},
  {Ford}, {Morbidelli}, {Raymond}, \& {Veras}}]{davies2014}
{Davies}, M.~B., {Adams}, F.~C., {Armitage}, P., {et~al.} 2014, Protostars and
  Planets VI, 787

\bibitem[{{Delorme} {et~al.}(2012){Delorme}, {Lagrange}, {Chauvin}, {Bonavita},
  {Lacour}, {Bonnefoy}, {Ehrenreich}, \& {Beust}}]{delorme2012}
{Delorme}, P., {Lagrange}, A.~M., {Chauvin}, G., {et~al.} 2012, \aap, 539, A72

\bibitem[{{Dent} {et~al.}(2005){Dent}, {Greaves}, \& {Coulson}}]{dent2005}
{Dent}, W.~R.~F., {Greaves}, J.~S., \& {Coulson}, I.~M. 2005, \mnras, 359, 663

\bibitem[{{Desidera} {et~al.}(2015){Desidera}, {Covino}, {Messina}, {Carson},
  {Hagelberg}, {Schlieder}, {Biazzo}, {Alcal{\'a}}, {Chauvin}, {Vigan},
  {Beuzit}, {Bonavita}, {Bonnefoy}, {Delorme}, {D'Orazi}, {Esposito}, {Feldt},
  {Girardi}, {Gratton}, {Henning}, {Lagrange}, {Lanzafame}, {Launhardt},
  {Marmier}, {Melo}, {Meyer}, {Mouillet}, {Moutou}, {Segransan}, {Udry}, \&
  {Zaidi}}]{desidera2015}
{Desidera}, S., {Covino}, E., {Messina}, S., {et~al.} 2015, \aap, 573, A126

\bibitem[{{Dobbie} {et~al.}(2010){Dobbie}, {Lodieu}, \& {Sharp}}]{dobbie2010}
{Dobbie}, P.~D., {Lodieu}, N., \& {Sharp}, R.~G. 2010, \mnras, 409, 1002

\bibitem[{{Draper} {et~al.}(2016){Draper}, {Duch{\^e}ne}, {Millar-Blanchaer},
  {Matthews}, {Wang}, {Kalas}, {Graham}, {Padgett}, {Ammons}, {Bulger}, {Chen},
  {Chilcote}, {Doyon}, {Fitzgerald}, {Follette}, {Gerard}, {Greenbaum},
  {Hibon}, {Hinkley}, {Macintosh}, {Ingraham}, {Lafreni{\`e}re}, {Marchis},
  {Marois}, {Nielsen}, {Oppenheimer}, {Patel}, {Patience}, {Perrin}, {Pueyo},
  {Rajan}, {Rameau}, {Sivaramakrishnan}, {Vega}, {Ward-Duong}, \&
  {Wolff}}]{draper2016}
{Draper}, Z.~H., {Duch{\^e}ne}, G., {Millar-Blanchaer}, M.~A., {et~al.} 2016,
  \apj, 826, 147

\bibitem[{{Ducourant} {et~al.}(2014){Ducourant}, {Teixeira}, {Galli}, {Le
  Campion}, {Krone-Martins}, {Zuckerman}, {Chauvin}, \& {Song}}]{ducourant2014}
{Ducourant}, C., {Teixeira}, R., {Galli}, P.~A.~B., {et~al.} 2014, \aap, 563,
  A121

\bibitem[{{Eiroa} {et~al.}(2013){Eiroa}, {Marshall}, {Mora}, {Montesinos},
  {Absil}, {Augereau}, {Bayo}, {Bryden}, {Danchi}, {del Burgo}, {Ertel},
  {Fridlund}, {Heras}, {Krivov}, {Launhardt}, {Liseau}, {L{\"o}hne},
  {Maldonado}, {Pilbratt}, {Roberge}, {Rodmann}, {Sanz-Forcada}, {Solano},
  {Stapelfeldt}, {Th{\'e}bault}, {Wolf}, {Ardila}, {Ar{\'e}valo}, {Beichmann},
  {Faramaz}, {Gonz{\'a}lez-Garc{\'{\i}}a}, {Guti{\'e}rrez}, {Lebreton},
  {Mart{\'{\i}}nez-Arn{\'a}iz}, {Meeus}, {Montes}, {Olofsson}, {Su}, {White},
  {Barrado}, {Fukagawa}, {Gr{\"u}n}, {Kamp}, {Lorente}, {Morbidelli},
  {M{\"u}ller}, {Mutschke}, {Nakagawa}, {Ribas}, \& {Walker}}]{eiroa2013}
{Eiroa}, C., {Marshall}, J.~P., {Mora}, A., {et~al.} 2013, \aap, 555, A11

\bibitem[{{Eiroa} {et~al.}(2016){Eiroa}, {Rebollido}, {Montesinos}, {Villaver},
  {Absil}, {Henning}, {Bayo}, {Canovas}, {Carmona}, {Chen}, {Ertel},
  {Iglesias}, {Launhardt}, {Maldonado}, {Meeus}, {Mo{\'o}r}, {Mora}, {Mustill},
  {Olofsson}, {Riviere-Marichalar}, \& {Roberge}}]{eiroa2016}
{Eiroa}, C., {Rebollido}, I., {Montesinos}, B., {et~al.} 2016, \aap, 594, L1

\bibitem[{{Eisner}(2015)}]{eisner2015}
{Eisner}, J.~A. 2015, \apjl, 803, L4

\bibitem[{{Fairlamb} {et~al.}(2015){Fairlamb}, {Oudmaijer},
  {Mendigut{\'{\i}}a}, {Ilee}, \& {van den Ancker}}]{fairlamb2015}
{Fairlamb}, J.~R., {Oudmaijer}, R.~D., {Mendigut{\'{\i}}a}, I., {Ilee}, J.~D.,
  \& {van den Ancker}, M.~E. 2015, \mnras, 453, 976

\bibitem[{{Fedele} {et~al.}(2017){Fedele}, {Carney}, {Hogerheijde}, {Walsh},
  {Miotello}, {Klaassen}, {Bruderer}, {Henning}, \& {van
  Dishoeck}}]{fedele2017}
{Fedele}, D., {Carney}, M., {Hogerheijde}, M.~R., {et~al.} 2017, \aap, 600, A72

\bibitem[{{Feroz} {et~al.}(2009){Feroz}, {Hobson}, \& {Bridges}}]{feroz2009}
{Feroz}, F., {Hobson}, M.~P., \& {Bridges}, M. 2009, \mnras, 398, 1601

\bibitem[{{Fortney} {et~al.}(2008){Fortney}, {Marley}, {Saumon}, \&
  {Lodders}}]{fortney2008}
{Fortney}, J.~J., {Marley}, M.~S., {Saumon}, D., \& {Lodders}, K. 2008, \apj,
  683, 1104

\bibitem[{{Fukagawa} {et~al.}(2006){Fukagawa}, {Tamura}, {Itoh}, {Kudo},
  {Imaeda}, {Oasa}, {Hayashi}, \& {Hayashi}}]{fukagawa2006}
{Fukagawa}, M., {Tamura}, M., {Itoh}, Y., {et~al.} 2006, \apjl, 636, L153

\bibitem[{{Gaia Collaboration} {et~al.}(2018){Gaia Collaboration}, {Brown},
  {Vallenari}, {Prusti}, {de Bruijne}, {Babusiaux}, {Bailer-Jones}, {Biermann},
  {Evans}, {Eyer}, \& et~al.}]{gaia_dr2}
{Gaia Collaboration}, {Brown}, A.~G.~A., {Vallenari}, A., {et~al.} 2018, \aap,
  616, A1

\bibitem[{{Gaia Collaboration} {et~al.}(2016){Gaia Collaboration}, {Prusti},
  {de Bruijne}, {Brown}, {Vallenari}, {Babusiaux}, {Bailer-Jones}, {Bastian},
  {Biermann}, {Evans}, \& et~al.}]{gaia_mission}
{Gaia Collaboration}, {Prusti}, T., {de Bruijne}, J.~H.~J., {et~al.} 2016,
  \aap, 595, A1

\bibitem[{{Galicher} {et~al.}(2016){Galicher}, {Marois}, {Macintosh},
  {Zuckerman}, {Barman}, {Konopacky}, {Song}, {Patience}, {Lafreni{\`e}re},
  {Doyon}, \& {Nielsen}}]{galicher2016}
{Galicher}, R., {Marois}, C., {Macintosh}, B., {et~al.} 2016, \aap, 594, A63

\bibitem[{{Geers} {et~al.}(2007){Geers}, {van Dishoeck}, {Visser},
  {Pontoppidan}, {Augereau}, {Habart}, \& {Lagrange}}]{geers2007}
{Geers}, V.~C., {van Dishoeck}, E.~F., {Visser}, R., {et~al.} 2007, \aap, 476,
  279

\bibitem[{{Gillon} {et~al.}(2017){Gillon}, {Triaud}, {Demory}, {Jehin}, {Agol},
  {Deck}, {Lederer}, {de Wit}, {Burdanov}, {Ingalls}, {Bolmont}, {Leconte},
  {Raymond}, {Selsis}, {Turbet}, {Barkaoui}, {Burgasser}, {Burleigh}, {Carey},
  {Chaushev}, {Copperwheat}, {Delrez}, {Fernandes}, {Holdsworth}, {Kotze}, {Van
  Grootel}, {Almleaky}, {Benkhaldoun}, {Magain}, \& {Queloz}}]{gillon2017}
{Gillon}, M., {Triaud}, A.~H.~M.~J., {Demory}, B.-O., {et~al.} 2017, \nat, 542,
  456

\bibitem[{{Grady} {et~al.}(2004){Grady}, {Woodgate}, {Torres}, {Henning},
  {Apai}, {Rodmann}, {Wang}, {Stecklum}, {Linz}, {Williger}, {Brown},
  {Wilkinson}, {Harper}, {Herczeg}, {Danks}, {Vieira}, {Malumuth}, {Collins},
  \& {Hill}}]{grady2004}
{Grady}, C.~A., {Woodgate}, B., {Torres}, C.~A.~O., {et~al.} 2004, \apj, 608,
  809

\bibitem[{{Greaves} {et~al.}(2016){Greaves}, {Holland}, {Matthews}, {Marshall},
  {Dent}, {Woitke}, {Wyatt}, {Matr{\`a}}, \& {Jackson}}]{greaves2016}
{Greaves}, J.~S., {Holland}, W.~S., {Matthews}, B.~C., {et~al.} 2016, \mnras,
  461, 3910

\bibitem[{{Hagelberg} {et~al.}(2016{\natexlab{a}}){Hagelberg}, {S{\'e}gransan},
  {Udry}, \& {Wildi}}]{2016MNRAS.455.2178H}
{Hagelberg}, J., {S{\'e}gransan}, D., {Udry}, S., \& {Wildi}, F.
  2016{\natexlab{a}}, \mnras, 455, 2178

\bibitem[{{Hagelberg} {et~al.}(2016{\natexlab{b}}){Hagelberg}, {S{\'e}gransan},
  {Udry}, \& {Wildi}}]{hagelberg2016}
{Hagelberg}, J., {S{\'e}gransan}, D., {Udry}, S., \& {Wildi}, F.
  2016{\natexlab{b}}, \mnras, 455, 2178

\bibitem[{{Hamidouche}(2010)}]{hami2010}
{Hamidouche}, M. 2010, \apj, 722, 204

\bibitem[{{Hashimoto} {et~al.}(2012){Hashimoto}, {Dong}, {Kudo}, {Honda},
  {McClure}, {Zhu}, {Muto}, {Wisniewski}, {Abe}, {Brandner}, {Brandt},
  {Carson}, {Egner}, {Feldt}, {Fukagawa}, {Goto}, {Grady}, {Guyon}, {Hayano},
  {Hayashi}, {Hayashi}, {Henning}, {Hodapp}, {Ishii}, {Iye}, {Janson},
  {Kandori}, {Knapp}, {Kusakabe}, {Kuzuhara}, {Kwon}, {Matsuo}, {Mayama},
  {McElwain}, {Miyama}, {Morino}, {Moro-Martin}, {Nishimura}, {Pyo}, {Serabyn},
  {Suenaga}, {Suto}, {Suzuki}, {Takahashi}, {Takami}, {Takato}, {Terada},
  {Thalmann}, {Tomono}, {Turner}, {Watanabe}, {Yamada}, {Takami}, {Usuda}, \&
  {Tamura}}]{hashimoto2012}
{Hashimoto}, J., {Dong}, R., {Kudo}, T., {et~al.} 2012, \apjl, 758, L19

\bibitem[{{Hauser} {et~al.}(1989){Hauser}, {Kelsall}, {Moseley}, {Silverberg},
  {Murdock}, \& {Wright}}]{hauser1989}
{Hauser}, M.~G., {Kelsall}, T., {Moseley}, S.~H., {et~al.} 1989, in \baas,
  Vol.~21, Bulletin of the American Astronomical Society, 1219

\bibitem[{{Heinze} {et~al.}(2010){Heinze}, {Hinz}, {Sivanandam}, {Kenworthy},
  {Meyer}, \& {Miller}}]{heinze2010}
{Heinze}, A.~N., {Hinz}, P.~M., {Sivanandam}, S., {et~al.} 2010, \apj, 714,
  1551

\bibitem[{{Hinkley} {et~al.}(2011){Hinkley}, {Oppenheimer}, {Zimmerman},
  {Brenner}, {Parry}, {Crepp}, {Vasisht}, {Ligon}, {King}, {Soummer},
  {Sivaramakrishnan}, {Beichman}, {Shao}, {Roberts}, {Bouchez}, {Dekany},
  {Pueyo}, {Roberts}, {Lockhart}, {Zhai}, {Shelton}, \&
  {Burruss}}]{hinkley2011}
{Hinkley}, S., {Oppenheimer}, B.~R., {Zimmerman}, N., {et~al.} 2011,
  Publications of the Astronomical Society of the Pacific, 123, 74

\bibitem[{{Holland} {et~al.}(2017){Holland}, {Matthews}, {Kennedy}, {Greaves},
  {Wyatt}, {Booth}, {Bastien}, {Bryden}, {Butner}, {Chen}, {Chrysostomou},
  {Davies}, {Dent}, {Di Francesco}, {Duch{\^e}ne}, {Gibb}, {Friberg}, {Ivison},
  {Jenness}, {Kavelaars}, {Lawler}, {Lestrade}, {Marshall}, {Moro-Martin},
  {Pani{\'c}}, {Phillips}, {Serjeant}, {Schieven}, {Sibthorpe}, {Vican},
  {Ward-Thompson}, {van der Werf}, {White}, {Wilner}, \&
  {Zuckerman}}]{holland2017}
{Holland}, W.~S., {Matthews}, B.~C., {Kennedy}, G.~M., {et~al.} 2017, \mnras,
  470, 3606

\bibitem[{{Hu{\'e}lamo} {et~al.}(2018){Hu{\'e}lamo}, {Chauvin}, {Schmid},
  {Quanz}, {Whelan}, {Lillo-Box}, {Barrado}, {Montesinos}, {Alcal{\'a}},
  {Benisty}, {de Gregorio-Monsalvo}, {Mendigut{\'{\i}}a}, {Bouy},
  {Mer{\'{\i}}n}, {de Boer}, {Garufi}, \& {Pantin}}]{huelamo2018}
{Hu{\'e}lamo}, N., {Chauvin}, G., {Schmid}, H.~M., {et~al.} 2018, \aap, 613, L5

\bibitem[{{Hung} {et~al.}(2015){Hung}, {Duch{\^e}ne}, {Arriaga}, {Fitzgerald},
  {Maire}, {Marois}, {Millar-Blanchaer}, {Bruzzone}, {Rajan}, {Pueyo}, {Kalas},
  {De Rosa}, {Graham}, {Konopacky}, {Wolff}, {Ammons}, {Chen}, {Chilcote},
  {Draper}, {Esposito}, {Gerard}, {Goodsell}, {Greenbaum}, {Hibon}, {Hinkley},
  {Macintosh}, {Marchis}, {Metchev}, {Nielsen}, {Oppenheimer}, {Patience},
  {Perrin}, {Rantakyr{\"o}}, {Sivaramakrishnan}, {Wang}, {Ward-Duong}, \&
  {Wiktorowicz}}]{hung2015}
{Hung}, L.-W., {Duch{\^e}ne}, G., {Arriaga}, P., {et~al.} 2015, \apjl, 815, L14

\bibitem[{{Hunziker} {et~al.}(2018){Hunziker}, {Quanz}, {Amara}, \&
  {Meyer}}]{2018A&A...611A..23H}
{Hunziker}, S., {Quanz}, S.~P., {Amara}, A., \& {Meyer}, M.~R. 2018, \aap, 611,
  A23

\bibitem[{{Husser} {et~al.}(2013){Husser}, {Wende-von Berg}, {Dreizler},
  {Homeier}, {Reiners}, {Barman}, \& {Hauschildt}}]{husser2013}
{Husser}, T.-O., {Wende-von Berg}, S., {Dreizler}, S., {et~al.} 2013, \aap,
  553, A6

\bibitem[{{Ida} \& {Lin}(2004)}]{idalin2004}
{Ida}, S. \& {Lin}, D.~N.~C. 2004, \apj, 616, 567

\bibitem[{{Isella} {et~al.}(2010){Isella}, {Carpenter}, \&
  {Sargent}}]{isella2010}
{Isella}, A., {Carpenter}, J.~M., \& {Sargent}, A.~I. 2010, \apj, 714, 1746

\bibitem[{{Janson} {et~al.}(2013){Janson}, {Brandt}, {Moro-Mart{\'{\i}}n},
  {Usuda}, {Thalmann}, {Carson}, {Goto}, {Currie}, {McElwain}, {Itoh},
  {Fukagawa}, {Crepp}, {Kuzuhara}, {Hashimoto}, {Kudo}, {Kusakabe}, {Abe},
  {Brandner}, {Egner}, {Feldt}, {Grady}, {Guyon}, {Hayano}, {Hayashi},
  {Hayashi}, {Henning}, {Hodapp}, {Ishii}, {Iye}, {Kandori}, {Knapp}, {Kwon},
  {Matsuo}, {Miyama}, {Morino}, {Nishimura}, {Pyo}, {Serabyn}, {Suenaga},
  {Suto}, {Suzuki}, {Takahashi}, {Takami}, {Takato}, {Terada}, {Tomono},
  {Turner}, {Watanabe}, {Wisniewski}, {Yamada}, {Takami}, \&
  {Tamura}}]{janson2013a}
{Janson}, M., {Brandt}, T.~D., {Moro-Mart{\'{\i}}n}, A., {et~al.} 2013, \apj,
  773, 73

\bibitem[{{Janson} {et~al.}(2012){Janson}, {Carson}, {Lafreni{\`e}re},
  {Spiegel}, {Bent}, \& {Wong}}]{janson2012}
{Janson}, M., {Carson}, J.~C., {Lafreni{\`e}re}, D., {et~al.} 2012, \apj, 747,
  116

\bibitem[{{Johansen} \& \mbox{Lacerda}(2010)}]{johansen2010}
{Johansen}, A. \& \mbox{Lacerda}, P. 2010, \mnras, 404, 475

\bibitem[{{Kalas} {et~al.}(2007){Kalas}, {Fitzgerald}, \& {Graham}}]{kalas2007}
{Kalas}, P., {Fitzgerald}, M.~P., \& {Graham}, J.~R. 2007, \apjl, 661, L85

\bibitem[{{Kalas} {et~al.}(2008){Kalas}, {Graham}, {Chiang}, {Fitzgerald},
  {Clampin}, {Kite}, {Stapelfeldt}, {Marois}, \& {Krist}}]{kalas2008}
{Kalas}, P., {Graham}, J.~R., {Chiang}, E., {et~al.} 2008, Science, 322, 1345

\bibitem[{{Kalas} {et~al.}(2006){Kalas}, {Graham}, {Clampin}, \&
  {Fitzgerald}}]{kalas2006}
{Kalas}, P., {Graham}, J.~R., {Clampin}, M.~C., \& {Fitzgerald}, M.~P. 2006,
  \apjl, 637, L57

\bibitem[{{Kasper} {et~al.}(2007){Kasper}, {Apai}, {Janson}, \&
  {Brandner}}]{kasper2007}
{Kasper}, M., {Apai}, D., {Janson}, M., \& {Brandner}, W. 2007, \aap, 472, 321

\bibitem[{{Kastner} {et~al.}(2014){Kastner}, {Hily-Blant}, {Rodriguez},
  {Punzi}, \& {Forveille}}]{kastner2014}
{Kastner}, J.~H., {Hily-Blant}, P., {Rodriguez}, D.~R., {Punzi}, K., \&
  {Forveille}, T. 2014, \apj, 793, 55

\bibitem[{{Kennedy} \& {Wyatt}(2014)}]{kennedy2014}
{Kennedy}, G.~M. \& {Wyatt}, M.~C. 2014, \mnras, 444, 3164

\bibitem[{{Keppler} {et~al.}(2018){Keppler}, {Benisty}, {M{\"u}ller},
  {Henning}, {van Boekel}, {Cantalloube}, {Ginski}, {van Holstein}, {Maire},
  {Pohl}, {Samland}, {Avenhaus}, {Baudino}, {Boccaletti}, {de Boer},
  {Bonnefoy}, {Chauvin}, {Desidera}, {Langlois}, {Lazzoni}, {Marleau},
  {Mordasini}, {Pawellek}, {Stolker}, {Vigan}, {Zurlo}, {Birnstiel},
  {Brandner}, {Feldt}, {Flock}, {Girard}, {Gratton}, {Hagelberg}, {Isella},
  {Janson}, {Juhasz}, {Kemmer}, {Kral}, {Lagrange}, {Launhardt}, {Matter},
  {M{\'e}nard}, {Milli}, {Molli{\`e}re}, {Olofsson}, {P{\'e}rez}, {Pinilla},
  {Pinte}, {Quanz}, {Schmidt}, {Udry}, {Wahhaj}, {Williams}, {Buenzli},
  {Cudel}, {Dominik}, {Galicher}, {Kasper}, {Lannier}, {Mesa}, {Mouillet},
  {Peretti}, {Perrot}, {Salter}, {Sissa}, {Wildi}, {Abe}, {Antichi},
  {Augereau}, {Baruffolo}, {Baudoz}, {Bazzon}, {Beuzit}, {Blanchard}, {Brems},
  {Buey}, {De Caprio}, {Carbillet}, {Carle}, {Cascone}, {Cheetham}, {Claudi},
  {Costille}, {Delboulb{\'e}}, {Dohlen}, {Fantinel}, {Feautrier}, {Fusco},
  {Giro}, {Gluck}, {Gry}, {Hubin}, {Hugot}, {Jaquet}, {Le Mignant}, {Llored},
  {Madec}, {Magnard}, {Martinez}, {Maurel}, {Meyer}, {M{\"o}ller-Nilsson},
  {Moulin}, {Mugnier}, {Orign{\'e}}, {Pavlov}, {Perret}, {Petit}, {Pragt},
  {Puget}, {Rabou}, {Ramos}, {Rigal}, {Rochat}, {Roelfsema}, {Rousset}, {Roux},
  {Salasnich}, {Sauvage}, {Sevin}, {Soenke}, {Stadler}, {Suarez}, {Turatto}, \&
  {Weber}}]{keppler2018}
{Keppler}, M., {Benisty}, M., {M{\"u}ller}, A., {et~al.} 2018, \aap, 617, A44

\bibitem[{{Kral} {et~al.}(2016){Kral}, {Wyatt}, {Carswell}, {Pringle},
  {Matr{\`a}}, \& {Juh{\'a}sz}}]{kral2016}
{Kral}, Q., {Wyatt}, M., {Carswell}, R.~F., {et~al.} 2016, \mnras, 461, 845

\bibitem[{{Kraus} \& {Ireland}(2012)}]{kraus2012}
{Kraus}, A.~L. \& {Ireland}, M.~J. 2012, \apj, 745, 5

\bibitem[{{Kraus} {et~al.}(2008){Kraus}, {Preibisch}, \& {Ohnaka}}]{kraus2008}
{Kraus}, S., {Preibisch}, T., \& {Ohnaka}, K. 2008, \apj, 676, 490

\bibitem[{{Kreplin} {et~al.}(2013){Kreplin}, {Weigelt}, {Kraus}, {Grinin},
  {Hofmann}, {Kishimoto}, {Schertl}, {Tambovtseva}, {Clausse}, {Massi},
  {Perraut}, \& {Stee}}]{kreplin2013}
{Kreplin}, A., {Weigelt}, G., {Kraus}, S., {et~al.} 2013, \aap, 551, A21

\bibitem[{{Krist} {et~al.}(2012){Krist}, {Stapelfeldt}, {Bryden}, \&
  {Plavchan}}]{krist2012}
{Krist}, J.~E., {Stapelfeldt}, K.~R., {Bryden}, G., \& {Plavchan}, P. 2012,
  \aj, 144, 45

\bibitem[{{Kuzuhara} {et~al.}(2013){Kuzuhara}, {Tamura}, {Kudo}, {Janson},
  {Kandori}, {Brandt}, {Thalmann}, {Spiegel}, {Biller}, {Carson}, {Hori},
  {Suzuki}, {Burrows}, {Henning}, {Turner}, {McElwain}, {Moro-Mart{\'{\i}}n},
  {Suenaga}, {Takahashi}, {Kwon}, {Lucas}, {Abe}, {Brandner}, {Egner}, {Feldt},
  {Fujiwara}, {Goto}, {Grady}, {Guyon}, {Hashimoto}, {Hayano}, {Hayashi},
  {Hayashi}, {Hodapp}, {Ishii}, {Iye}, {Knapp}, {Matsuo}, {Mayama}, {Miyama},
  {Morino}, {Nishikawa}, {Nishimura}, {Kotani}, {Kusakabe}, {Pyo}, {Serabyn},
  {Suto}, {Takami}, {Takato}, {Terada}, {Tomono}, {Watanabe}, {Wisniewski},
  {Yamada}, {Takami}, \& {Usuda}}]{kuzuhara2013}
{Kuzuhara}, M., {Tamura}, M., {Kudo}, T., {et~al.} 2013, \apj, 774, 11

\bibitem[{{Lachaume} {et~al.}(1999){Lachaume}, {Dominik}, {Lanz}, \&
  {Habing}}]{lachaume1999}
{Lachaume}, R., {Dominik}, C., {Lanz}, T., \& {Habing}, H.~J. 1999, \aap, 348,
  897

\bibitem[{{Lafreni{\`e}re} {et~al.}(2007){Lafreni{\`e}re}, {Doyon}, {Marois},
  {Nadeau}, {Oppenheimer}, {Roche}, {Rigaut}, {Graham}, {Jayawardhana},
  {Johnstone}, {Kalas}, {Macintosh}, \& {Racine}}]{lafreniere2007}
{Lafreni{\`e}re}, D., {Doyon}, R., {Marois}, C., {et~al.} 2007, \apj, 670, 1367

\bibitem[{{Lagrange} {et~al.}(2010){Lagrange}, {Bonnefoy}, {Chauvin}, {Apai},
  {Ehrenreich}, {Boccaletti}, {Gratadour}, {Rouan}, {Mouillet}, {Lacour}, \&
  {Kasper}}]{lagrange2010}
{Lagrange}, A.-M., {Bonnefoy}, M., {Chauvin}, G., {et~al.} 2010, Science, 329,
  57

\bibitem[{{Lagrange} {et~al.}(2019){Lagrange}, {Meunier}, {Rubini}, {Keppler},
  {Galland}, {Chapellier}, {Michel}, {Balona}, {Beust}, {Guillot}, {Grandjean},
  {Borgniet}, {M{\'e}karnia}, {Wilson}, {Kiefer}, {Bonnefoy}, {Lillo-Box},
  {Pantoja}, {Jones}, {Iglesias}, {Rodet}, {Diaz}, {Zapata}, {Abe}, \&
  {Schmider}}]{lagrange2019}
{Lagrange}, A.~M., {Meunier}, N., {Rubini}, P., {et~al.} 2019, Nature
  Astronomy, 421

\bibitem[{{Lannier} {et~al.}(2016){Lannier}, {Delorme}, {Lagrange}, {Borgniet},
  {Rameau}, {Schlieder}, {Gagn{\'e}}, {Bonavita}, {Malo}, {Chauvin},
  {Bonnefoy}, \& {Girard}}]{lannier2016}
{Lannier}, J., {Delorme}, P., {Lagrange}, A.~M., {et~al.} 2016, \aap, 596, A83

\bibitem[{{Larkin} {et~al.}(1997){Larkin}, {Oldfield}, \& {Klemm}}]{larkin1997}
{Larkin}, K.~G., {Oldfield}, M.~A., \& {Klemm}, H. 1997, Optics Communications,
  139, 99

\bibitem[{{Lenzen} {et~al.}(2003){Lenzen}, {Hartung}, {Brandner}, {Finger},
  {Hubin}, {Lacombe}, {Lagrange}, {Lehnert}, {Moorwood}, \&
  {Mouillet}}]{lenzen2003}
{Lenzen}, R., {Hartung}, M., {Brandner}, W., {et~al.} 2003, in \procspie, Vol.
  4841, Instrument Design and Performance for Optical/Infrared Ground-based
  Telescopes, ed. M.~{Iye} \& A.~F.~M. {Moorwood}, 944--952

\bibitem[{{Lestrade} {et~al.}(2012){Lestrade}, {Matthews}, {Sibthorpe},
  {Kennedy}, {Wyatt}, {Bryden}, {Greaves}, {Thilliez}, {Moro-Mart{\'{\i}}n},
  {Booth}, {Dent}, {Duch{\^e}ne}, {Harvey}, {Horner}, {Kalas}, {Kavelaars},
  {Phillips}, {Rodriguez}, {Su}, \& {Wilner}}]{lestrade2012}
{Lestrade}, J.-F., {Matthews}, B.~C., {Sibthorpe}, B., {et~al.} 2012, \aap,
  548, A86

\bibitem[{{Lieman-Sifry} {et~al.}(2016){Lieman-Sifry}, {Hughes}, {Carpenter},
  {Gorti}, {Hales}, \& {Flaherty}}]{lieman2016}
{Lieman-Sifry}, J., {Hughes}, A.~M., {Carpenter}, J.~M., {et~al.} 2016, \apj,
  828, 25

\bibitem[{{Ligi} {et~al.}(2018){Ligi}, {Vigan}, {Gratton}, {de Boer},
  {Benisty}, {Boccaletti}, {Quanz}, {Meyer}, {Ginski}, {Sissa}, {Gry},
  {Henning}, {Beuzit}, {Biller}, {Bonnefoy}, {Chauvin}, {Cheetham}, {Cudel},
  {Delorme}, {Desidera}, {Feldt}, {Galicher}, {Girard}, {Janson}, {Kasper},
  {Kopytova}, {Lagrange}, {Langlois}, {Lecoroller}, {Maire}, {M{\'e}nard},
  {Mesa}, {Peretti}, {Perrot}, {Pinilla}, {Pohl}, {Rouan}, {Stolker},
  {Samland}, {Wahhaj}, {Wildi}, {Zurlo}, {Buey}, {Fantinel}, {Fusco}, {Jaquet},
  {Moulin}, {Ramos}, {Suarez}, \& {Weber}}]{ligi2018}
{Ligi}, R., {Vigan}, A., {Gratton}, R., {et~al.} 2018, \mnras, 473, 1774

\bibitem[{{Liu} {et~al.}(2010){Liu}, {Wahhaj}, {Biller}, {Nielsen}, {Chun},
  {Close}, {Ftaclas}, {Hartung}, {Hayward}, {Clarke}, {Reid}, {Shkolnik},
  {Tecza}, {Thatte}, {Alencar}, {Artymowicz}, {Boss}, {Burrows}, {de Gouveia
  Dal Pino}, {Gregorio-Hetem}, {Ida}, {Kuchner}, {Lin}, \& {Toomey}}]{liu2010}
{Liu}, M.~C., {Wahhaj}, Z., {Biller}, B.~A., {et~al.} 2010, in Society of
  Photo-Optical Instrumentation Engineers (SPIE) Conference Series, Vol. 7736,
  Adaptive Optics Systems II, 77361K

\bibitem[{{Liu} {et~al.}(2011){Liu}, {Zhang}, {Wu}, {Qin}, \&
  {Miller}}]{liu2011}
{Liu}, T., {Zhang}, H., {Wu}, Y., {Qin}, S.-L., \& {Miller}, M. 2011, \apj,
  734, 22

\bibitem[{{Macintosh} {et~al.}(2018){Macintosh}, {Chilcote}, {Bailey}, {de
  Rosa}, {Nielsen}, {Norton}, {Poyneer}, {Wang}, {Ruffio}, {Graham}, {Marois},
  {Savransky}, \& {Veran}}]{macintosh2018}
{Macintosh}, B., {Chilcote}, J.~K., {Bailey}, V.~P., {et~al.} 2018, in Society
  of Photo-Optical Instrumentation Engineers (SPIE) Conference Series, Vol.
  10703, \procspie, 107030K

\bibitem[{{Macintosh} {et~al.}(2015){Macintosh}, {Graham}, {Barman}, {De Rosa},
  {Konopacky}, {Marley}, {Marois}, {Nielsen}, {Pueyo}, {Rajan}, {Rameau},
  {Saumon}, {Wang}, {Patience}, {Ammons}, {Arriaga}, {Artigau}, {Beckwith},
  {Brewster}, {Bruzzone}, {Bulger}, {Burningham}, {Burrows}, {Chen}, {Chiang},
  {Chilcote}, {Dawson}, {Dong}, {Doyon}, {Draper}, {Duch{\^e}ne}, {Esposito},
  {Fabrycky}, {Fitzgerald}, {Follette}, {Fortney}, {Gerard}, {Goodsell},
  {Greenbaum}, {Hibon}, {Hinkley}, {Cotten}, {Hung}, {Ingraham},
  {Johnson-Groh}, {Kalas}, {Lafreniere}, {Larkin}, {Lee}, {Line}, {Long},
  {Maire}, {Marchis}, {Matthews}, {Max}, {Metchev}, {Millar-Blanchaer},
  {Mittal}, {Morley}, {Morzinski}, {Murray-Clay}, {Oppenheimer}, {Palmer},
  {Patel}, {Perrin}, {Poyneer}, {Rafikov}, {Rantakyr{\"o}}, {Rice}, {Rojo},
  {Rudy}, {Ruffio}, {Ruiz}, {Sadakuni}, {Saddlemyer}, {Salama}, {Savransky},
  {Schneider}, {Sivaramakrishnan}, {Song}, {Soummer}, {Thomas}, {Vasisht},
  {Wallace}, {Ward-Duong}, {Wiktorowicz}, {Wolff}, \&
  {Zuckerman}}]{macintosh2015}
{Macintosh}, B., {Graham}, J.~R., {Barman}, T., {et~al.} 2015, Science, 350, 64

\bibitem[{{Maldonado} {et~al.}(2015){Maldonado}, {Eiroa}, {Villaver},
  {Montesinos}, \& {Mora}}]{maldonado2015}
{Maldonado}, J., {Eiroa}, C., {Villaver}, E., {Montesinos}, B., \& {Mora}, A.
  2015, \aap, 579, A20

\bibitem[{{Maldonado} {et~al.}(2010){Maldonado}, {Mart{\'{\i}}nez-Arn{\'a}iz},
  {Eiroa}, {Montes}, \& {Montesinos}}]{maldonado2010}
{Maldonado}, J., {Mart{\'{\i}}nez-Arn{\'a}iz}, R.~M., {Eiroa}, C., {Montes},
  D., \& {Montesinos}, B. 2010, \aap, 521, A12

\bibitem[{{Mamajek} {et~al.}(2013){Mamajek}, {Bartlett}, {Seifahrt}, {Henry},
  {Dieterich}, {Lurie}, {Kenworthy}, {Jao}, {Riedel}, {Subasavage}, {Winters},
  {Finch}, {Ianna}, \& {Bean}}]{mama2013}
{Mamajek}, E.~E., {Bartlett}, J.~L., {Seifahrt}, A., {et~al.} 2013, \aj, 146,
  154

\bibitem[{{Mamajek} \& {Bell}(2014)}]{mama2014}
{Mamajek}, E.~E. \& {Bell}, C.~P.~M. 2014, \mnras, 445, 2169

\bibitem[{{Mamajek} \& {Hillenbrand}(2008)}]{MH2008}
{Mamajek}, E.~E. \& {Hillenbrand}, L.~A. 2008, \apj, 687, 1264

\bibitem[{{Manoj} {et~al.}(2006){Manoj}, {Bhatt}, {Maheswar}, \&
  {Muneer}}]{manoj2006}
{Manoj}, P., {Bhatt}, H.~C., {Maheswar}, G., \& {Muneer}, S. 2006, \apj, 653,
  657

\bibitem[{{Marchis} {et~al.}(2016){Marchis}, {Kalas}, {Perrin}, {Konopacky},
  {Savransky}, {Macintosh}, {Marois}, \& {Graham}}]{marchis2016}
{Marchis}, F., {Kalas}, P.~G., {Perrin}, M.~D., {et~al.} 2016, in \procspie,
  Vol. 9910, Observatory Operations: Strategies, Processes, and Systems VI,
  99102D

\bibitem[{{Mari{\~n}as} {et~al.}(2011){Mari{\~n}as}, {Telesco}, {Fisher}, \&
  {Packham}}]{marinas2011}
{Mari{\~n}as}, N., {Telesco}, C.~M., {Fisher}, R.~S., \& {Packham}, C. 2011,
  \apj, 737, 57

\bibitem[{{Marino} {et~al.}(2016){Marino}, {Matr{\`a}}, {Stark}, {Wyatt},
  {Casassus}, {Kennedy}, {Rodriguez}, {Zuckerman}, {Perez}, {Dent}, {Kuchner},
  {Hughes}, {Schneider}, {Steele}, {Roberge}, {Donaldson}, \&
  {Nesvold}}]{marino2016}
{Marino}, S., {Matr{\`a}}, L., {Stark}, C., {et~al.} 2016, \mnras, 460, 2933

\bibitem[{{Marino} {et~al.}(2017){Marino}, {Wyatt}, {Pani{\'c}}, {Matr{\`a}},
  {Kennedy}, {Bonsor}, {Kral}, {Dent}, {Duchene}, {Wilner}, {Lisse},
  {Lestrade}, \& {Matthews}}]{marino2017}
{Marino}, S., {Wyatt}, M.~C., {Pani{\'c}}, O., {et~al.} 2017, \mnras, 465, 2595

\bibitem[{{Marois} {et~al.}(2006){Marois}, {Lafreni{\`e}re}, {Doyon},
  {Macintosh}, \& {Nadeau}}]{marois2006}
{Marois}, C., {Lafreni{\`e}re}, D., {Doyon}, R., {Macintosh}, B., \& {Nadeau},
  D. 2006, \apj, 641, 556

\bibitem[{{Marois} {et~al.}(2008){Marois}, {Macintosh}, {Barman}, {Zuckerman},
  {Song}, {Patience}, {Lafreni{\`e}re}, \& {Doyon}}]{marois2008}
{Marois}, C., {Macintosh}, B., {Barman}, T., {et~al.} 2008, Science, 322, 1348

\bibitem[{{Marois} {et~al.}(2010){Marois}, {Zuckerman}, {Konopacky},
  {Macintosh}, \& {Barman}}]{marois2010}
{Marois}, C., {Zuckerman}, B., {Konopacky}, Q.~M., {Macintosh}, B., \&
  {Barman}, T. 2010, \nat, 468, 1080

\bibitem[{{Marshall} {et~al.}(2014){Marshall}, {Kirchschlager}, {Ertel},
  {Augereau}, {Kennedy}, {Booth}, {Wolf}, {Montesinos}, {Eiroa}, \&
  {Matthews}}]{marshall2014}
{Marshall}, J.~P., {Kirchschlager}, F., {Ertel}, S., {et~al.} 2014, \aap, 570,
  A114

\bibitem[{{Masciadri} {et~al.}(2005){Masciadri}, {Mundt}, {Henning}, {Alvarez},
  \& {Barrado y Navascu{\'e}s}}]{masciadri2005}
{Masciadri}, E., {Mundt}, R., {Henning}, T., {Alvarez}, C., \& {Barrado y
  Navascu{\'e}s}, D. 2005, The Astrophysical Journal, 625, 1004

\bibitem[{{Matr{\`a}} {et~al.}(2019){Matr{\`a}}, {{\"O}berg}, {Wilner},
  {Olofsson}, \& {Bayo}}]{matra2019}
{Matr{\`a}}, L., {{\"O}berg}, K.~I., {Wilner}, D.~J., {Olofsson}, J., \&
  {Bayo}, A. 2019, \aj, 157, 117

\bibitem[{{Matthews} {et~al.}(2010){Matthews}, {Sibthorpe}, {Kennedy},
  {Phillips}, {Churcher}, {Duch{\^e}ne}, {Greaves}, {Lestrade}, {Moro-Martin},
  \& {Wyatt}}]{matthews2010}
{Matthews}, B.~C., {Sibthorpe}, B., {Kennedy}, G., {et~al.} 2010, \aap, 518,
  L135

\bibitem[{{Mawet} {et~al.}(2013){Mawet}, {Absil}, {Delacroix}, {Girard},
  {Milli}, {O'Neal}, {Baudoz}, {Boccaletti}, {Bourget}, \&
  {Christiaens}}]{mawet2013}
{Mawet}, D., {Absil}, O., {Delacroix}, C., {et~al.} 2013, \aap, 552, L13

\bibitem[{{Mawet} {et~al.}(2012){Mawet}, {Absil}, {Montagnier}, {Riaud},
  {Surdej}, {Ducourant}, {Augereau}, {R{\"o}ttinger}, {Girard}, {Krist}, \&
  {Stapelfeldt}}]{mawet2012}
{Mawet}, D., {Absil}, O., {Montagnier}, G., {et~al.} 2012, \aap, 544, A131

\bibitem[{{Mawet} {et~al.}(2017){Mawet}, {Choquet}, {Absil}, {Huby}, {Bottom},
  {Serabyn}, {Femenia}, {Lebreton}, {Matthews}, {Gomez Gonzalez}, {Wertz},
  {Carlomagno}, {Christiaens}, {Defr{\`e}re}, {Delacroix}, {Forsberg},
  {Habraken}, {Jolivet}, {Karlsson}, {Milli}, {Pinte}, {Piron}, {Reggiani},
  {Surdej}, \& {Vargas Catalan}}]{mawet2017}
{Mawet}, D., {Choquet}, {\'E}., {Absil}, O., {et~al.} 2017, \aj, 153, 44

\bibitem[{{Mawet} {et~al.}(2014){Mawet}, {Milli}, {Wahhaj}, {Pelat}, {Absil},
  {Delacroix}, {Boccaletti}, {Kasper}, {Kenworthy}, {Marois}, {Mennesson}, \&
  {Pueyo}}]{mawet2014}
{Mawet}, D., {Milli}, J., {Wahhaj}, Z., {et~al.} 2014, \apj, 792, 97

\bibitem[{{Mawet} {et~al.}(2005){Mawet}, {Riaud}, {Absil}, \&
  {Surdej}}]{2005ApJ...633.1191M}
{Mawet}, D., {Riaud}, P., {Absil}, O., \& {Surdej}, J. 2005, \apj, 633, 1191

\bibitem[{{Mayor} \& {Queloz}(1995)}]{MQ1995}
{Mayor}, M. \& {Queloz}, D. 1995, \nat, 378, 355

\bibitem[{{McLaughlin} {et~al.}(2006){McLaughlin}, {Anderson}, {Meylan},
  {Gebhardt}, {Pryor}, {Minniti}, \& {Phinney}}]{2006ApJS..166..249M}
{McLaughlin}, D.~E., {Anderson}, J., {Meylan}, G., {et~al.} 2006, \apjs, 166,
  249

\bibitem[{{Meeus} {et~al.}(2012){Meeus}, {Montesinos}, {Mendigut{\'{\i}}a},
  {Kamp}, {Thi}, {Eiroa}, {Grady}, {Mathews}, {Sandell}, {Martin-Za{\"i}di},
  {Brittain}, {Dent}, {Howard}, {M{\'e}nard}, {Pinte}, {Roberge},
  {Vandenbussche}, \& {Williams}}]{meeus2012}
{Meeus}, G., {Montesinos}, B., {Mendigut{\'{\i}}a}, I., {et~al.} 2012, \aap,
  544, A78

\bibitem[{{Menu} {et~al.}(2015){Menu}, {van Boekel}, {Henning}, {Leinert},
  {Waelkens}, \& {Waters}}]{menu2015}
{Menu}, J., {van Boekel}, R., {Henning}, T., {et~al.} 2015, \aap, 581, A107

\bibitem[{{Mer{\'{\i}}n} {et~al.}(2004){Mer{\'{\i}}n}, {Montesinos}, {Eiroa},
  {Solano}, {Mora}, {D'Alessio}, {Calvet}, {Oudmaijer}, {de Winter}, {Davies},
  {Harris}, {Collier Cameron}, {Deeg}, {Ferlet}, {Garz{\'o}n}, {Grady},
  {Horne}, {Miranda}, {Palacios}, {Penny}, {Quirrenbach}, {Rauer}, {Schneider},
  \& {Wesselius}}]{merin2004}
{Mer{\'{\i}}n}, B., {Montesinos}, B., {Eiroa}, C., {et~al.} 2004, \aap, 419,
  301

\bibitem[{{Mesa} {et~al.}(2019){Mesa}, {Bonnefoy}, {Gratton}, {Van Der Plas},
  {D'Orazi}, {Sissa}, {Zurlo}, {Rigliaco}, {Schmidt}, {Langlois}, {Vigan},
  {Ubeira Gabellini}, {Desidera}, {Antoniucci}, {Barbieri}, {Benisty},
  {Boccaletti}, {Claudi}, {Fedele}, {Gasparri}, {Henning}, {Kasper},
  {Lagrange}, {Lazzoni}, {Lodato}, {Maire}, {Manara}, {Meyer}, {Reggiani},
  {Samland}, {Van den Ancker}, {Chauvin}, {Cheetham}, {Feldt}, {Hugot},
  {Janson}, {Ligi}, {M{\"o}ller-Nilsson}, {Petit}, {Rickman}, {Rigal}, \&
  {Wildi}}]{mesa2019}
{Mesa}, D., {Bonnefoy}, M., {Gratton}, R., {et~al.} 2019, \aap, 624, A4

\bibitem[{{Meshkat} {et~al.}(2015){Meshkat}, {Bailey}, {Su}, {Kenworthy},
  {Mamajek}, {Hinz}, \& {Smith}}]{meshkat2015}
{Meshkat}, T., {Bailey}, V.~P., {Su}, K.~Y.~L., {et~al.} 2015, \apj, 800, 5

\bibitem[{{Meshkat} {et~al.}(2017){Meshkat}, {Mawet}, {Bryan}, {Hinkley},
  {Bowler}, {Stapelfeldt}, {Batygin}, {Padgett}, {Morales}, {Serabyn},
  {Christiaens}, {Brandt}, \& {Wahhaj}}]{meshkat2017}
{Meshkat}, T., {Mawet}, D., {Bryan}, M.~L., {et~al.} 2017, \aj, 154, 245

\bibitem[{{Moerchen} {et~al.}(2010){Moerchen}, {Telesco}, \&
  {Packham}}]{moerchen2010}
{Moerchen}, M.~M., {Telesco}, C.~M., \& {Packham}, C. 2010, \apj, 723, 1418

\bibitem[{{Montes} {et~al.}(2001){Montes}, {L{\'o}pez-Santiago}, {G{\'a}lvez},
  {Fern{\'a}ndez-Figueroa}, {De Castro}, \& {Cornide}}]{montes2001}
{Montes}, D., {L{\'o}pez-Santiago}, J., {G{\'a}lvez}, M.~C., {et~al.} 2001,
  \mnras, 328, 45

\bibitem[{{Mo{\'o}r} {et~al.}(2011){Mo{\'o}r}, {{\'A}brah{\'a}m}, {Juh{\'a}sz},
  {Kiss}, {Pascucci}, {K{\'o}sp{\'a}l}, {Apai}, {Henning}, {Csengeri}, \&
  {Grady}}]{moor2011}
{Mo{\'o}r}, A., {{\'A}brah{\'a}m}, P., {Juh{\'a}sz}, A., {et~al.} 2011, \apjl,
  740, L7

\bibitem[{{Mo{\'o}r} {et~al.}(2017){Mo{\'o}r}, {Cur{\'e}}, {K{\'o}sp{\'a}l},
  {{\'A}brah{\'a}m}, {Csengeri}, {Eiroa}, {Gunawan}, {Henning}, {Hughes},
  {Juh{\'a}sz}, {Pawellek}, \& {Wyatt}}]{moor2017}
{Mo{\'o}r}, A., {Cur{\'e}}, M., {K{\'o}sp{\'a}l}, {\'A}., {et~al.} 2017, \apj,
  849, 123

\bibitem[{{Mo{\'o}r} {et~al.}(2015{\natexlab{a}}){Mo{\'o}r}, {Henning},
  {Juh{\'a}sz}, {{\'A}brah{\'a}m}, {Balog}, {K{\'o}sp{\'a}l}, {Pascucci},
  {Szab{\'o}}, {Vavrek}, {Cur{\'e}}, {Csengeri}, {Grady}, {G{\"u}sten}, \&
  {Kiss}}]{moor2015b}
{Mo{\'o}r}, A., {Henning}, T., {Juh{\'a}sz}, A., {et~al.} 2015{\natexlab{a}},
  \apj, 814, 42

\bibitem[{{Mo{\'o}r} {et~al.}(2013){Mo{\'o}r}, {Juh{\'a}sz}, {K{\'o}sp{\'a}l},
  {{\'A}brah{\'a}m}, {Apai}, {Csengeri}, {Grady}, {Henning}, {Hughes}, {Kiss},
  {Pascucci}, {Schmalzl}, \& {Gab{\'a}nyi}}]{moor2013}
{Mo{\'o}r}, A., {Juh{\'a}sz}, A., {K{\'o}sp{\'a}l}, {\'A}., {et~al.} 2013,
  \apjl, 777, L25

\bibitem[{{Mo{\'o}r} {et~al.}(2015{\natexlab{b}}){Mo{\'o}r}, {K{\'o}sp{\'a}l},
  {{\'A}brah{\'a}m}, {Apai}, {Balog}, {Grady}, {Henning}, {Juh{\'a}sz}, {Kiss},
  {Krivov}, {Pawellek}, \& {Szab{\'o}}}]{moor2015a}
{Mo{\'o}r}, A., {K{\'o}sp{\'a}l}, {\'A}., {{\'A}brah{\'a}m}, P., {et~al.}
  2015{\natexlab{b}}, \mnras, 447, 577

\bibitem[{{Mo{\'o}r} {et~al.}(2016){Mo{\'o}r}, {K{\'o}sp{\'a}l},
  {{\'A}brah{\'a}m}, {Balog}, {Csengeri}, {Henning}, {Juh{\'a}sz}, \&
  {Kiss}}]{moor2016}
{Mo{\'o}r}, A., {K{\'o}sp{\'a}l}, {\'A}., {{\'A}brah{\'a}m}, P., {et~al.} 2016,
  \apj, 826, 123

\bibitem[{{Morales} {et~al.}(2016){Morales}, {Bryden}, {Werner}, \&
  {Stapelfeldt}}]{morales2016}
{Morales}, F.~Y., {Bryden}, G., {Werner}, M.~W., \& {Stapelfeldt}, K.~R. 2016,
  \apj, 831, 97

\bibitem[{{Morbidelli}(2018)}]{morbidelli2018}
{Morbidelli}, A. 2018, {Dynamical Evolution of Planetary Systems}, 145

\bibitem[{{Mordasini} {et~al.}(2012){Mordasini}, {Alibert}, {Klahr}, \&
  {Henning}}]{mordasini2012}
{Mordasini}, C., {Alibert}, Y., {Klahr}, H., \& {Henning}, T. 2012, \aap, 547,
  A111

\bibitem[{{M{\"u}ller} {et~al.}(2018){M{\"u}ller}, {Keppler}, {Henning},
  {Samland}, {Chauvin}, {Beust}, {Maire}, {Molaverdikhani}, {van Boekel},
  {Benisty}, {Boccaletti}, {Bonnefoy}, {Cantalloube}, {Charnay}, {Baudino},
  {Gennaro}, {Long}, {Cheetham}, {Desidera}, {Feldt}, {Fusco}, {Girard},
  {Gratton}, {Hagelberg}, {Janson}, {Lagrange}, {Langlois}, {Lazzoni}, {Ligi},
  {M{\'e}nard}, {Mesa}, {Meyer}, {Molli{\`e}re}, {Mordasini}, {Moulin},
  {Pavlov}, {Pawellek}, {Quanz}, {Ramos}, {Rouan}, {Sissa}, {Stadler}, {Vigan},
  {Wahhaj}, {Weber}, \& {Zurlo}}]{mueller2018}
{M{\"u}ller}, A., {Keppler}, M., {Henning}, T., {et~al.} 2018, \aap, 617, L2

\bibitem[{{Musso Barcucci} {et~al.}(2019){Musso Barcucci}, {Launhardt},
  {Kennedy}, {Avenhaus}, {Brems}, {van Boekel}, {Cantalloube}, {Cheetham},
  {Cugno}, \& {Girard}}]{musso2019}
{Musso Barcucci}, A., {Launhardt}, R., {Kennedy}, G.~M., {et~al.} 2019, \aap,
  627, A77

\bibitem[{{Nielsen} {et~al.}(2019){Nielsen}, {De Rosa}, {Macintosh}, {Wang},
  {Ruffio}, {Chiang}, {Marley}, {Saumon}, {Savransky}, \&
  {Ammons}}]{nielsen2019}
{Nielsen}, E.~L., {De Rosa}, R.~J., {Macintosh}, B., {et~al.} 2019, \aj, 158,
  13

\bibitem[{{Nielsen} {et~al.}(2016){Nielsen}, {Liu}, {Wahhaj}, {Biller},
  {Hayward}, {Close}, {Close}, {Macintosh}, {Savransky}, {Wang}, {Graham}, {De
  Rosa}, {Rajan}, \& {Rajan}}]{nielsen2016}
{Nielsen}, E.~L., {Liu}, M.~C., {Wahhaj}, Z., {et~al.} 2016, in IAU Symposium,
  Vol. 314, Young Stars \& Planets Near the Sun, ed. J.~H. {Kastner},
  B.~{Stelzer}, \& S.~A. {Metchev}, 220--225

\bibitem[{{Nielsen} {et~al.}(2013){Nielsen}, {Liu}, {Wahhaj}, {Biller},
  {Hayward}, {Close}, {Males}, {Skemer}, {Chun}, {Ftaclas}, {Alencar},
  {Artymowicz}, {Boss}, {Clarke}, {de Gouveia Dal Pino}, {Gregorio-Hetem},
  {Hartung}, {Ida}, {Kuchner}, {Lin}, {Reid}, {Shkolnik}, {Tecza}, {Thatte}, \&
  {Toomey}}]{nielsen2013}
{Nielsen}, E.~L., {Liu}, M.~C., {Wahhaj}, Z., {et~al.} 2013, \apj, 776, 4

\bibitem[{{Ochsenbein} {et~al.}(2000){Ochsenbein}, {Bauer}, \&
  {Marcout}}]{vizier2000}
{Ochsenbein}, F., {Bauer}, P., \& {Marcout}, J. 2000, \aaps, 143, 23

\bibitem[{{Olofsson} {et~al.}(2012){Olofsson}, {Juh{\'a}sz}, {Henning},
  {Mutschke}, {Tamanai}, {Mo{\'o}r}, \& {{\'A}brah{\'a}m}}]{olofsson2012}
{Olofsson}, J., {Juh{\'a}sz}, A., {Henning}, T., {et~al.} 2012, \aap, 542, A90

\bibitem[{{Olofsson} {et~al.}(2016){Olofsson}, {Samland}, {Avenhaus},
  {Caceres}, {Henning}, {Mo{\'o}r}, {Milli}, {Canovas}, {Quanz}, {Schreiber},
  {Augereau}, {Bayo}, {Bazzon}, {Beuzit}, {Boccaletti}, {Buenzli}, {Casassus},
  {Chauvin}, {Dominik}, {Desidera}, {Feldt}, {Gratton}, {Janson}, {Lagrange},
  {Langlois}, {Lannier}, {Maire}, {Mesa}, {Pinte}, {Rouan}, {Salter},
  {Thalmann}, \& {Vigan}}]{olofsson2016}
{Olofsson}, J., {Samland}, M., {Avenhaus}, H., {et~al.} 2016, \aap, 591, A108

\bibitem[{{Ormel} \& {Klahr}(2010)}]{ormel2010}
{Ormel}, C.~W. \& {Klahr}, H.~H. 2010, \aap, 520, A43

\bibitem[{{Palla} \& {Stahler}(2002)}]{palla2002}
{Palla}, F. \& {Stahler}, S.~W. 2002, \apj, 581, 1194

\bibitem[{{Pascual} {et~al.}(2016){Pascual}, {Montesinos}, {Meeus}, {Marshall},
  {Mendigut{\'{\i}}a}, \& {Sandell}}]{pascual2016}
{Pascual}, N., {Montesinos}, B., {Meeus}, G., {et~al.} 2016, \aap, 586, A6

\bibitem[{{Pawellek}(2016)}]{pawellek2016}
{Pawellek}, N. 2016, PhD thesis, PhD Thesis, University of Jena, (2016)

\bibitem[{{Pawellek} \& {Krivov}(2015)}]{pawellek2015}
{Pawellek}, N. \& {Krivov}, A.~V. 2015, \mnras, 454, 3207

\bibitem[{{Pearce} {et~al.}(2015){Pearce}, {Wyatt}, \& {Kennedy}}]{pearce2015}
{Pearce}, T.~D., {Wyatt}, M.~C., \& {Kennedy}, G.~M. 2015, \mnras, 448, 3679

\bibitem[{{Pecaut} \& {Mamajek}(2016)}]{pm2016}
{Pecaut}, M.~J. \& {Mamajek}, E.~E. 2016, \mnras, 461, 794

\bibitem[{{Perets} \& {Kouwenhoven}(2012)}]{perets2012}
{Perets}, H.~B. \& {Kouwenhoven}, M.~B.~N. 2012, \apj, 750, 83

\bibitem[{{P{\'e}rez} {et~al.}(2016){P{\'e}rez}, {Carpenter}, {Andrews},
  {Ricci}, {Isella}, {Linz}, {Sargent}, {Wilner}, {Henning}, {Deller},
  {Chandler}, {Dullemond}, {Lazio}, {Menten}, {Corder}, {Storm}, {Testi},
  {Tazzari}, {Kwon}, {Calvet}, {Greaves}, {Harris}, \& {Mundy}}]{perez2016}
{P{\'e}rez}, L.~M., {Carpenter}, J.~M., {Andrews}, S.~M., {et~al.} 2016,
  Science, 353, 1519

\bibitem[{{Perrot} {et~al.}(2016){Perrot}, {Boccaletti}, {Pantin}, {Augereau},
  {Lagrange}, {Galicher}, {Maire}, {Mazoyer}, {Milli}, {Rousset}, {Gratton},
  {Bonnefoy}, {Brandner}, {Buenzli}, {Langlois}, {Lannier}, {Mesa}, {Peretti},
  {Salter}, {Sissa}, {Chauvin}, {Desidera}, {Feldt}, {Vigan}, {Di Folco},
  {Dutrey}, {P{\'e}ricaud}, {Baudoz}, {Benisty}, {De Boer}, {Garufi}, {Girard},
  {Menard}, {Olofsson}, {Quanz}, {Mouillet}, {Christiaens}, {Casassus},
  {Beuzit}, {Blanchard}, {Carle}, {Fusco}, {Giro}, {Hubin}, {Maurel},
  {Moeller-Nilsson}, {Sevin}, \& {Weber}}]{perrot2016}
{Perrot}, C., {Boccaletti}, A., {Pantin}, E., {et~al.} 2016, \aap, 590, L7

\bibitem[{{Perryman} {et~al.}(2014){Perryman}, {Hartman}, {Bakos}, \&
  {Lindegren}}]{perryman2014}
{Perryman}, M., {Hartman}, J., {Bakos}, G.~{\'A}., \& {Lindegren}, L. 2014,
  \apj, 797, 14

\bibitem[{{Pi{\'e}tu} {et~al.}(2007){Pi{\'e}tu}, {Dutrey}, \&
  {Guilloteau}}]{pietu2007}
{Pi{\'e}tu}, V., {Dutrey}, A., \& {Guilloteau}, S. 2007, \aap, 467, 163

\bibitem[{{Plavchan} {et~al.}(2009){Plavchan}, {Werner}, {Chen}, {Stapelfeldt},
  {Su}, {Stauffer}, \& {Song}}]{plavchan2009}
{Plavchan}, P., {Werner}, M.~W., {Chen}, C.~H., {et~al.} 2009, \apj, 698, 1068

\bibitem[{{Pollack} {et~al.}(1996){Pollack}, {Hubickyj}, {Bodenheimer},
  {Lissauer}, {Podolak}, \& {Greenzweig}}]{pollack1996}
{Pollack}, J.~B., {Hubickyj}, O., {Bodenheimer}, P., {et~al.} 1996, \icarus,
  124, 62

\bibitem[{{Quanz}(2015)}]{quanz2015a}
{Quanz}, S.~P. 2015, \apss, 357, 148

\bibitem[{{Quanz} {et~al.}(2015){Quanz}, {Amara}, {Meyer}, {Girard},
  {Kenworthy}, \& {Kasper}}]{quanz2015b}
{Quanz}, S.~P., {Amara}, A., {Meyer}, M.~R., {et~al.} 2015, \apj, 807, 64

\bibitem[{{Quanz} {et~al.}(2013){Quanz}, {Amara}, {Meyer}, {Kenworthy},
  {Kasper}, \& {Girard}}]{quanz2013}
{Quanz}, S.~P., {Amara}, A., {Meyer}, M.~R., {et~al.} 2013, \apjl, 766, L1

\bibitem[{{Rameau} {et~al.}(2013{\natexlab{a}}){Rameau}, {Chauvin}, {Lagrange},
  {Klahr}, {Bonnefoy}, {Mordasini}, {Bonavita}, {Desidera}, {Dumas}, \&
  {Girard}}]{rameau2013a}
{Rameau}, J., {Chauvin}, G., {Lagrange}, A.-M., {et~al.} 2013{\natexlab{a}},
  \aap, 553, A60

\bibitem[{{Rameau} {et~al.}(2013{\natexlab{b}}){Rameau}, {Chauvin}, {Lagrange},
  {Meshkat}, {Boccaletti}, {Quanz}, {Currie}, {Mawet}, {Girard}, {Bonnefoy}, \&
  {Kenworthy}}]{rameau2013c}
{Rameau}, J., {Chauvin}, G., {Lagrange}, A.-M., {et~al.} 2013{\natexlab{b}},
  \apjl, 779, L26

\bibitem[{{Rameau} {et~al.}(2012){Rameau}, {Chauvin}, {Lagrange},
  {Th{\'e}bault}, {Milli}, {Girard}, \& {Bonnefoy}}]{rameau2012}
{Rameau}, J., {Chauvin}, G., {Lagrange}, A.-M., {et~al.} 2012, \aap, 546, A24

\bibitem[{{Rameau} {et~al.}(2017){Rameau}, {Follette}, {Pueyo}, {Marois},
  {Macintosh}, {Millar-Blanchaer}, {Wang}, {Vega}, {Doyon}, {Lafreni{\`e}re},
  {Nielsen}, {Bailey}, {Chilcote}, {Close}, {Esposito}, {Males}, {Metchev},
  {Morzinski}, {Ruffio}, {Wolff}, {Ammons}, {Barman}, {Bulger}, {Cotten}, {De
  Rosa}, {Duchene}, {Fitzgerald}, {Goodsell}, {Graham}, {Greenbaum}, {Hibon},
  {Hung}, {Ingraham}, {Kalas}, {Konopacky}, {Larkin}, {Maire}, {Marchis},
  {Oppenheimer}, {Palmer}, {Patience}, {Perrin}, {Poyneer}, {Rajan},
  {Rantakyr{\"o}}, {Marley}, {Savransky}, {Schneider}, {Sivaramakrishnan},
  {Song}, {Soummer}, {Thomas}, {Wallace}, {Ward-Duong}, \&
  {Wiktorowicz}}]{rameau2017}
{Rameau}, J., {Follette}, K.~B., {Pueyo}, L., {et~al.} 2017, \aj, 153, 244

\bibitem[{{Reggiani} {et~al.}(2018){Reggiani}, {Christiaens}, {Absil}, {Mawet},
  {Huby}, {Choquet}, {Gomez Gonzalez}, {Ruane}, {Femenia}, {Serabyn},
  {Matthews}, {Barraza}, {Carlomagno}, {Defr{\`e}re}, {Delacroix}, {Habraken},
  {Jolivet}, {Karlsson}, {Orban de Xivry}, {Piron}, {Surdej}, {Vargas Catalan},
  \& {Wertz}}]{reggiani2018}
{Reggiani}, M., {Christiaens}, V., {Absil}, O., {et~al.} 2018, \aap, 611, A74

\bibitem[{{Reggiani} {et~al.}(2016){Reggiani}, {Meyer}, {Chauvin}, {Vigan},
  {Quanz}, {Biller}, {Bonavita}, {Desidera}, {Delorme}, {Hagelberg}, {Maire},
  {Boccaletti}, {Beuzit}, {Buenzli}, {Carson}, {Covino}, {Feldt}, {Girard},
  {Gratton}, {Henning}, {Kasper}, {Lagrange}, {Mesa}, {Messina}, {Montagnier},
  {Mordasini}, {Mouillet}, {Schlieder}, {Segransan}, {Thalmann}, \&
  {Zurlo}}]{reggiani2016}
{Reggiani}, M., {Meyer}, M.~R., {Chauvin}, G., {et~al.} 2016, \aap, 586, A147

\bibitem[{{Reggiani} {et~al.}(2014){Reggiani}, {Quanz}, {Meyer}, {Pueyo},
  {Absil}, {Amara}, {Anglada}, {Avenhaus}, {Girard}, {Carrasco Gonzalez},
  {Graham}, {Mawet}, {Meru}, {Milli}, {Osorio}, {Wolff}, \&
  {Torrelles}}]{reggiani2014}
{Reggiani}, M., {Quanz}, S.~P., {Meyer}, M.~R., {et~al.} 2014, \apjl, 792, L23

\bibitem[{{Rhee} {et~al.}(2007){Rhee}, {Song}, {Zuckerman}, \&
  {McElwain}}]{rhee2007}
{Rhee}, J.~H., {Song}, I., {Zuckerman}, B., \& {McElwain}, M. 2007, \apj, 660,
  1556

\bibitem[{{Ribas} {et~al.}(2018){Ribas}, {Tuomi}, {Reiners}, {Butler},
  {Morales}, {Perger}, {Dreizler}, {Rodr{\'{\i}}guez-L{\'o}pez}, {Gonz{\'a}lez
  Hern{\'a}ndez}, {Rosich}, {Feng}, {Trifonov}, {Vogt}, {Caballero}, {Hatzes},
  {Herrero}, {Jeffers}, {Lafarga}, {Murgas}, {Nelson}, {Rodr{\'{\i}}guez},
  {Strachan}, {Tal-Or}, {Teske}, {Toledo-Padr{\'o}n}, {Zechmeister},
  {Quirrenbach}, {Amado}, {Azzaro}, {B{\'e}jar}, {Barnes}, {Berdi{\~n}as},
  {Burt}, {Coleman}, {Cort{\'e}s-Contreras}, {Crane}, {Engle}, {Guinan},
  {Haswell}, {Henning}, {Holden}, {Jenkins}, {Jones}, {Kaminski}, {Kiraga},
  {K{\"u}rster}, {Lee}, {L{\'o}pez-Gonz{\'a}lez}, {Montes}, {Morin}, {Ofir},
  {Pall{\'e}}, {Rebolo}, {Reffert}, {Schweitzer}, {Seifert}, {Shectman},
  {Staab}, {Street}, {Su{\'a}rez Mascare{\~n}o}, {Tsapras}, {Wang}, \&
  {Anglada-Escud{\'e}}}]{ribas2018}
{Ribas}, I., {Tuomi}, M., {Reiners}, A., {et~al.} 2018, \nat, 563, 365

\bibitem[{{Rodigas} {et~al.}(2014){Rodigas}, {Debes}, {Hinz}, {Mamajek},
  {Pecaut}, {Currie}, {Bailey}, {Defrere}, {De Rosa}, {Hill}, {Leisenring},
  {Schneider}, {Skemer}, {Skrutskie}, {Vaitheeswaran}, \&
  {Ward-Duong}}]{rodigas2014a}
{Rodigas}, T.~J., {Debes}, J.~H., {Hinz}, P.~M., {et~al.} 2014, \apj, 783, 21

\bibitem[{{Rosenfeld} {et~al.}(2013){Rosenfeld}, {Andrews}, {Wilner},
  {Kastner}, \& {McClure}}]{rosenfeld2013}
{Rosenfeld}, K.~A., {Andrews}, S.~M., {Wilner}, D.~J., {Kastner}, J.~H., \&
  {McClure}, M.~K. 2013, \apj, 775, 136

\bibitem[{{Rousset} {et~al.}(2003){Rousset}, {Lacombe}, {Puget}, {Hubin},
  {Gendron}, {Fusco}, {Arsenault}, {Charton}, {Feautrier}, {Gigan}, {Kern},
  {Lagrange}, {Madec}, {Mouillet}, {Rabaud}, {Rabou}, {Stadler}, \&
  {Zins}}]{rousset2003}
{Rousset}, G., {Lacombe}, F., {Puget}, P., {et~al.} 2003, in \procspie, Vol.
  4839, Adaptive Optical System Technologies II, ed. P.~L. {Wizinowich} \&
  D.~{Bonaccini}, 140--149

\bibitem[{{Ruffio} {et~al.}(2017){Ruffio}, {Macintosh}, {Wang}, {Pueyo},
  {Nielsen}, {De Rosa}, {Czekala}, {Marley}, {Arriaga}, {Bailey}, {Barman},
  {Bulger}, {Chilcote}, {Cotten}, {Doyon}, {Duch{\^e}ne}, {Fitzgerald},
  {Follette}, {Gerard}, {Goodsell}, {Graham}, {Greenbaum}, {Hibon}, {Hung},
  {Ingraham}, {Kalas}, {Konopacky}, {Larkin}, {Maire}, {Marchis}, {Marois},
  {Metchev}, {Millar-Blanchaer}, {Morzinski}, {Oppenheimer}, {Palmer},
  {Patience}, {Perrin}, {Poyneer}, {Rajan}, {Rameau}, {Rantakyr{\"o}},
  {Savransky}, {Schneider}, {Sivaramakrishnan}, {Song}, {Soummer}, {Thomas},
  {Wallace}, {Ward-Duong}, {Wiktorowicz}, \& {Wolff}}]{ruffio2017}
{Ruffio}, J.-B., {Macintosh}, B., {Wang}, J.~J., {et~al.} 2017, \apj, 842, 14

\bibitem[{{Saar} \& {Donahue}(1997)}]{saar1997}
{Saar}, S.~H. \& {Donahue}, R.~A. 1997, \apj, 485, 319

\bibitem[{{Schegerer} {et~al.}(2009){Schegerer}, {Wolf}, {Hummel}, {Quanz}, \&
  {Richichi}}]{schegerer2009}
{Schegerer}, A.~A., {Wolf}, S., {Hummel}, C.~A., {Quanz}, S.~P., \& {Richichi},
  A. 2009, \aap, 502, 367

\bibitem[{{Schneider} {et~al.}(2005){Schneider}, {Silverstone}, \&
  {Hines}}]{schneider2005}
{Schneider}, G., {Silverstone}, M.~D., \& {Hines}, D.~C. 2005, \apjl, 629, L117

\bibitem[{{Schneider} {et~al.}(2006){Schneider}, {Silverstone}, {Hines},
  {Augereau}, {Pinte}, {M{\'e}nard}, {Krist}, {Clampin}, {Grady}, {Golimowski},
  {Ardila}, {Henning}, {Wolf}, \& {Rodmann}}]{schneider2006}
{Schneider}, G., {Silverstone}, M.~D., {Hines}, D.~C., {et~al.} 2006, \apj,
  650, 414

\bibitem[{{Schneider} {et~al.}(2011){Schneider}, {Dedieu}, {Le Sidaner},
  {Savalle}, \& {Zolotukhin}}]{schneider2011}
{Schneider}, J., {Dedieu}, C., {Le Sidaner}, P., {Savalle}, R., \&
  {Zolotukhin}, I. 2011, \aap, 532, A79

\bibitem[{{Schrijver} \& {Zwaan}(2008)}]{schrijver2008}
{Schrijver}, C.~J. \& {Zwaan}, C. 2008, {Solar and Stellar Magnetic Activity}

\bibitem[{{Sicilia-Aguilar} {et~al.}(2011){Sicilia-Aguilar}, {Henning},
  {Kainulainen}, \& {Roccatagliata}}]{sicilia2011}
{Sicilia-Aguilar}, A., {Henning}, T., {Kainulainen}, J., \& {Roccatagliata}, V.
  2011, \apj, 736, 137

\bibitem[{{Siess} {et~al.}(2000){Siess}, {Dufour}, \& {Forestini}}]{siess2000}
{Siess}, L., {Dufour}, E., \& {Forestini}, M. 2000, \aap, 358, 593

\bibitem[{{Sissa} {et~al.}(2018){Sissa}, {Gratton}, {Garufi}, {Rigliaco},
  {Zurlo}, {Mesa}, {Langlois}, {de Boer}, {Desidera}, {Ginski}, {Lagrange},
  {Maire}, {Vigan}, {Dima}, {Antichi}, {Baruffolo}, {Bazzon}, {Benisty},
  {Beuzit}, {Biller}, {Boccaletti}, {Bonavita}, {Bonnefoy}, {Brandner},
  {Bruno}, {Buenzli}, {Cascone}, {Chauvin}, {Cheetham}, {Claudi}, {Cudel}, {De
  Caprio}, {Dominik}, {Fantinel}, {Farisato}, {Feldt}, {Fontanive}, {Galicher},
  {Giro}, {Hagelberg}, {Incorvaia}, {Janson}, {Kasper}, {Keppler}, {Kopytova},
  {Lagadec}, {Lannier}, {Lazzoni}, {LeCoroller}, {Lessio}, {Ligi}, {Marzari},
  {Menard}, {Meyer}, {Mouillet}, {Peretti}, {Perrot}, {Potiron}, {Rouan},
  {Salasnich}, {Salter}, {Samland}, {Schmidt}, {Scuderi}, \&
  {Wildi}}]{sissa2018}
{Sissa}, E., {Gratton}, R., {Garufi}, A., {et~al.} 2018, \aap, 619, A160

\bibitem[{{Skemer} {et~al.}(2014){Skemer}, {Hinz}, {Esposito}, {Skrutskie},
  {Defr{\`e}re}, {Bailey}, {Leisenring}, {Apai}, {Biller}, {Bonnefoy},
  {Brandner}, {Buenzli}, {Close}, {Crepp}, {De Rosa}, {Desidera}, {Eisner},
  {Fortney}, {Henning}, {Hofmann}, {Kopytova}, {Maire}, {Males},
  {Millan-Gabet}, {Morzinski}, {Oza}, {Patience}, {Rajan}, {Rieke}, {Schertl},
  {Schlieder}, {Su}, {Vaz}, {Ward-Duong}, {Weigelt}, {Woodward}, \&
  {Zimmerman}}]{skemer2014}
{Skemer}, A.~J., {Hinz}, P., {Esposito}, S., {et~al.} 2014, in \procspie, Vol.
  9148, Adaptive Optics Systems IV, 91480L

\bibitem[{{Smith} {et~al.}(2009){Smith}, {Wyatt}, \& {Haniff}}]{smith2009}
{Smith}, R., {Wyatt}, M.~C., \& {Haniff}, C.~A. 2009, \aap, 503, 265

\bibitem[{{Soummer} {et~al.}(2012){Soummer}, {Pueyo}, \&
  {Larkin}}]{soummer2012}
{Soummer}, R., {Pueyo}, L., \& {Larkin}, J. 2012, \apjl, 755, L28

\bibitem[{{Spiegel} \& {Burrows}(2012)}]{sb2012}
{Spiegel}, D.~S. \& {Burrows}, A. 2012, \apj, 745, 174

\bibitem[{{Stone} {et~al.}(2018){Stone}, {Skemer}, {Hinz}, {Bonavita},
  {Kratter}, {Maire}, {Defrere}, {Bailey}, {Spalding}, {Leisenring},
  {Desidera}, {Bonnefoy}, {Biller}, {Woodward}, {Henning}, {Skrutskie},
  {Eisner}, {Crepp}, {Patience}, {Weigelt}, {De Rosa}, {Schlieder}, {Brandner},
  {Apai}, {Su}, {Ertel}, {Ward-Duong}, {Morzinski}, {Schertl}, {Hofmann},
  {Close}, {Brems}, {Fortney}, {Oza}, {Buenzli}, \& {Bass}}]{stone2018}
{Stone}, J.~M., {Skemer}, A.~J., {Hinz}, P.~M., {et~al.} 2018, \aj, 156, 286

\bibitem[{{Szul{\'a}gyi} {et~al.}(2019){Szul{\'a}gyi}, {Dullemond}, {Pohl}, \&
  {Quanz}}]{szulagyi2019}
{Szul{\'a}gyi}, J., {Dullemond}, C.~P., {Pohl}, A., \& {Quanz}, S.~P. 2019,
  \mnras, 487, 1248

\bibitem[{{Tamura}(2016)}]{tamura2016}
{Tamura}, M. 2016, Proceeding of the Japan Academy, Series B, 92, 45

\bibitem[{{Tatulli} {et~al.}(2011){Tatulli}, {Benisty}, {M{\'e}nard},
  {Varni{\`e}re}, {Martin-Za{\"i}di}, {Thi}, {Pinte}, {Massi}, {Weigelt},
  {Hofmann}, \& {Petrov}}]{tatulli2011}
{Tatulli}, E., {Benisty}, M., {M{\'e}nard}, F., {et~al.} 2011, \aap, 531, A1

\bibitem[{{The} {et~al.}(1994){The}, {de Winter}, \& {Perez}}]{the1994}
{The}, P.~S., {de Winter}, D., \& {Perez}, M.~R. 1994, \aaps, 104, 315

\bibitem[{{Tucci Maia} {et~al.}(2016){Tucci Maia}, {Ram{\'{\i}}rez},
  {Mel{\'e}ndez}, {Bedell}, {Bean}, \& {Asplund}}]{tucci2016}
{Tucci Maia}, M., {Ram{\'{\i}}rez}, I., {Mel{\'e}ndez}, J., {et~al.} 2016,
  \aap, 590, A32

\bibitem[{{van Leeuwen}(2007)}]{leeuwen2007}
{van Leeuwen}, F. 2007, \aap, 474, 653

\bibitem[{{Verhoeff} {et~al.}(2010){Verhoeff}, {Min}, {Acke}, {van Boekel},
  {Pantin}, {Waters}, {Tielens}, {van den Ancker}, {Mulders}, {de Koter}, \&
  {Bouwman}}]{verhoeff2010}
{Verhoeff}, A.~P., {Min}, M., {Acke}, B., {et~al.} 2010, \aap, 516, A48

\bibitem[{{Vican}(2012)}]{vican2012}
{Vican}, L. 2012, \aj, 143, 135

\bibitem[{{Vigan} {et~al.}(2017){Vigan}, {Bonavita}, {Biller}, {Forgan},
  {Rice}, {Chauvin}, {Desidera}, {Meunier}, {Delorme}, {Schlieder}, {Bonnefoy},
  {Carson}, {Covino}, {Hagelberg}, {Henning}, {Janson}, {Lagrange}, {Quanz},
  {Zurlo}, {Beuzit}, {Boccaletti}, {Buenzli}, {Feldt}, {Girard}, {Gratton},
  {Kasper}, {Le Coroller}, {Mesa}, {Messina}, {Meyer}, {Montagnier},
  {Mordasini}, {Mouillet}, {Moutou}, {Reggiani}, {Segransan}, \&
  {Thalmann}}]{vigan2017}
{Vigan}, A., {Bonavita}, M., {Biller}, B., {et~al.} 2017, \aap, 603, A3

\bibitem[{{Vigan} {et~al.}(2016){Vigan}, {Bonnefoy}, {Ginski}, {Beust},
  {Galicher}, {Janson}, {Baudino}, {Buenzli}, {Hagelberg}, {D'Orazi},
  {Desidera}, {Maire}, {Gratton}, {Sauvage}, {Chauvin}, {Thalmann}, {Malo},
  {Salter}, {Zurlo}, {Antichi}, {Baruffolo}, {Baudoz}, {Blanchard},
  {Boccaletti}, {Beuzit}, {Carle}, {Claudi}, {Costille}, {Delboulb{\'e}},
  {Dohlen}, {Dominik}, {Feldt}, {Fusco}, {Gluck}, {Girard}, {Giro}, {Gry},
  {Henning}, {Hubin}, {Hugot}, {Jaquet}, {Kasper}, {Lagrange}, {Langlois}, {Le
  Mignant}, {Llored}, {Madec}, {Martinez}, {Mawet}, {Mesa}, {Milli},
  {Mouillet}, {Moulin}, {Moutou}, {Orign{\'e}}, {Pavlov}, {Perret}, {Petit},
  {Pragt}, {Puget}, {Rabou}, {Rochat}, {Roelfsema}, {Salasnich}, {Schmid},
  {Sevin}, {Siebenmorgen}, {Smette}, {Stadler}, {Suarez}, {Turatto}, {Udry},
  {Vakili}, {Wahhaj}, {Weber}, \& {Wildi}}]{vigan2016}
{Vigan}, A., {Bonnefoy}, M., {Ginski}, C., {et~al.} 2016, \aap, 587, A55

\bibitem[{{Vigan} {et~al.}(2012){Vigan}, {Patience}, {Marois}, {Bonavita}, {De
  Rosa}, {Macintosh}, {Song}, {Doyon}, {Zuckerman}, {Lafreni{\`e}re}, \&
  {Barman}}]{vigan2012}
{Vigan}, A., {Patience}, J., {Marois}, C., {et~al.} 2012, \aap, 544, A9

\bibitem[{{Wade} {et~al.}(2005){Wade}, {Drouin}, {Bagnulo}, {Landstreet},
  {Mason}, {Silvester}, {Alecian}, {B{\"o}hm}, {Bouret}, {Catala}, \&
  {Donati}}]{wade2005}
{Wade}, G.~A., {Drouin}, D., {Bagnulo}, S., {et~al.} 2005, \aap, 442, L31

\bibitem[{{Wagner} {et~al.}(2019){Wagner}, {Stone}, {Spalding}, {Apai}, {Dong},
  {Ertel}, {Leisenring}, \& {Webster}}]{wagner2019}
{Wagner}, K., {Stone}, J.~M., {Spalding}, E., {et~al.} 2019, \apj, 882, 20

\bibitem[{{Wahhaj} {et~al.}(2013{\natexlab{a}}){Wahhaj}, {Liu}, {Biller},
  {Nielsen}, {Close}, {Hayward}, {Hartung}, {Chun}, {Ftaclas}, \&
  {Toomey}}]{wahhaj2013b}
{Wahhaj}, Z., {Liu}, M.~C., {Biller}, B.~A., {et~al.} 2013{\natexlab{a}}, The
  Astrophysical Journal, 779, 80

\bibitem[{{Wahhaj} {et~al.}(2013{\natexlab{b}}){Wahhaj}, {Liu}, {Nielsen},
  {Biller}, {Hayward}, {Close}, {Males}, {Skemer}, {Ftaclas}, {Chun}, {Thatte},
  {Tecza}, {Shkolnik}, {Kuchner}, {Reid}, {de Gouveia Dal Pino}, {Alencar},
  {Gregorio-Hetem}, {Boss}, {Lin}, \& {Toomey}}]{wahhaj2013a}
{Wahhaj}, Z., {Liu}, M.~C., {Nielsen}, E.~L., {et~al.} 2013{\natexlab{b}},
  \apj, 773, 179

\bibitem[{{Wahhaj} {et~al.}(2016){Wahhaj}, {Milli}, {Kennedy}, {Ertel},
  {Matr{\`a}}, {Boccaletti}, {del Burgo}, {Wyatt}, {Pinte}, {Lagrange},
  {Absil}, {Choquet}, {G{\'o}mez Gonz{\'a}lez}, {Kobayashi}, {Mawet},
  {Mouillet}, {Pueyo}, {Dent}, {Augereau}, \& {Girard}}]{wahhaj2016}
{Wahhaj}, Z., {Milli}, J., {Kennedy}, G., {et~al.} 2016, \aap, 596, L4

\bibitem[{{Walsh} {et~al.}(2016){Walsh}, {Juh{\'a}sz}, {Meeus}, {Dent}, {Maud},
  {Aikawa}, {Millar}, \& {Nomura}}]{walsh2016}
{Walsh}, C., {Juh{\'a}sz}, A., {Meeus}, G., {et~al.} 2016, \apj, 831, 200

\bibitem[{{Weise} {et~al.}(2010){Weise}, {Launhardt}, {Setiawan}, \&
  {Henning}}]{weise2010}
{Weise}, P., {Launhardt}, R., {Setiawan}, J., \& {Henning}, T. 2010, \aap, 517,
  A88

\bibitem[{{Wheelwright} {et~al.}(2013){Wheelwright}, {Weigelt}, {Caratti o
  Garatti}, \& {Garcia Lopez}}]{wheelwright2013}
{Wheelwright}, H.~E., {Weigelt}, G., {Caratti o Garatti}, A., \& {Garcia
  Lopez}, R. 2013, \aap, 558, A116

\bibitem[{{Wolszczan} \& {Frail}(1992)}]{wolszczan1992}
{Wolszczan}, A. \& {Frail}, D.~A. 1992, \nat, 355, 145

\bibitem[{{Wyatt}(2008)}]{wyatt2008}
{Wyatt}, M.~C. 2008, \araa, 46, 339

\bibitem[{{Zechmeister} {et~al.}(2019){Zechmeister}, {Dreizler}, {Ribas},
  {Reiners}, {Caballero}, {Bauer}, {B{\'e}jar}, {Gonz{\'a}lez-Cuesta},
  {Herrero}, \& {Lalitha}}]{zechmeister2019}
{Zechmeister}, M., {Dreizler}, S., {Ribas}, I., {et~al.} 2019, \aap, 627, A49

\bibitem[{{Zhu}(2015)}]{zhu2015}
{Zhu}, Z. 2015, \apj, 799, 16

\bibitem[{{Zuckerman} {et~al.}(1995){Zuckerman}, {Kim}, \&
  {Liu}}]{zuckerman1995}
{Zuckerman}, B., {Kim}, S.~S., \& {Liu}, T. 1995, \apjl, 446, L79

\bibitem[{{Zuckerman} {et~al.}(2011){Zuckerman}, {Rhee}, {Song}, \&
  {Bessell}}]{zuckerman2011}
{Zuckerman}, B., {Rhee}, J.~H., {Song}, I., \& {Bessell}, M.~S. 2011, \apj,
  732, 61

\bibitem[{{Zuckerman} \& {Song}(2004)}]{ZS2004}
{Zuckerman}, B. \& {Song}, I. 2004, \apj, 603, 738

\bibitem[{{Zuckerman} \& {Song}(2012)}]{ZS2012}
{Zuckerman}, B. \& {Song}, I. 2012, \apj, 758, 77

\end{thebibliography}
\bibliographystyle{aa}


\begin{appendix}

\section{Observing statistics histograms}
\label{app:histo}

\begin{figure}[!ht]
\centering
\subfloat[]{\label{fig:histtot}\includegraphics[width=0.23\textwidth]{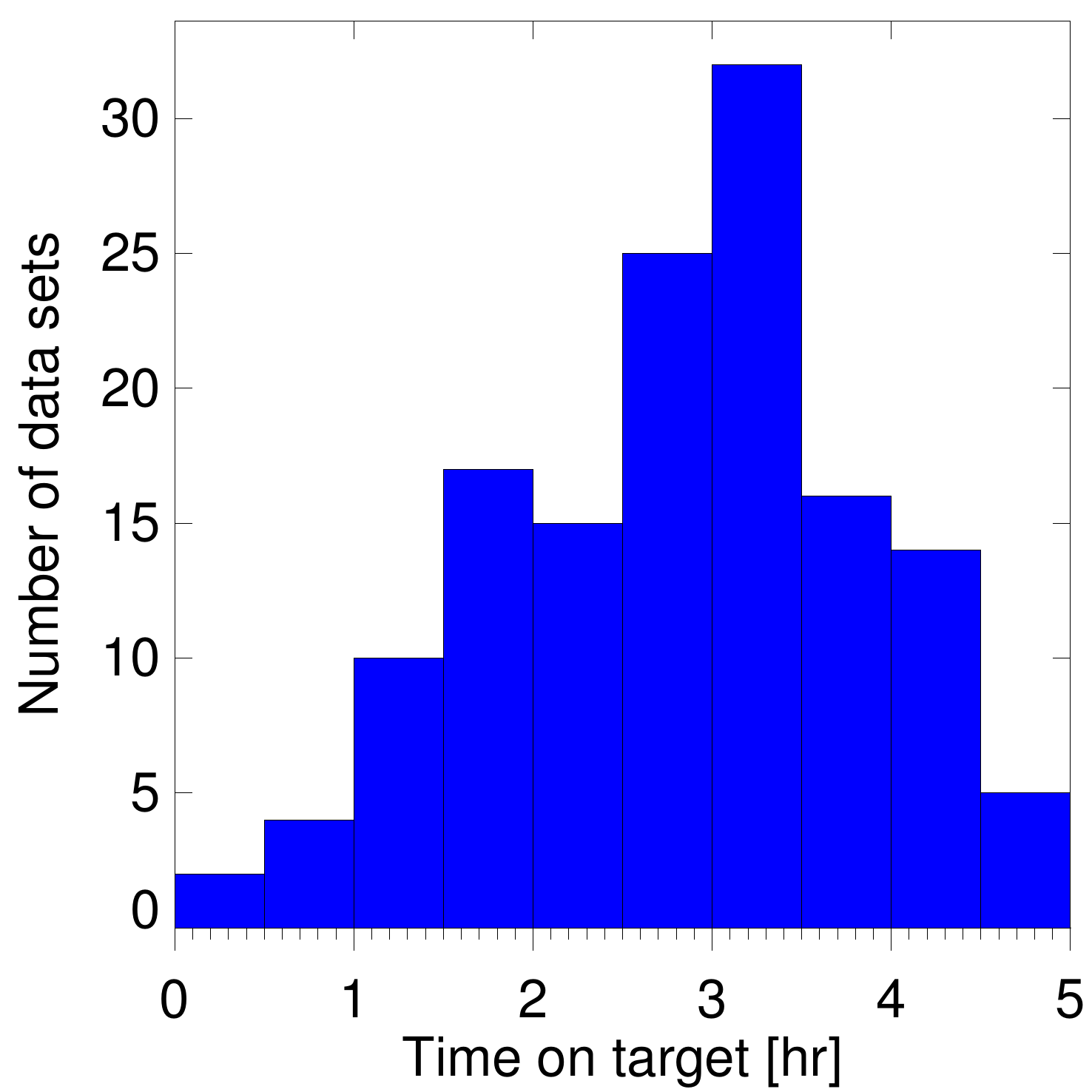}}
\subfloat[]{\label{fig:histrot}\includegraphics[width=0.23\textwidth]{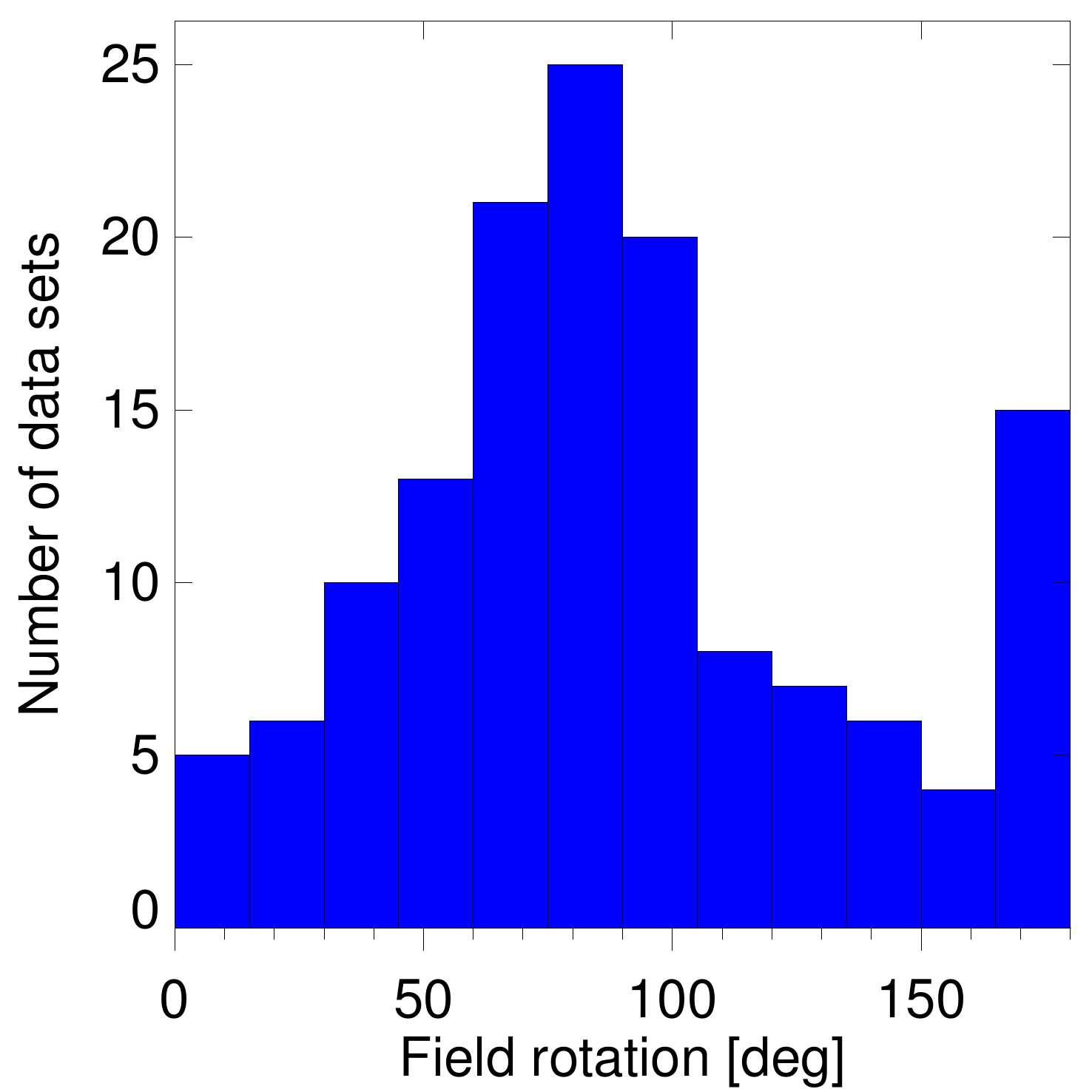}}
\quad
\subfloat[]{\label{fig:histseeing}\includegraphics[width=0.23\textwidth]{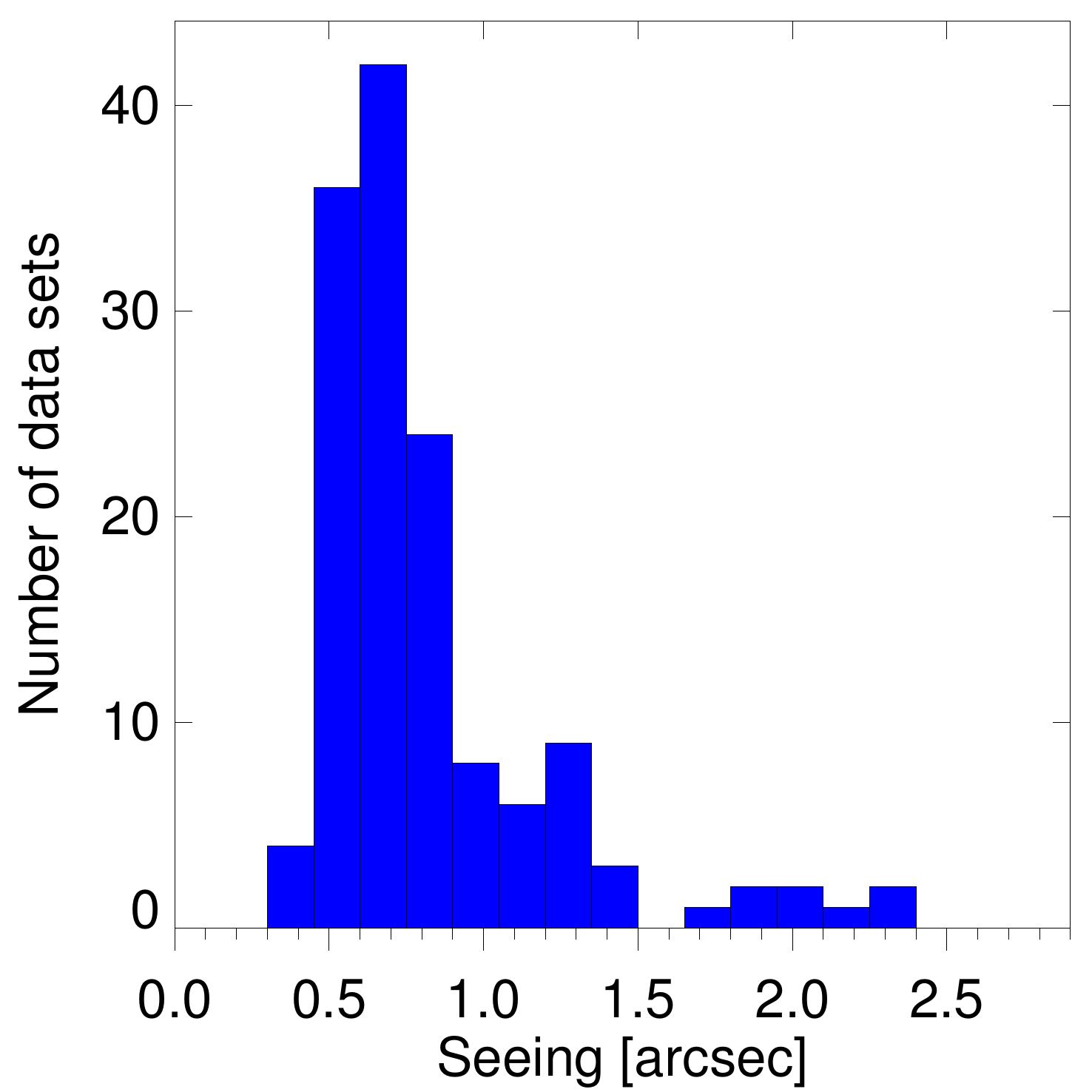}}
\subfloat[]{\label{fig:histt0}\includegraphics[width=0.23\textwidth]{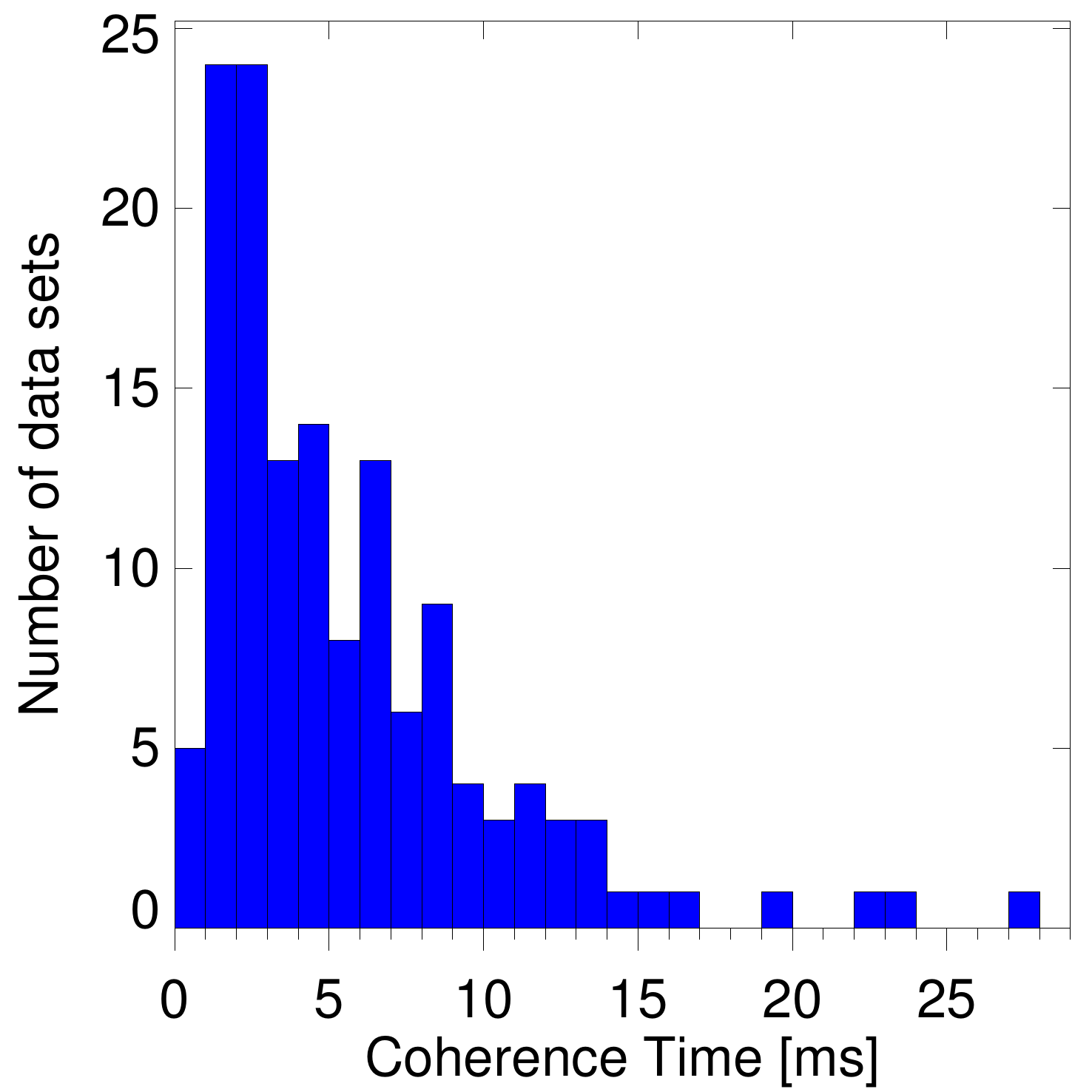}}
\quad
\caption{Distribution of the main observing characteristics for all targets observed during P96 to P100. (a) Total time  spent on target. For the majority of our targets we spend between 2 and 4\,hours to maximise field rotation (b).}
\label{fig:obshist}
\end{figure}


\section{Off-axis transmission of the annular groove phase mask vector vortex coronagraph}
\label{app:agpmtrans}

\begin{figure}[htb]
 \centering
 \includegraphics[width=9cm]{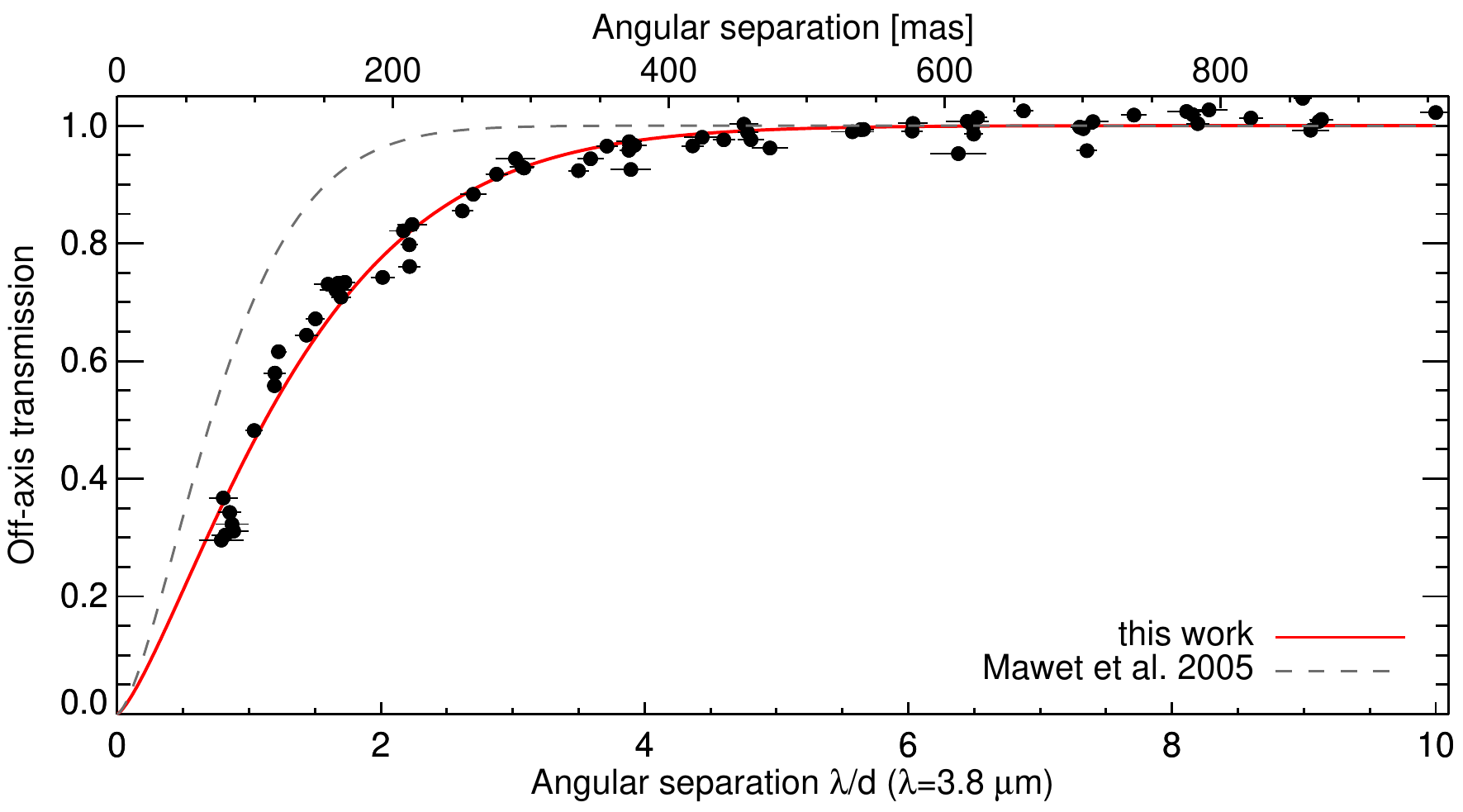}
 \caption{\label{fig:agpmtransmission} Measured off-axis transmission (black points) of the AGPM and the exponential fit (red line) to the data as a function of radial distance from the AGPM centre in resolution elements. The data of both stars are combined after normalising the stellar fluxes (see text for explanation). The error bars of the flux measurements are smaller than the size of the symbols. The exponential fit on simulated data by \mbox{\citet{2005ApJ...633.1191M}} is shown as a dashed grey line for comparison.}
\end{figure}

To measure the on-sky off-axis AGPM transmission, two bright stars were observed for comparison. HD\,146624 was observed during the night of May 19, 2017, and HIP98421 was observed during the night of August 31, 2017. The instrument setup and observation are similar to standard AGPM observations. The star was placed manually in vertical and horizontal direction around the AGPM with the goal of a one-pixel sampling. Sampling at different position angles is not necessary because of the perfect circular symmetry of the  AGPM. For HD\,146624 and  HIP\,98421, 45 and 30 data cubes were recorded, respectively, each containing 400 frames. Data cubes recorded with the star centred behind the AGPM or of bad data quality were rejected in the analysis, which leaves 40 useful cubes for HD\,146624 and 29 for HIP\,98421. A sky observation was taken after every five cubes.

The sky frames were cosmetically reduced including dark subtraction, flat field, and bad pixel correction. This was necessary for determining the position of the AGPM centre on the detector. In the sky frames, the AGPM centre appears similar to a (faint) stellar source due to thermal emission. The AGPM centre was fitted by a Moffat function for each frame in each cube. The final position of the AGPM centre per cube was retrieved by computing the median and the standard deviation of the individual measurements. 

The data cubes with the object were cosmetically reduced including sky subtraction, flat field, and bad pixel correction. The star in each frame and cube was fitted by a Moffat function to get its position and ultimately its radial distance from the AGPM centre. To measure the flux of a star, aperture photometry was performed using an aperture radius of 4$\times$FWHM\footnote{FWHM is the full width at half maximum and is $\approx3.5$\,px for NaCo observations in L' at \SI[]{3.8}{\micro\meter}} for the source, and an annulus of 6-10$\times$FWHM for the background. The final flux value for each cube is the weighted average of the individual flux measurements.

The off-axis transmission of the AGPM, $T(r)$, has a theoretical profile of the form \mbox{$1-\exp(-2\cdot r^{2})$}, where $r$\ is the radial distance of the AGPM centre in resolution elements $\lambda/D$ \mbox{\citep{2005ApJ...633.1191M}}. 
For measuring the empirical transmission curve, $T^{\ast}(r)$, we therefore use the form
\begin{equation} \label{eq:agpmtrans}
T^{\ast}(r) = 1-\exp(-a\cdot r^{b}) ~~.
\end{equation}
As both the position and flux have uncertainties, an MC approach was used to determine the fit parameters $a$\ and $b$. For both stars, the measured flux as a function of radial distance from the AGPM centre was fitted using eq.\,\ref{eq:agpmtrans} with the additional flux-scaling parameter $c$, but without weighting of the data points or taking into account uncertainties. Instead, 10$^{5}$ new data sets were generated by randomly drawing new measurements from each data point taking its uncertainty into account. 
The final fit value for each variable was obtained by computing the bootstrap mean of the $10^5$\ measurements by generating $10^5$ bootstrap samples. The uncertainty in the fit parameters is the standard deviation of all measurements. The explained fitting procedure on the individual stars was used only to determine the fit parameter $c$, which denotes the stellar flux. This is required before combining both data sets because the stars have different fluxes and $c$\ normalises the fluxes of both stars to one. The fitting procedure was then applied to the combined normalised data set (Fig.\,\ref{fig:agpmtransmission}). The final values and uncertainties of the fit parameters are: $a=0.586\pm0.022$, $b=1.330\pm0.059$.


\section{Astrometric correction}
\label{app:astrometry}

\begin{table}[!ht]
  \caption{Parameters and values used for the AGPM and satPSF instrument setup. \label{tbl:inssetup}}
  \centering
  \begin{tabular}{lcc}
  \hline\hline
  Parameter & AGPM & satPSF\\
\hline
  naxis [px] & 768 & 1024\\
  DIT [s] & 0.35 & 0.2\\
  NDIT & 100 & 100\\
  $N$ cubes/object & 5 & 5\\
  $N$ cubes/sky & 6 & 5\\
\hline
  \end{tabular}
\end{table}

\begin{table}[!ht]
  \caption{Weighted averaged values for plate scale and true north for different instrument setups. \label{tbl:astrometryvalues}}
  \centering
  \begin{tabular}{lcc}
  \hline\hline
  Mode & Pixel scale [mas] & True north [deg]\\
  \hline
  AGPM & 27.208$\pm$0.088 & 0.572$\pm$0.178 \\
  satPSF & 27.193$\pm$0.059 & 0.568$\pm$0.115 \\
  \hline
  \end{tabular}
\end{table}

Measurements of the plate scale and the true north orientation on a regular basis are essential to warrant accurate and precise astrometric values of identified candidate companions. The main stellar field used for astrometric calibration is 47\,Tuc as it usually provides more than 100 stars under good atmospheric conditions in the field of view of NaCo. An alternative field is the Orion Trapezium Cluster, but with significantly fewer sources visible in L'.

Astrometric fields are observed since the start of the survey in December 2015 with and without (satPSF) the AGPM. Table\,\ref{tbl:inssetup} lists typical values used for the instrument setup. The actual values can vary because of the brightness of the sky background. In addition, we refined the observing template with time. The goal during each observing run is to observe an astrometric field with and without the AGPM in both field and pupil tracking mode, respectively.

The image reduction steps for the AGPM and satPSF mode are similar and include: bad pixel and flat field correction, and sky subtraction. Since we are recording up to six sky cubes, the frames and cubes are median-combined to a single image to remove any stellar sources present. Each object cube (the actual astrometric field) is treated individually. The final analysis is based on the average image for each individual cube.  If the data were recorded in pupil tracking mode, the frames are derotated before averaging them.

After the cosmetic reduction, stellar sources in the image are identified with a version of DAOPHOT implemented in IDL. Marginal Gaussians are used to find centroids. To refine the result and reject bad detections, each initially found source is fitted with a Gaussian function. The fitted FWHM values of both directions and the fitted position are used to reject non-stellar sources. 
For the remaining sources, the x and y pixel coordinates are converted into RA and DEC using the WCS information written in the fits header and the separation and position angles between all stars are computed. 
To compare these values with a reference, we use the catalogue of \mbox{\citet{2006ApJS..166..249M}} for 47\,Tuc and the catalogue of \mbox{\citet{2012ApJ...749..180C}} for the Trapezium. 
The catalogue entries (RA, DEC) are corrected for proper motion if present and converted to x,y-coordinates using the WCS information written in the fits header and the separation and position angles between all catalogue stars are computed. 
This allows us to identify the detected stars in the image with their corresponding catalogue entry. The derived value for the plate scale between one pair of stars is then the distance from the catalogue pair in milliarcseconds divided by the measured distance of the same pair detected in the image. The true north value is computed from the difference between measured position angle of a stellar pair and its respective catalogue entry. For 100 stars detected in the image, there are therefore over 4400 individual measurements available.
The final values for plate scale and true north including their uncertainties are listed in Table\,\ref{tbl:astrometryvalues} and were produced by computing the two-sigma-clipped resistant mean.

\end{appendix}

\clearpage
\onecolumn

\setcounter{table}{1}
\begin{longtable}{llllccllll}
\caption{\label{tab:slist2} DEB targets observed between December 2015 and March 2018 (P\,96 through P\,100)}\\
\hline\hline
Name\tablefootmark{a}            & 
RA(J2000)                        & 
DEC(J2000)                       & 
dist.\tablefootmark{b}           & 
$V$                              & 
log($L_{\ast}$)\tablefootmark{c} & 
$T_{\rm eff}$\tablefootmark{c}   & 
age\tablefootmark{d}             & 
$R_{\rm disc}$\tablefootmark{c}  & 
Ref.\tablefootmark{e}            \\
                 & 
[hh:mm:ss]       & 
[dd:mm:ss]       & 
[pc]             & 
[mag]            & 
[L$_{\odot}$]    & 
[K]              & 
[Myr]            &  
[au]             &
                 \\
\hline
\endfirsthead
\caption{continued.}\\
\hline\hline
Name\tablefootmark{a}   & 
RA(J2000)               & 
DEC(J2000)              & 
dist.\tablefootmark{b}  & 
$V$                               & 
log($L_{\ast}$)\tablefootmark{c}  & 
$T_{\rm eff}$\tablefootmark{c}    & 
age\tablefootmark{d}              & 
$R_{\rm disc}$\tablefootmark{c}   & 
Ref.\tablefootmark{e}             \\
                 & 
[hh:mm:ss]       & 
[dd:mm:ss]       & 
[pc]             & 
[mag]            & 
[L$_{\odot}$]    & 
[K]              & 
[Myr]            &  
[au]             &
                 \\
\hline
\endhead
\hline
\endfoot
HD\,203    & 00:06:50.02   & -23:06:26.7   & 39.9 & 6.2 & 0.63 & 6800 & 12$\pm$2  & 47$\pm$3           &  1 \\
HD\,377    & 00:08:25.69   & ~06:37:00.5  & 38.5 & 7.6 & 0.07 & 5820 & 90$\pm$65 & 136$\pm$11\tablefootmark{g} (86) & 2, 50 \\
HD\,2262  & 00:26:12.20  & -43:40:47.4   & 22.3 & 3.9 & 1.01 & 7750 & 200$\pm$100 & 45$\pm$5       & 4 \\
HD\,3003  & 00:32:43.79 & -63:01:53.0   & 45.8 & 5.1 & 1.28 & 9260 & 30$\pm$15   & 24$\pm$2          & 1 \\
HD\,3670  & 00:38:56.70 & -52:32:03.4   & 77.4 & 8.2 & 0.40 & 6390 & 30$\pm$15   & 160$\pm$8        & 4 \\
HIP\,6276  & 01:20:32.20 & -11:28:02.5   & 35.3 & 8.4 & -0.28 & 5320 & 70$\pm$25   & 28$\pm$4     & 5 \\
HD\,9672  & 01:34:37.78  & -15:40:34.9  & 57.0 & 5.6 & 1.18 & 8750 & 40$\pm$10 & 208$\pm$19\tablefootmark{g} (420) & 6, 51 \\
HD\,10008 & 01:37:35.37 & -06:45:36.7  & 24.0 & 7.7 & -0.33 & 5280 & 3000$\pm$2400 & 56$\pm$6    & 7 \\
HD\,10472 & 01:40:24.07 & -60:59:56.6  & 71.0 & 7.6 & 0.56 & 6710 & 30$\pm$15     & 180$\pm$13\tablefootmark{g}      & 1 \\
HD\,12039 & 01:57:48.91 & -21:54:04.9  & 41.4 & 8.1 & -0.05 & 5660 & 30$\pm$15    & 20$\pm$1         & 3 \\
HD\,13246 & 02:07:26.02 & -59:40:45.8  & 45.5 & 7.5 & 0.24 & 6130 & 30$\pm$15    & 12$\pm$1         & 3 \\
HD\,15115 & 02:26:16.25 & ~06:17:33.2  & 48.9 & 6.8 & 0.56 & 6720 & 12$\pm$2 & 212$\pm$13\tablefootmark{g} (315-550)  & 8, 52 \\
HD\,16743 & 02:39:07.56 & -52:56:05.3  & 57.8 & 6.8 & 0.71 & 6900 & 200$\pm$100 & 233$\pm$16\tablefootmark{g} (370) & 4, 53 \\
HD\,17925 & 02:52:31.89 & -12:46:09.3  & 10.4 & 6.1 & -0.40 & 5180 & 90$\pm$10     & 20$\pm$2        & 3 \\
HD\,19668 & 03:09:42.23 & -09:34:45.6  & 38.7 & 8.5 & -0.24 & 5420 & 100$\pm$50   & 16$\pm$2        & 9 \\
HD\,21997 & 03:31:53.65 & -25:36:50.9  & 69.5 & 6.4 & 1.03 & 8380 & 30$\pm$15 & 180$\pm$18 (150) & 4, 54 \\
HD\,22049 & 03:32:56.42 & -09:27:29.9  & 3.20 & 3.7 & -0.47 & 5073 & 1000$\pm$500 & 32$\pm$2 (67) & 10, 11, 51  \\
HD\,25457 & 04:02:36.66 & -00:16:05.9  & 18.8 & 5.4 & 0.31 & 6270 & 75$\pm$25      & 56$\pm$4       & 2 \\
HD\,27290 & 04:16:01.59 & -51:29:11.9  & 20.4 & 4.2 & 0.82 & 7140 & 300$\pm$250 & 218$\pm$18 (190) & 12, 55 \\
HD\,30422 & 04:46:25.75 & -28:05:14.8  & 57.4 & 6.2 & 0.93 & 7700 & 13$\pm$7 & 128$\pm$9 (123) & 4, 56 \\
HD\,30495 & 04:47:36.21 & -16:56:05.5  & 13.2 & 5.5 & -0.02 & 5850 & 45$\pm$10  & 84$\pm$5         & 13 \\
HD\,32297 & 05:02:27.44 & ~07:27:39.7  & 132.3 & 8.1 & 0.92 & 7790 & 20$\pm$10    & 107$\pm$8\tablefootmark{g} (110) & 12, 57 \\
HD\,35850 & 05:27:04.75 & -11:54:03.0  & 26.9   & 6.3 & 0.27 & 6040 & 12$\pm$15 & 68$\pm$4          & 3 \\
HD\,37484 & 05:37:39.63 & -28:37:34.7  & 59.0 & 7.3 & 0.54 & 6680 & 30$\pm$10   & 53$\pm$3          & 14 \\
SAO\,150676 & 05:40:20.74 & -19:40:10.8 & 73.0 & 9.0 & 0.06 & 5620 & 63$\pm$10 & 3$\pm$1           & 15 \\
HD\,38206 & 05:43:21.67 & -18:33:26.9  & 71.3 & 5.7 & 1.49 & 10380 & 30$\pm$15 & 194$\pm$22\tablefootmark{g} (170) & 12, 56 \\
HD\,38678 & 05:46:57.34 & -14:49:19.0  & 21.9 & 3.5 & 1.22 & 8790 & 12$\pm$2   & 24$\pm$2 (3)        & 4, 62 \\
HD\,43989 & 06:19:08.05 & -03:26:20.0  & 51.9 & 8.0 & 0.19 & 5940 & 30$\pm$15 & 8$\pm$2              & 5 \\
HD\,46190 & 06:27:48.62 & -62:8:59.7   & 84.3 & 6.6 & 1.13 & 8400 & 5$\pm$5     & 13$\pm$7            & 4 \\
HD\,48370 & 06:43:01.02 & -02:53:19.3  & 36.0 & 7.9 & -0.11 & 5580 & 30$\pm$10 & 168$\pm$7 (89)  & 16, 28 \\
HD\,53143 & 06:59:59.85 & -61:20:12.6  & 18.4 & 6.8 & -0.23 & 5410 & 300$\pm$100 & 60$\pm$4 (55-110) & 12, 59 \\
HD\,59967 & 07:30:42.51 & -37:20:21.7  & 21.8 & 6.6 & -0.05 & 5790 & 353$\pm$68  & 68$\pm$8         & 17 \\
HD\,61005 & 07:35:47.50 & -32:12:14.7  & 36.4 & 8.2 & -0.20 & 5400 & 30$\pm$15   & 67$\pm$2 (60)   & 3, 60 \\
HD\,69830 & 08:18:23.78 & -12:37:47.2  & 12.6 & 6.0 & -0.22 & 5420 & 4000$\pm$1000 & 2.6$\pm$0.1 (2.4) & 18, 61 \\
HD\,70573 & 08:22:49.90 & ~01:51:33.6  & 59.2 & 8.7 & 0.00 & 5750  & 70$\pm$15  & ...\tablefootmark{f} & 3 \\
HD\,71043 & 08:22:55.16 & -52:07:25.4  & 73.1 & 5.9 & 1.38 & 9260 & 30$\pm$15    & 16$\pm$2        & 4  \\
HD\,71155 & 08:25:39.63 & -03:54:23.1  & 37.4 & 3.9 & 1.60 & 9890 & 170$\pm$70 & 65$\pm$14 (88) & 19, 62 \\ 
HD\,72660 & 08:34:01.62 & -02:09:05.6  & 98.8 & 5.8 & 1.67 & 9500 & 200$\pm$100 & ...\tablefootmark{f} & 4 \\
HD\,76151 & 08:54:17.95 & -05:26:04.1  & 16.8 & 6.0 & -0.01 & 5830 & 1047$\pm$301 & 28$\pm$2  & 20 \\
HD\,92536 & 10:39:22.83 & -64:06:42.4  & 157.3 & 6.3 & 1.95 & 10800 & 46$\pm$6.5 & 13$\pm$3  & 21 \\
HD\,98058 & 11:16:39.70 & -03:39:05.8  & 57.3 & 4.5 & 1.62 & 7500 & 700$\pm$200 & ...\tablefootmark{f}  & 22 \\
HD\,102458 & 11:47:24.58 & -49:53:02.9 & 113.0 & 9.1 & 0.43 & 5670 & 8$\pm$5 & 18$\pm$2  & 3 \\
HD\,103703 & 11:56:26.56 & -58:49:16.8 & 107.1 & 8.5 & 0.57 & 6410 & 17$\pm$10 & 13$\pm$1  & 4 \\
MML\,8     & 12:12:35.77 & -55:20:27.3 & 112.9 & 10.5 & -0.04 & 4710 & 5$\pm$3 & 16$\pm$1  & 3 \\
HD\,107146 & 12:19:06.61 & ~16:32:55.2 & 27.4 & 7.0 & 0.00 & 5830 & 160$\pm$0 & 134$\pm$4\tablefootmark{g} (97-130) & 15, 63 \\
HD\,107649 & 12:22:24.86 & -51:01:34.4 & 108.0 & 8.8 & 0.45 & 6290 & 17$\pm$10 & 24$\pm$2  & 4 \\
HD\,109085 & 12:32:04.23 & -16:11:45.6 & 18.0 & 4.3 & 0.69 & 6940 & 950$\pm$350 & 230$\pm$15\tablefootmark{g} (145) & 23, 51 \\
HD\,110411 & 12:41:53.06 & ~10:14:08.3 & 38.1 & 4.9 & 1.18 & 8780 & 200$\pm$100 & 112$\pm$10 (111) & 12, 64 \\
HD\,111520 & 12:50:19.72 & -49:51:49.0 & 108.6 & 8.9 & 0.43 & 6140 & 17$\pm$10 & 71$\pm$4 (70) & 4, 65 \\
HD\,111631 & 12:50:43.58 & -00:46:01.8 & 10.6 & 8.5 & -1.00 & 3993 & 680$\pm$300 & 400$\pm$40  & 19 \\
MML\,28    & 13:01:50.70 & -53:04:58.1 & 117.5 & 11.1 & -0.29 & 4620 & 8$\pm$3 & 25$\pm$2  & 3 \\
HD\,113556 & 13:05:32.61 & -58:32:08.0 & 101.2 & 8.1 & 0.65 & 6840 & 17$\pm$10 & 167$\pm$16\tablefootmark{g}  & 4 \\
HD\,114082 & 13:09:16.19 & -60:18:30.1 & 95.4 & 8.2 & 0.58 & 6580 & 17$\pm$10 & 48$\pm$2 (27.7) & 4, 66 \\
HD\,115820 & 13:20:26.81 & -49:13:25.2 & 115.5 & 8.0 & 0.85 & 7470 & 17$\pm$10 & 39$\pm$3  & 4 \\
HD\,115892 & 13:20:35.82 & -36:42:44.2 & 17.8 & 2.7 & 1.36 & 9160 & 45$\pm$20 & 26$\pm$3  & 4 \\
MML\,36    & 13:37:57.30 & -41:34:42.0 & 99.5 & 10.1 & 0.00 & 4960 & 30$\pm$10 & 16$\pm$1  & 3 \\
HD\,125541 & 14:21:11.54 & -41:42:24.9 & 159.8 & 8.9 & 0.76 & 7160 & 16$\pm$0 & 29$\pm$2  & 4 \\
HD\,126062 & 14:24:37.00 & -47:10:39.9 & 132.0 & 7.4 & 1.23 & 9100 & 16$\pm$0 & 325$\pm$27\tablefootmark{g}  & 4 \\
HD\,131511 & 14:53:23.77 & ~19:09:10.1 & 11.4 & 6.0 & -0.29 & 5090 & 1000$\pm$300 & 120$\pm$13 (74) & 24, 67 \\
HD\,131835 & 14:56:54.47 & -35:41:43.7 & 133.3 & 7.9 & 0.98 & 7690 & 16$\pm$2 & 150$\pm$12\tablefootmark{g} (75-210) & 25, 68 \\
HD\,134888 & 15:13:27.96 & -33:08:50.2 & 112.0 & 8.7 & 0.50 & 6320 & 16$\pm$8 & 88$\pm$4  & 4 \\
GJ\,581    & 15:19:27.54 & -07:43:19.3 & 6.30 & 10.6 & -1.92 & 3238 & 1000$\pm$500 & 16$\pm$2 (25) & 26, 69 \\
HD\,156623 & 17:20:50.62 & -45:25:14.5 & 111.4 & 7.3 & 1.11 & 8610 & 16$\pm$2 & 75$\pm$10\tablefootmark{g} (150) & 25, 70 \\
HD\,172555 & 18:45:26.86 & -64:52:15.2 & 28.3 & 4.8 & 0.89 & 7620 & 23$\pm$3 & 21$\pm$2  & 27 \\
HD\,177171 & 19:06:19.96 & -52:20:27.3 & 58.8 & 5.2 & 1.40 & 6010 & 30$\pm$15 & ...\tablefootmark{f} & 4 \\
HD\,181296 & 19:22:51.18 & -54:25:25.4 & 47.3 & 5.0 & 1.43 & 9440 & 23$\pm$3 & 43$\pm$4 (24) & 27, 61 \\
HD\,181327 & 19:22:58.92 & -54:32:16.3 & 48.1 & 7.0 & 0.46 & 6450 & 23$\pm$3 & 82$\pm$4 (86.3) & 27, 71 \\
HD\,191089 & 20:09:05.22 & -26:13:26.5 & 50.1 & 7.2 & 0.44 & 6460 & 12$\pm$2 & 59$\pm$3 (73) & 1, 72 \\
HD\,193571 & 20:22:27.50 & -42:02:58.4 & 68.3 & 5.6 & 1.44 & 9740 & 170$\pm$70 & 119$\pm$15 & 4 \\
HD\,196544 & 20:37:49.12 & ~11:22:39.6 & 59.4 & 5.4 & 1.36 & 9130 & 225$\pm$40 & 66$\pm$7  & 9 \\
HD\,197481 & 20:45:09.34 & -31:20:24.1 & 9.72 & 8.6 & -1.02 & 3600 & 15$\pm$10 & 44$\pm$2 (70) & 3, 51 \\
HD\,202628 & 21:18:27.27 & -43:20:04.8 & 23.8 & 6.8 & -0.01 & 5780 & 604$\pm$445 & 160$\pm$24 (162-254) & 28, 73 \\
HD\,206860 & 21:44:31.19 & ~14:46:20.0 & 18.1 & 6.0 & 0.06 & 5970 & 85$\pm$65 & 40$\pm$2  & 29 \\
NLTT\,54872 & 22:48:04.47 & -24:22:07.5 & 7.67 & 12.6 & -2.33 & 3011 & 440$\pm$40 & 48$\pm$8  & 30 \\
HD\,218340 & 23:08:12.24 & -63:37:40.8 & 56.0 & 8.4 & 0.06 & 5810 & 45$\pm$10 & 200$\pm$16  & 4 \\
HD\,218511 & 23:09:41.44 & -67:43:56.3 & 14.8 & 8.3 & -0.80 & 4370 & 1260$\pm$600 & 34$\pm$2  & 19 \\
HD\,220825 & 23:26:55.96 & ~01:15:20.2 & 48.9 & 4.9 & 1.34 & 9400 & 70$\pm$25 & 28$\pm$4  & 5 \\
HD\,223340 & 23:48:50.50 & -28:07:15.7 & 44.2 & 9.3 & -0.35 & 5250 & 220$\pm$100 & 42$\pm$2  & 19 \\
\end{longtable}
\mbox{\tablefoottext{a}{Where available, we use the HD number as main source ID.}}
\tablefoottext{b}{All distances in this table are inferred from {\it Gaia}\,DR2 parallaxes \mbox{\citep{gaia_mission,gaia_dr2}} with the method described by \mbox{\citet{bailer2018} and retrieved through VizieR \mbox{\citep{vizier2000}}.} Uncertainties are typically <0.6\% and exceed in no case 2.6\%.}
\tablefoottext{c}{Stellar $L_{\ast}$, $T_{\rm eff}$, and $R_{\rm disc}$\ are derived from SED fitting (see text Sect.\,\ref{ssec:targetsample}). Additional $R_{\rm disc}$\ values in brackets refer to directly observed and spatially resolved disc radii compiled from the literature (references 50 through 73).}
\tablefoottext{d}{Age notes}
\tablefoottext{e}{References 1 through 30 refer to age, references 50 through 73 refer to observed disc radius.}
\tablefoottext{f}{disc excess is not significant.}
\tablefoottext{g}{SED fit indicates second hot inner dust belt.}
\tablebib{(1)~\mbox{\citet{ZS2004}};
(2)~\mbox{\citet{apai2008}};
(3)~\mbox{\citet{weise2010}};
(4)~\mbox{\citet{chen2014}};
(5)~\mbox{\citet{zuckerman2011}};
(6)~\mbox{\citet{ZS2012}};
(7)~\mbox{\citet{casa2011}};
(8)~\mbox{\citet{moor2011}};
(9)~\mbox{\citet{desidera2015}};
(10)~\mbox{\citet{MH2008}};
(11)~\mbox{\citet{bonfanti2015}};
(12)~\mbox{\citet{rhee2007}};
(13)~\mbox{\citet{maldonado2010}};
(14)~\mbox{\citet{meshkat2015}};
(15)~\mbox{\citet{carpenter2009}};
(16)~\mbox{\citet{moor2016}};
(17)~\mbox{\citet{plavchan2009}};
(18)~\mbox{\citet{olofsson2012}};
(19)~\mbox{\citet{vican2012}};
(20)~\mbox{\citet{lachaume1999}};
(21)~\mbox{\citet{dobbie2010}};
(22)~\mbox{\citet{eiroa2016}};
(23)~\mbox{\citet{lafreniere2007}};
(24)~\mbox{\citet{MH2008}};
(25)~\mbox{\citet{pm2016}};
(26)~\mbox{\citet{bryden2009}};
(27)~\mbox{\citet{mama2014}};
(28)~\mbox{\citet{tucci2016}};
(29)~\mbox{\citet{montes2001}};
(30)~\mbox{\citet{mama2013}};
(50)~\mbox{\citet{choquet2016}};
(51)~\mbox{\citet{holland2017}};
(52)~\mbox{\citet{kalas2007}};
(53)~\mbox{\citet{moor2015a}};
(54)~\mbox{\citet{moor2013}};
(55)~\mbox{\citet{broek2013}};
(56)~\mbox{\citet{morales2016}};
(57)~\mbox{\citet{boccaletti2012}};
(58)~\mbox{\citet{moor2016}};
(59)~\mbox{\citet{kalas2006}};
(60)~\mbox{\citet{olofsson2016}};
(61)~\mbox{\citet{smith2009}};
(62)~\mbox{\citet{moerchen2010}};
(63)~\mbox{\citet{corder2009}};
(64)~\mbox{\citet{booth2013}};
(65)~\mbox{\citet{draper2016}};
(66)~\mbox{\citet{wahhaj2016}};
(67)~\mbox{\citet{marshall2014}};
(68)~\mbox{\citet{hung2015}};
(69)~\mbox{\citet{lestrade2012}};
(70)~\mbox{\citet{lieman2016}};
(71)~\mbox{\citet{schneider2006}};
(72)~\mbox{\citet{churcher2011}};
(73)~\mbox{\citet{krist2012}};
}

\setcounter{table}{4}
\begin{longtable}{llcccccccccc}
\caption{\label{tab:obsslist2} Observations and achieved contrast values of DEB targets.}\\
\hline\hline
Name & obs date & Seeing  & field    & AGPM & L$^{\prime}$\tablefootmark{a} & \multicolumn{6}{c}{-----------------\ 5\,$\sigma$\ contrast\tablefootmark{b} at $r$\ -----------------} \\
           &               &               & rot.  &            &            &  0\asp25 & 0\asp5 & 0\asp75 & 1\asp0 & 2\asp0 & 3\asp0  \\
           &               & [arcsec] & [deg]  &            & [mag] & [mag] & [mag] & [mag] & [mag] & [mag] & [mag] \\
\hline
\endfirsthead
\caption{continued.}\\
\hline\hline
Name & obs date & Seeing  & field    & AGPM & L$^{\prime}$\tablefootmark{a} & \multicolumn{6}{c}{-----------------\ 5\,$\sigma$\ contrast\tablefootmark{b} at $r$\ -----------------} \\
           &               &               & rot.         &            &            &  0\asp25 & 0\asp5 & 0\asp75 & 1\asp0 & 2\asp0 & 3\asp0  \\
           &               & [arcsec] & [deg]  &            & [mag] & [mag] & [mag] & [mag] & [mag] & [mag] & [mag] \\
\hline
\endhead
\hline
\endfoot
HD\,203                 & 2017-10-29 & 0.6 & 156 & y & 5.09$\pm$0.22 & 6.0 & 8.2 & 8.9 & 9.5 & 9.8 & 9.7 \\
HD\,377                 & 2016-07-31 & 0.4 & 74  & y & 6.07$\pm$0.12 & 8.0 & 9.4 & 9.7 & 9.9 & 10.5 & 10.6 \\
HD\,2262                & 2017-10-31 & 0.8 & 72  & y & 3.35$\pm$0.63 & 7.6 & 10.1 & 11.0 & 11.8 & 13.1 & 13.1 \\
HD\,3003                & 2017-11-01 & 0.6 & 94  & y & 4.88$\pm$0.24 & 8.8 & 10.0 & 11.0 & 11.5 & 11.9 & 12.0 \\
HD\,3670                & 2016-08-01 & 0.6 & 104 & n & 7.02$\pm$0.06 & 7.3 & 9.2 & 9.6 & 10.0 & 10.3 & 10.2 \\
HIP\,6276               & 2016-12-11 & 1.4 & 40  & n & 6.53$\pm$0.08 & 5.8 & 8.0 & 9.1 & 9.6 & 10.0 & 10.0 \\
HD\,9672                & 2016-11-06 & 0.5 & 129 & y & 5.41$\pm$0.18 & 7.8 & 9.0 & 9.5 & 10 & 10.6 & 10.9 \\
HD\,10008               & 2016-11-12 & 0.5 & 106 & y & 5.73$\pm$0.14 & 7.2 & 8.9 & 9.5 & 9.8 & 10.7 & 10.7 \\
HD\,10472               & 2016-11-07 & 0.7 & 71  & y & 6.61$\pm$0.08 & 7.1 & 8.3 & 8.5 & 9.2 & 9.7 & 9.9 \\
HD\,12039               & 2017-11-03 & 0.7 & 147 & y & 6.40$\pm$0.08 & 6.5 & 8.2 & 8.6 & 8.9 & 9.5 & 9.6 \\
HD\,13246               & 2017-08-29 & 0.6 & 99  & y & 6.16$\pm$0.10 & 6.7 & 8.5 & 9.2 & 9.4 & 10.0 & 10.1 \\
HD\,15115               & 2016-11-08 & 0.4 & 91  & y & 5.75$\pm$0.07 & 8.0 & 9.1 & 9.9 & 10.4 & 10.9 & 10.9 \\
HD\,16743               & 2016-11-11 & 0.7 & 92  & y & 5.90$\pm$0.14 & 8.5 & 9.5 & 10.1 & 10.3 & 10.7 & 10.9 \\
HD\,17925               & 2015-12-15 & 1.0 & 94  & y & 3.89$\pm$0.42 & 7.6 & 9.9 & 10.8 & 11.5 & 12.5 & 12.5 \\
HD\,19668               & 2015-12-18 & 0.9 & 79  & n & 6.66$\pm$0.07 & 7.1 & 9.6 & 10.0 & 10.3 & 10.3 & 10.4 \\
HD\,21997               & 2017-11-02 & 0.6 & 182 & y & 6.08$\pm$0.11 & 7.0 & 8.2 & 9.0 & 9.4 & 10.0 & 10.2 \\
HD\,22049               & 2017-10-29 & 0.5 & 104 & y & 1.60$\pm$0.05 & 8.4 & 10.5 & 11.6 & 12.6 & 14.2 & 14.4 \\
HD\,25457               & 2015-12-17 & 0.8 & 79  & y & 4.06$\pm$0.30 & 8.4 & 10.1 & 10.7 & 11.4 & 12.5 & 12.5 \\
HD\,27290               & 2016-12-12 & 0.6 & 100 & y & 3.37$\pm$0.55 & 8.0 & 9.8 & 10.9 & 11.8 & 13.0 & 13.0 \\
HD\,30422               & 2016-11-11 & 0.9 & 169 & y & 5.72$\pm$0.12 & 7.9 & 9.0 & 9.8 & 10.2 & 10.8 & 10.8 \\
HD\,30495               & 2016-12-11 & 1.9 & 48  & y & 3.89$\pm$0.20 & 6.3 & 8.6 & 9.8 & 10.3 & 11.4 & 11.3 \\
HD\,32297               & 2017-11-03 & 0.5 & 85  & n & 7.61$\pm$0.04 & 6.0 & 7.9 & 8.9 & 9.4 & 9.7 & 9.7 \\
HD\,35850               & 2015-12-15 & 0.8 & 96  & y & 4.87$\pm$0.19 & 7.1 & 9.1 & 10.2 & 11.0 & 11.7 & 11.6 \\
HD\,37484               & 2016-11-12 & 0.7 & 183 & y & 6.23$\pm$0.09 & 7.0 & 8.1 & 8.6 & 9.1 & 9.8 & 10.1 \\
SAO\,150676             & 2015-12-18 & 0.6 & 133 & n & 7.41$\pm$0.04 & 7.6 & 8.5 & 9.0 & 9.3 & 9.5 & 9.5 \\
HD\,38206               & 2016-11-08 & 0.6 & 138 & y & 5.78$\pm$0.15 & 8.1 & 9.2 & 9.7 & 10.3 & 10.6 & 10.7 \\
HD\,38678               & 2015-12-16 & 0.8 & 86  & y & 3.21$\pm$0.55 & 6.5 & 9.4 & 10.4 & 11.4 & 12.7 & 12.9 \\
HD\,43989               & 2015-12-17 & 0.7 & 79  & y & 6.44$\pm$0.08 & 7.1 & 8.9 & 9.5 & 9.8 & 10.2 & 10.3 \\
HD\,46190               & 2016-11-06 & 0.4 & 81  & y & 6.33$\pm$0.10 & 8.0 & 9.1 & 9.8 & 10.0 & 10.3 & 10.4 \\
HD\,48370               & 2018-03-02 & 0.5 & 65  & y & 6.25$\pm$0.09 & 7.4 & 8.8 & 9.4 & 9.7 & 10.2 & 10.3 \\
HD\,53143               & 2015-12-16 & 0.6 & 59  & y & 4.86$\pm$0.25 & 7.5 & 9.8 & 10.7 & 11.1 & 11.7 & 11.8 \\
HD\,59967               & 2015-12-15 & 0.8 & 96  & y & 5.01$\pm$0.23 & 7.9 & 9.5 & 10.4 & 10.8 & 11.3 & 11.4 \\
HD\,61005               & 2018-02-25 & 0.9 & 142 & n & 6.42$\pm$0.09 & 7.5 & 9.1 & 9.8 & 10.1 & 10.6 & 10.6 \\
HD\,69830               & 2017-03-14 & 0.5 & 86  & y & 3.99$\pm$0.38 & 7.3 & 9.9 & 10.7 & 11.0 & 11.9 & 12.1 \\
HD\,70573               & 2015-12-18 & 0.5 & 78  & n & 7.12$\pm$0.04 & 6.5 & 8.7 & 9.7 & 9.8 & 9.9 & 10.0 \\
HD\,71043               & 2018-02-26 & 0.8 & 76  & y & 5.82$\pm$0.15 & 6.8 & 8.6 & 9.5 & 10.1 & 10.6 & 10.7 \\
HD\,71155               & 2017-03-17 & 0.6 & 48  & y & 3.79$\pm$0.41 & 7.6 & 10.3 & 10.5 & 11.5 & 12.3 & 12.3 \\
HD\,72660               & 2016-02-17 & 1.7 & 53  & y & 5.76$\pm$0.13 & 5.7 & 8.2 & 9.3 & 9.9 & 10.2 & 10.3 \\
HD\,76151               & 2018-02-24 & 0.4 & 53  & y & 4.38$\pm$0.36 & 7.7 & 10.2 & 10.7 & 11.2 & 12.3 & 12.4 \\
HD\,92536               & 2016-02-18 & 1.3 & 70  & y & 6.46$\pm$0.08 & 6.3 & 8.5 & 8.3 & 9.0 & 10.0 & 10.0 \\
HD\,98058               & 2017-05-01 & 0.3 & 92  & y & 4.04$\pm$0.31 & 6.9 & 9.4 & 10.3 & 11.2 & 12.3 & 12.4 \\
HD\,102458              & 2017-05-03 & 0.6 & 97  & n & 7.41$\pm$0.04 & 6.7 & 8.4 & 8.9 & 8.9 & 9.4 & 9.4 \\
HD\,103703              & 2017-05-15 & 0.9 & 77  & n & 7.31$\pm$0.04 & 7.2 & 7.6 & 6.7 & 9.0 & 9.9 & 9.9 \\
MML\,8                  & 2018-02-23 & 0.6 & 110 & n & 8.07$\pm$0.03 & 7.3 & 8.8 & 9.2 & 9.4 & 9.6 & 9.6 \\
HD\,107146              & 2018-02-26 & 0.6 & 75  & y & 5.49$\pm$0.02 & 6.9 & 9.2 & 10.1 & 10.4 & 11.2 & 11.2 \\
HD\,107649              & 2018-02-25 & 0.4 & 112 & n & 7.60$\pm$0.06 & 7.9 & 9.1 & 9.6 & 9.8 & 10.1 & 10.1 \\
HD\,109085              & 2018-02-24 & 0.7 & 133 & y & 3.03$\pm$0.45 & 6.3 & 8.7 & 9.9 & 10.9 & 12.5 & 12.8 \\
HD\,110411              & 2016-02-18 & 1.7 & 66  & y & 4.52$\pm$0.32 & 7.7 & 9.8 & 10.2 & 11.0 & 11.7 & 11.7 \\
HD\,111520              & 2017-05-17 & 1.3 & 58  & n & 7.52$\pm$0.03 & 6.8 & 8.5 & 9.0 & 9.1 & 9.3 & 9.2 \\
HD\,111631              & 2018-03-01 & 0.5 & 92  & y & 4.81$\pm$0.26 & 8.8 & 10.2 & 10.8 & 11.3 & 12.1 & 12.3 \\
MML\,28                 & 2016-05-31 & 0.7 & 51  & n & 8.70$\pm$0.03 & 4.1 & 6.2 & 6.6 & 6.6 & 7.0 & 7.1 \\
HD\,113556              & 2016-06-02 & 0.7 & 77  & n & 7.20$\pm$0.04 & 6.3 & 8.3 & 9.0 & 9.4 & 9.5 & 9.6 \\
HD\,114082              & 2018-03-02 & 1.1 & 89  & n & 7.13$\pm$0.04 & 7.3 & 9.2 & 9.9 & 9.9 & 10.2 & 10.2 \\
HD\,115820              & 2017-05-02 & 1.4 & 74  & n & 7.29$\pm$0.04 & 6.0 & 8.2 & 8.6 & 8.7 & 9.1 & 9.1 \\
HD\,115892              & 2016-02-17 & 2.0 & 44  & y & 2.72$\pm$0.20 & 6.4 & 8.0 & 10.3 & 11.1 & 12.9 & 13.0 \\
MML\,36                 & 2017-06-17 & 0.9 & 102 & n & 7.80$\pm$0.04 & 6.9 & 8.6 & 8.9 & 9.3 & 9.4 & 9.5 \\
HD\,125541              & 2017-03-17 & 0.6 & 97  & n & 8.08$\pm$0.03 & 7.5 & 8.3 & 8.5 & 8.7 & 8.9 & 9.0 \\
HD\,126062              & 2017-03-14 & 0.6 & 93  & n & 7.32$\pm$0.40 & 6.8 & 8.8 & 8.9 & 9.1 & 9.5 & 9.4 \\
HD\,131511              & 2017-07-01 & 0.8 & 58  & y & 3.77$\pm$0.47 & 6.9 & 9.6 & 10.6 & 11.3 & 12.4 & 12.4 \\
HD\,131835              & 2016-05-03 & 0.4 & 111 & n & 7.52$\pm$0.04 & 7.4 & 9.0 & 9.3 & 9.4 & 9.7 & 9.8 \\
HD\,134888              & 2017-05-03 & 0.6 & 149 & n & 7.62$\pm$0.04 & 7.1 & 8.9 & 9.1 & 9.1 & 9.4 & 9.5 \\
GJ\,581                 & 2017-07-13 & 0.8 & 107 & y & 5.62$\pm$0.10 & 7.7 & 9.7 & 10.4 & 10.6 & 11.1 & 11.2 \\
HD\,156623              & 2017-05-02 & 2.0 & 84  & n & 6.95$\pm$0.06 & 6.5 & 9.1 & 9.5 & 9.7 & 9.9 & 9.9 \\
HD\,172555              & 2017-05-16 & 1.4 & 87  & y & 4.23$\pm$0.30 & 7.7 & 10.2 & 11.1 & 11.5 & 12.3 & 12.4 \\
HD\,177171              & 2016-05-31 & 1.3 & 63  & y & 3.67$\pm$0.44 & 7.2 & 9.2 & 10.2 & 11.0 & 12.1 & 12.0 \\
HD\,181296              & 2017-05-01 & 1.2 & 78  & y & 4.97$\pm$0.16 & 7.4 & 9.4 & 10.2 & 10.7 & 11.2 & 11.2 \\
HD\,181327              & 2016-07-31 & 0.9 & 84  & y & 5.86$\pm$0.12 & 7.7 & 9.3 & 9.7 & 9.9 & 10.7 & 10.6 \\
HD\,191089              & 2016-08-01 & 0.6 & 184 & y & 5.99$\pm$0.11 & 7.7 & 8.8 & 9.3 & 9.8 & 10.3 & 10.4 \\
HD\,193571              & 2016-05-30 & 0.7 & 78  & y & 5.55$\pm$0.15 & 5.3 & 7.9 & 8.9 & 9.1 & 9.7 & 9.8 \\
HD\,196544              & 2017-08-29 & 0.5 & 60  & y & 5.25$\pm$0.16 & 7.4 & 9.1 & 9.4 & 10.0 & 10.6 & 10.8 \\
HD\,197481              & 2017-07-13 & 0.8 & 131 & y & 4.28$\pm$0.32 & 6.2 & 9.6 & 10.6 & 11.1 & 11.9 & 12.1 \\
HD\,202628              & 2017-08-31 & 0.6 & 102 & y & 5.20$\pm$0.15 & 8.2 & 9.5 & 10.1 & 10.3 & 10.9 & 11.1 \\
HD\,206860              & 2017-06-17 & 0.7 & 47  & y & 4.43$\pm$0.26 & 6.8 & 9.4 & 10.1 & 10.6 & 11.7 & 11.8 \\
NLTT\,54872             & 2017-07-13 & 1.0 & 171 & n & 6.87$\pm$0.05 & 7.1 & 8.3 & 8.6 & 8.8 & 9.3 & 9.0 \\
HD\,218340              & 2016-07-30 & 0.8 & 91  & n & 6.94$\pm$0.07 & 6.9 & 8.5 & 9.4 & 9.9 & 9.9 & 10.1 \\
HD\,218511              & 2017-07-01 & 1.3 & 44  & y & 5.23$\pm$0.20 & 7.3 & 8.9 & 9.7 & 10.2 & 10.8 & 10.9 \\
HD\,220825              & 2017-11-03 & 0.8 & 53  & y & 4.83$\pm$0.29 & 6.9 & 9.7 & 10.5 & 10.9 & 11.6 & 11.6 \\
HD\,223340              & 2017-10-30 & 0.7 & 132 & y & 7.04$\pm$0.04 & 5.9 & 7.7 & 8.3 & 8.8 & 9.1 & 9.2 \\
\end{longtable}
\mbox{\tablefoottext{a}{$L^{\prime}$\ magnitudes and uncertainties are derived by interpolating between WISE bands W1 (\SI[]{3.35}{\micro\meter}) and W2 (\SI[]{4.6}{\micro\meter}) to \SI[]{3.8}{\micro\meter}.}}
\tablefoottext{b}{Corrected for AGPM throughput where applicable (see Sect.\,\ref{sec:obs}).}

\end{document}